# The Solar Wind Around Pluto (SWAP) Instrument Aboard *New Horizons*

D. McComas[1,*], F. Allegrini[1], F. Bagenal[2], P. Casey[1], P. Delamere[2], D. Demkee[1], G. Dunn[1], H. Elliott[1], J. Hanley[1], K. Johnson[1], J. Langle[3,1], G. Miller[1], S. Pope[1], M. Reno[1], B. Rodriguez[1], N. Schwadron[4,1], P. Valek[1], S. Weidner[1]

[1]*Southwest Research Institute®, San Antonio, Texas 78228-0510, U.S.A.*

[2]*University of Colorado, Boulder, Colorado 80309, U.S.A.*

[3]*Midwest Research Institute, Kansas City Missouri 64110, U.S.A.*

[4]*Boston University, Boston Massachusetts 02215, U.S.A.*

*( [*]author for correspondence, e-mail: DMcComas@swri.edu)*

Abstract. The Solar Wind Around Pluto (SWAP) instrument on *New Horizons* will measure the interaction between the solar wind and ions created by atmospheric loss from Pluto. These measurements provide a characterization of the total loss rate and allow us to examine the complex plasma interactions at Pluto for the first time. Constrained to fit within minimal resources, SWAP is optimized to make plasma-ion measurements at all rotation angles as the *New Horizons* spacecraft scans to image Pluto and Charon during the flyby. In order to meet these unique requirements, we combined a cylindrically symmetric retarding potential analyzer (RPA) with small deflectors, a top-hat analyzer, and a redundant/coincidence detection scheme. This configuration allows for highly sensitive measurements and a controllable energy passband at all scan angles of the spacecraft.

*Keywords: solar wind, Pluto plasma interaction, pickup ions*

## 1.0 Introduction

The *New Horizons* mission [Stern et al., 2007] will make the first up-close and detailed observations of Pluto and its moons. These observations include measurements of the solar wind interaction with Pluto. The Solar Wind Around Pluto (SWAP) instrument is designed to measure the tenuous solar wind out at ~32 AU and its interaction with Pluto. SWAP directly addresses the Group 1 science objective for the *New Horizons* mission to “characterize the neutral atmosphere of Pluto and its escape rate.” In addition, SWAP makes measurements critical for the Group 2 objective of characterizing Pluto's ionosphere and solar wind interaction. Finally, SWAP makes observations relevant to two Group 3 science objectives for *New Horizons*, both in support of characterizing the energetic particle environment of Pluto and Charon and in searching for magnetic fields of Pluto and Charon.

As the solar wind approaches Pluto, it interacts with ions produced when Pluto's thin upper atmosphere escapes and streams away as neutral particles that subsequently become ionized. This pickup process slows the solar wind, and the type of interaction varies greatly depending on the atmospheric escape rate, all the way from Comet-like for larger escape rates to Venus-like for low escape rates. SWAP measurements should provide the best estimate of the overall atmospheric escape rate at Pluto and allow the first detailed examination of its plasma interactions with the solar wind.

The SWAP observations are extremely challenging because the solar wind flux, which falls off roughly as the square with heliocentric distance, is approximately three orders of magnitude lower at Pluto compared to typical solar wind fluxes observed near Earth's orbit. In addition, because the solar wind continues to cool as it propagates out through the heliosphere, the solar wind beam becomes narrow in both angle and energy.

The SWAP design was strongly driven by three constraints: 1) very low use of spacecraft resources (mass, power, telemetry, etc.) as this is a non-core instrument on a relatively small planetary mission; 2) very high sensitivity to measure the solar wind and its interaction with Pluto out at ~32 AU, where the density is down by a factor of ~1000 compared to 1 AU; and 3) the need to make observations over a very large range of angles as the spacecraft constantly repoints its body-mounted cameras during the flyby. Given these not-entirely-consistent design drivers, we developed an entirely new design that combines elements of several different previous plasma instruments. Together these components comprise the SWAP instrument, which will measure the speed, density, and temperature of the distant solar wind and its interaction with Pluto.

## 2.0 Scientific Background and Objectives

Only partly in jest, Dessler and Russell (1980) suggested that Pluto might act like a colossal comet. The 1988 stellar occultation showed that Pluto's tenuous atmosphere could indeed be escaping (Hubbard et al., 1988; Elliot et al., 1989). Applying basic cometary theories to Pluto, Bagenal and McNutt (1989) showed that photoionization of escaping neutral molecules from Pluto's atmosphere could significantly alter the solar wind flow around Pluto for sufficiently large escape rates. For large atmospheric escape rates, the interaction may be best described as "comet-like," with significant mass-loading over an extensive region; for small escape rates the interaction is probably confined to a much smaller region, creating a more "Venus-like" interaction (Luhmann et al., 1991), where electrical currents in the gravitationally bound ionosphere deflect the solar wind flow. Figure 1 compares these two types of interactions schematically. At aphelion (50 AU), should Pluto's atmosphere completely collapse and freeze onto the surface, then the interaction becomes "Moon-like" with the solar wind suffering minimal deflection and directly bombarding the bare, icy dayside surface. Not having a detectable atmosphere, Charon almost certainly has such a "Moon-like" interaction, remaining primarily in the solar wind if Pluto's interaction is weak but becoming totally engulfed if Pluto's interaction is strongly "comet-like" and extends beyond Charon's orbit at 17 $R_P$ (Pluto radii, ~1150 km). For a review of early studies of the solar wind interaction with Pluto see Bagenal et al. (1997), which also reviews the implications of the unlikely possibility of Pluto having an intrinsic magnetic field.

The solar wind is supersonic so that when the flow impinges on a magnetic obstacle (such as the magnetosphere of the Earth or other planet) an upstream bow shock must form to slow and deflect the supersonic (actually superfast-mode magnetosonic) plasma. The weak interplanetary magnetic field (IMF) at 30 AU (see Table I for typical solar wind properties near Pluto) and heavy ions formed by photoionizing the heavy molecules of Pluto's escaping atmosphere have very large gyroradi (~500 $R_P$). The net results of these non-fluid or kinetic effects are to make the bow shock a thick transition region and to make the shape of the interaction region asymmetric where the direction of asymmetry is governed by the direction of the IMF. Recent simulations of the solar wind interaction with a strongly escaping atmosphere have necessarily been 3D and have either taken a multi-fluid approach (solar wind proton fluid and pick-up ion fluid) or a hybrid approach (electron fluid, ion particles) (Harnett et al. 2005; Delamere and Bagenal, 2004).

**Table I. Typical Solar wind Interaction Properties at Pluto (at 30 AU), taken from Bagenal et al. (1997) using Rp=1150 km.**

| | |
|---|---|
| Magnetic field strength | 0.2 nT |
| Proton density | 0.01 $cm^{-3}$ |
| Solar wind speed | 450 km/s |
| Proton temperature | 1.3 eV |
| Alfvén Mach number | ~45 |
| Sonic Mach number | ~40 |
| Proton gyroradius | 23,000 km (~20 *Rp*) |
| $N_2$+ gyroradius | 658,000 km (~550 *Rp*) |
| Ion inertial length | 2280 km (~2 *Rp*) |
| Electron inertial length | 53 km |

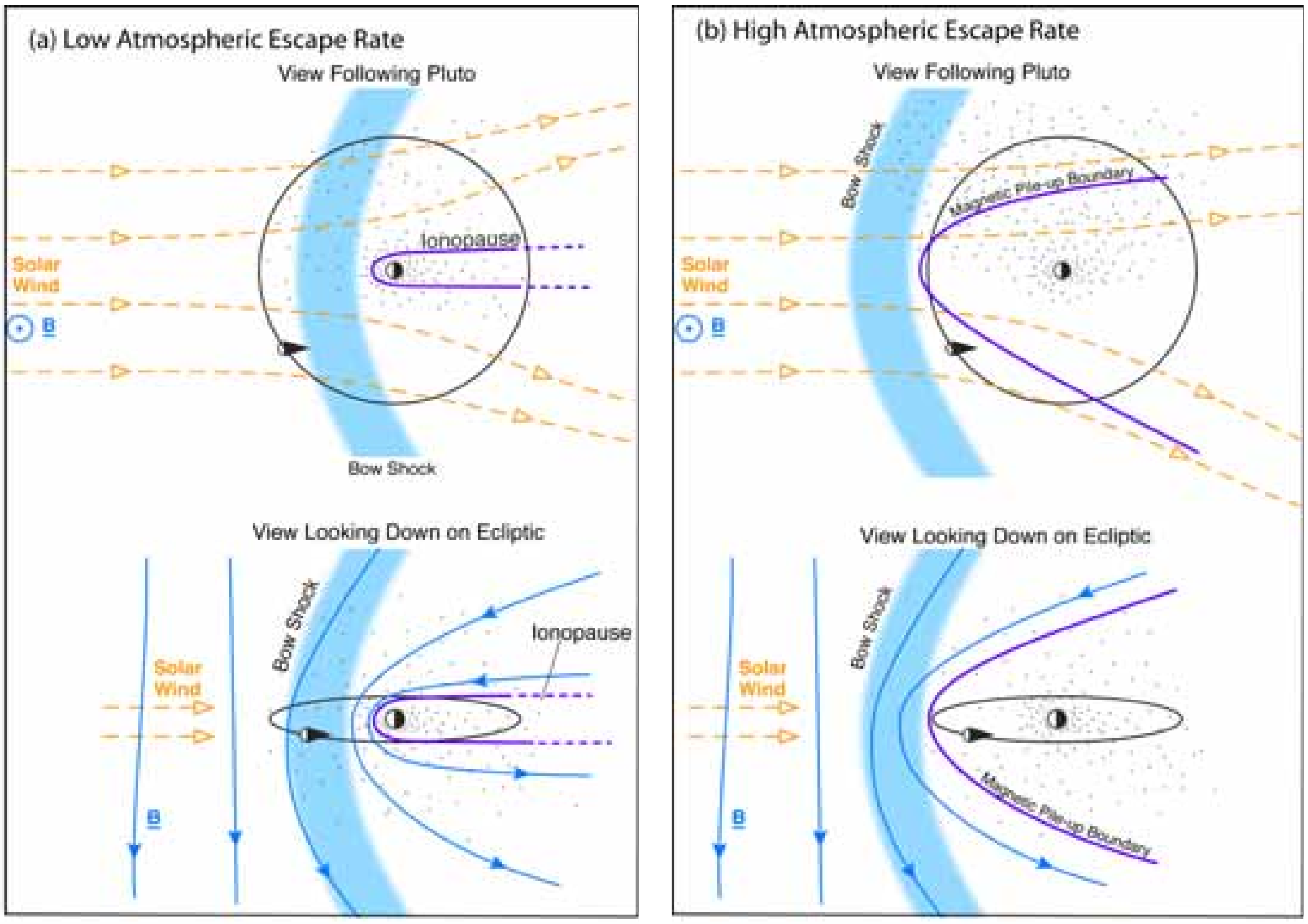

Figure 1. Solar wind interaction with Pluto for (a) low atmospheric escape rate and (b) high atmospheric escape rate. The dots indicate ions produced by the photoionization of Pluto’s escaping atmosphere. These pick-up ions will move perpendicular to both the magnetic field (B)

and solar wind (upwards in the top two diagrams). Note that while the IMF tends to be close to tangential to Sun direction, the sign of the direction varies on timescales of days. The asymmetries of the interaction will flip as the magnetic field changes direction.

Below we discuss how the current understanding of Pluto's atmosphere leads us to expect a more comet-like interaction at the time of the *New Horizons* flyby in 2015. Through measurements of bulk properties of the solar wind (flow, density, temperature) as well as the energy distribution of solar wind and pick-up ions, the SWAP instrument will not only characterize the solar wind interaction with Pluto but will also allow us to determine the global rate of atmospheric escape.

In the comet-like scenario, variations on the scale of the interaction region can be substantial over periods of days, and a factor of ~10 variations in the solar wind flux can change the size of the interaction region from a few to more than 20 $R_P$. It is therefore critical to measure the solar wind for several solar rotations (~26 days per rotation) before and after the flyby in order to characterize the most likely external solar wind properties during the actual encounter period. Furthermore, since the strong asymmetry of the interaction depends on the direction of the IMF, our analysis of SWAP data will need assistance from increasingly capable models of solar wind structure based on plasma and magnetometer data from spacecraft elsewhere in the solar system.

## *2.1 Atmospheric Escape*

The exact nature of Pluto's plasma interaction is critically dependent on the hydrodynamic escape rate of the atmosphere from its weak gravity. Escaping neutrals are photoionized by solar UV (or, less likely, suffer an ionizing collision). Freshly ionized particles experience an electric field due to their motion relative to the IMF (that is carried away from the Sun by the solar wind) and are accelerated by this motional electric field, extracting momentum from the solar wind flow. This electrodynamic interaction modifies the solar wind flow. Estimates of Pluto's atmospheric escape rate, Q, vary substantially: McNutt (1989) estimated 2.3-5.5 x $10^{27}$ $s^{-1}$ for $CH_4$-dominated outflow, Krasnopolsky (1999) found a hydrodynamic outflow of $N_2$ of 2.0-2.6 x $10^{27}$ $s^{-1}$, while Tian and Toon (2005) derived values for $N_2$ escape as high as 2 x $10^{28}$ $s^{-1}$. The nature of the plasma interaction varies considerably over this range of escape rates since the scale of the interaction is proportional to Q (Bagenal and McNutt, 1989).

Hydrodynamic escape of an atmosphere occurs when the atmospheric gases in the upper atmosphere (in the vicinity of the exobase) are heated significantly (so that the thermal speed is comparable to the local sound speed). Krasnopolsky and Cruikshank (1999) reviewed the photochemistry of Pluto's atmosphere while Krasnopolsky (1999) summarized approaches taken in modeling the complexities of hydrodynamic escape at Pluto. Earlier models approximated all the heating that occurs in a thin layer of the atmosphere. Recently, however, Tian and Toon (2005) developed a model that includes a distributed heating function appropriate for EUV absorption by the dominant molecule $N_2$ relatively high in Pluto's atmosphere, which leads to larger escape rates. These authors derive an exobase height of 10-13 $R_P$ and transonic point of ~30 $R_P$. The fact that even with significant heating the escape speed (<100 m/s) is subsonic at the exobase is consistent with what Krasnopolsky (1999) called "slow hydrodynamic escape." Nevertheless, an exobase at 10-13 $R_P$ implies a very extended atmosphere, and the present *New Horizons* trajectory with a currently planned closest approach of ~9 $R_P$, may well briefly dip below the exobase.

Tian and Toon (2005) modeled the effects of variations in (i) EUV flux over the solar cycle, and (ii) Pluto's distance from the Sun, in order to estimate the variability of atmospheric escape over the full Pluto orbit. They found escape rates varying from 2 x $10^{28}$ $s^{-1}$ (for Pluto at 30 AU and solar maximum activity) to 1 x $10^{28}$ $s^{-1}$(for Pluto at 40 AU and solar minimum). These authors, however, were not able to find a stable solution for atmospheric escape at the low heating levels appropriate for aphelion (50 AU). Whether Pluto retains a stable atmosphere through aphelion or not remains an open issue. Recent occultation measurements suggest that Pluto's atmosphere has been expanding rather than collapsing as Pluto recedes from the Sun (Elliot et al., 2003). The possibility that the atmospheric pressure could be increasing on Pluto even after perihelion was predicted [Stern et al.,1988; Hansen and Paige, 1996]. These authors suggested that there would be a phase lag between perihelion and the maximum atmospheric pressure. There is also the possibility that the polar obliquity is now putting some new, fresh frost into sunlight, further driving atmospheric expansion.

Assuming the current estimates for the escaping atmosphere have persisted over the age of the solar system, Pluto could have lost hundreds of meters to kilometers of material to space via "escape erosion." In this case, the ancient topography made of volatiles ($N_2$ and CO particularly, but also $CH_4$) would have escaped to space. This fantastic possibility appears to be something wholly unique to Pluto, since even Triton is in the Jeans escape regime. Thus, escape erosion might have caused Pluto to have a young surface, and if so, imaging from New Horizons will give us information on the recent Kuiper Belt impactor size-frequency distribution, in contrast to the time-integral distribution provided by imaging of Charon, where such erosion is not occurring.

## *2.2 Solar wind Interaction at High Atmospheric Escape*

Galeev et al. (1985) derived a size for the interaction region (over which significant momentum would be extracted from the solar wind) for comets that is proportional to Q and inversely proportional to the solar wind flux, $n_{sw}V_{sw}$. Bagenal and McNutt (1989) applied this simple scaling law to Pluto and found that for typical solar wind conditions and photoionization of $CH_4$ or $N_2$, the scale size of the dayside interaction region, $R_{SO}$ (the "stand-off" distance) is $R_{SO}/R_P = Q/Q_0$) where $Q_0$ is 1.5 x $10^{27}$ $s^{-1}$. Applying the upper range of atmospheric escape rates discussed above, one finds $R_{SO}$ = 6-13 $R_P$ for 1-2 x $10^{27}$ $s^{-1}$. (assuming an escape speed of 100 m/s). Note that these simple calculations assume the exobase is close to the planet. Tian and Toon (2005) suggest that the exobase may be as high as 10-13 Rp, in which case the standoff distance would be expanded accordingly. Nevertheless, the linear dependence on solar wind flux means that larger (factor of ~10) variations in solar wind density would produce a similar variation in $R_{SO}$, often extending the interaction region beyond the orbit of Charon at 17 $R_P$.

The above simple scaling is based on a fluid approach that is appropriate for cometary interactions in the inner solar system. The weak IMF at 30 AU (see Table I) means that ion kinetic effects are crucial at Pluto due to the large pickup-ion gyroradius (Kecskemety and Cravens, 1993) and the large turning distance for solar wind protons (Bagenal et al., 1997).

Previous models of weakly outgassing (i.e. production rates, $Q \sim 10^{27}$ $s^{-1}$) bodies used two-dimensional, two-ion fluid and/or hybrid simulations (Bogdanov et al., 1996; Sauer et al., 1997; Hopcroft and Chapman, 2001). All of these models showed the formation of asymmetric plasma structures. Sauer et al. (1997) specifically investigated the Pluto plasma interaction with two-dimensional models for $Q > 10^{27}$ $s^{-1}$. While the two-dimensional models provide a good qualitative description of the interaction, the quantitative details of the plasma coupling (i.e. momentum transfer) require three-dimensional models. More recently, Lipatov et al. (2002) performed a three-dimensional hybrid code simulation of the solar wind interaction with weak comets and illustrated the dependence of gas production rates ($6.2 \times 10^{26}$ $s^{-1} < Q < 10^{28}$ $s^{-1}$) on plasma structure. However, these results are applicable to 1 AU where the ions are more strongly magnetized compared to the situation at 30 AU where the IMF is very weak and the solar wind is tenuous.

Delamere and Bagenal (2004) modeled the kinetic interaction of the solar wind with Pluto's escaping atmosphere using a hybrid simulation that treats the pick-up ions and solar wind protons as particles and the electrons as a massless fluid. A hybrid code is a reasonable approach for a system with scale sizes ranging from the ion inertial length of the solar wind protons (~2 $R_P$) to the pick-up ion ($N_2^+$) gyroradius (~500 $R_P$). Figure 2 shows a schematic of the ion motion near Pluto in this kinetic regime. Note the asymmetry of the interaction imposed by the direction of the magnetic field in the solar wind ($B_{sw}$).

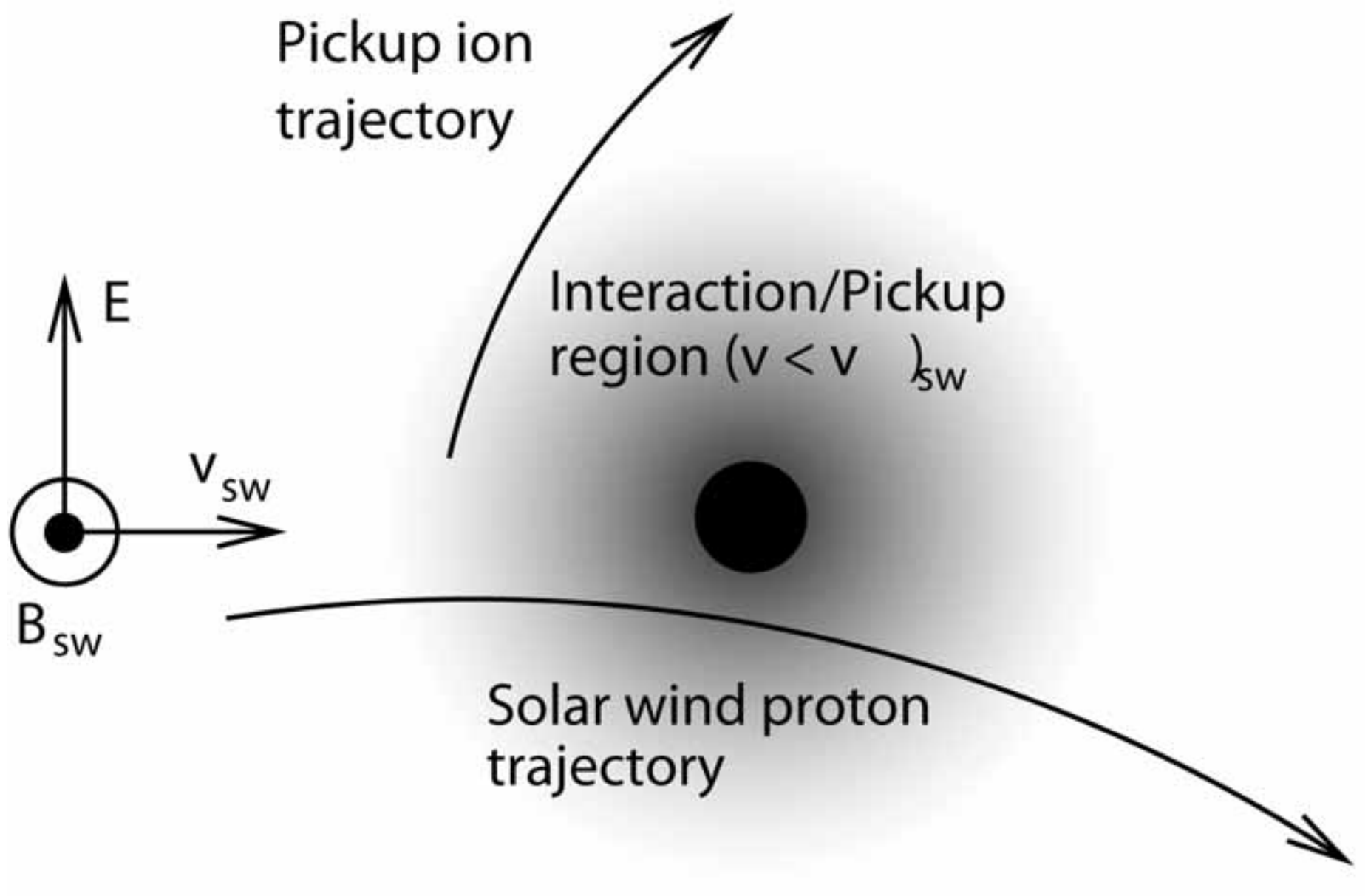

Figure 2. Schematic of ion motion near Pluto's interaction region. Pickup ions move initially in the direction of the solar wind convection electric field (+*y*) and the solar wind flow is deflected in the -*y* direction, consistent with momentum conservation.

Figures 3, 4 and 5 show the density, temperature, and bulk flow resulting from hybrid simulations for an atmospheric escape rate of Q = 1 x $10^{27}$ $s^{-1}$ |B| = 0.2 nT and an atmospheric escape speed of 100 m/s (Delamere and Bagenal, 2004, 2007) in the same geometry depicted in Figure 2 (upstream solar wind flow from the left and IMF pointing out of the page). The main features of the interaction are a broad region of proton heating extending ~20 $R_P$ upstream of Pluto consistent with the expected location of a bow shock. The kinetic energy of the solar wind protons is not fully converted to thermal energy, and the shock structure forms only in the upper half of the interaction region. In the lower half of the interaction region the protons are compressed (illustrated by higher density in Figure 3, higher temperature in Figure 4, slightly reduced and deflected flow in Figure 5) behind what is effectively a magnetic pile-up boundary. Since the gyroradii of the pick-up ions is huge compared with the scale of the interaction, they effectively move upwards in the geometry illustrated in these figures. The density of pick-up ions is relatively small but, because of their high mass, the momentum they extract from the solar wind is significant. The solar wind is slowed down (leading to compression and enhanced magnetic field) and deflected in the opposite direction to the pick-up ions. Superimposed on the density contours in Figure 5 are trajectories of solar wind protons. The large gyrations reflect the weak

magnetic field's inability to turn the protons. The smaller gyrations indicate stronger magnetic fields in the pile-up region. Harnett et al. (2005) showed that multi-fluid simulations produced qualitatively similar, though less pronounced, asymmetries in the interaction region.

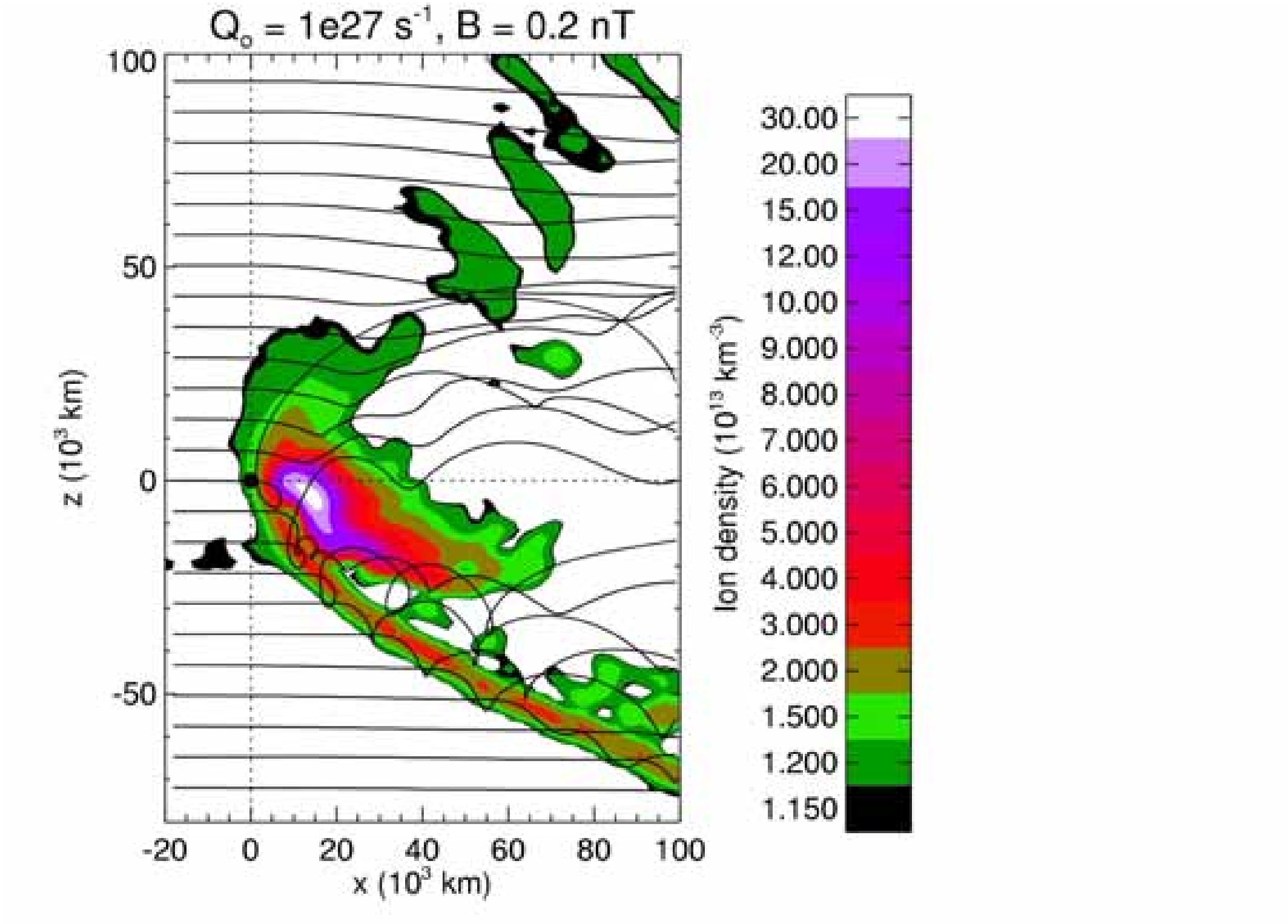

Figure 3. Sample trajectories of solar wind protons near Pluto (origin of plot) through the interaction region with total ion density ($N_2^+$ plus solar wind proton). The trajectories are altered at the magnetic pileup boundary, but only partially thermalized as illustrated in Figure 4. The magnetic pileup boundary is very abrupt in the -z half plane.

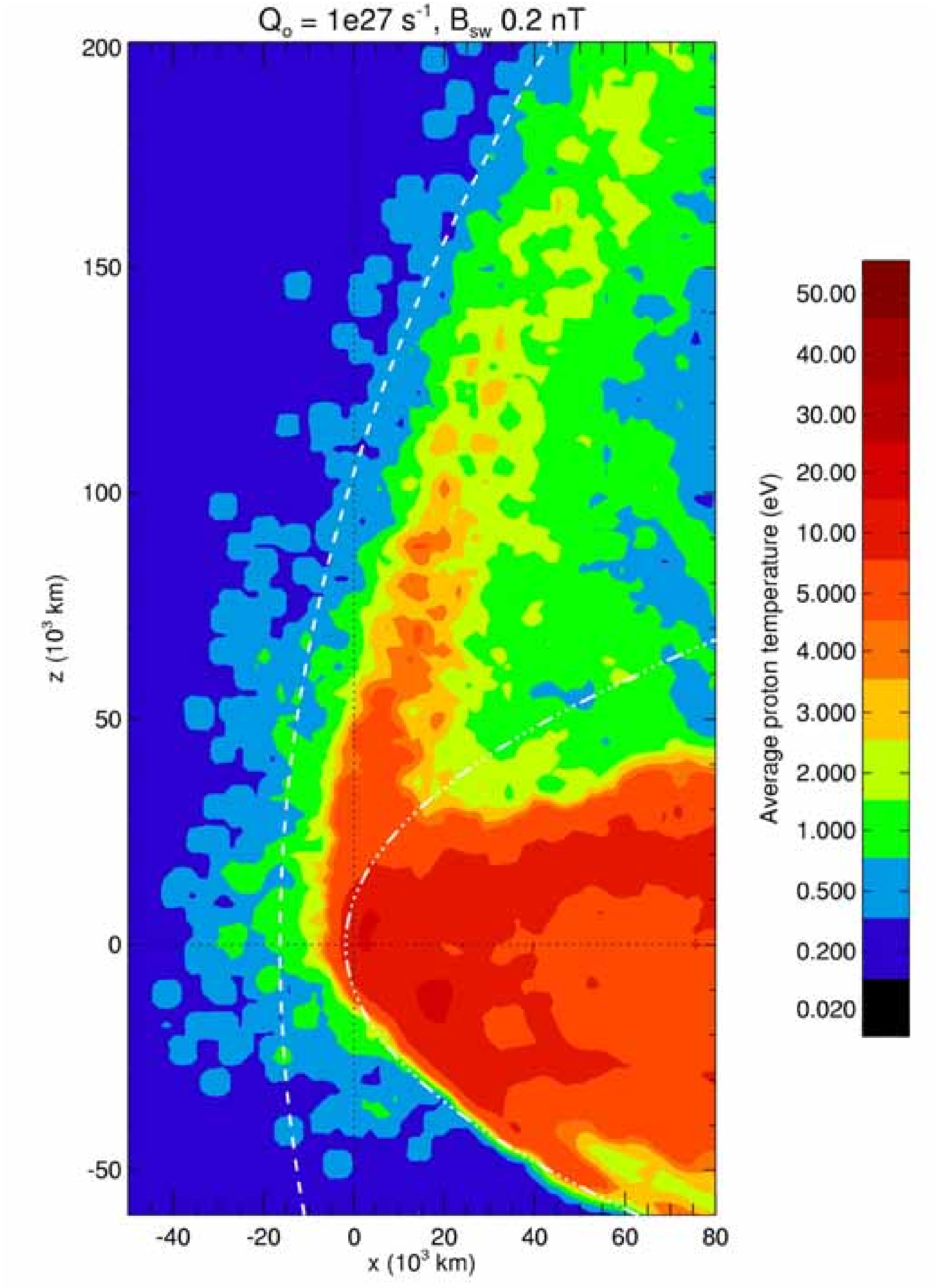

Figure 4. Column averaged proton temperature in the plane perpendicular to $B_{sw}$. The maximum temperature for the solar wind protons (i.e. fully thermalized) moving initially at 340 km/s is 360 eV. The interaction only partially thermalizes the solar wind with maximum temperatures of a few 10s of eV. The dashed line is the expected location (i.e. based on upstream standoff distance) of the bow shock and the dash-dot-dot-dot line is the expected location of the magnetic pileup boundary based on calculations for solar wind stagnation near comets for $Q = 10^{27}$ $s^{-1}$ and $B_{sw}$ = 0.2 nT (Biermann (1967), Galeev et al., (1985)). Given the extreme asymmetry of the interaction, it is not possible to fit the bow shock and the magnetic pileup boundary with symmetric parabolic functions. We based our fit for the bow shock on the +z half plane and the fit for the magnetic pileup boundary on the -z half plane.

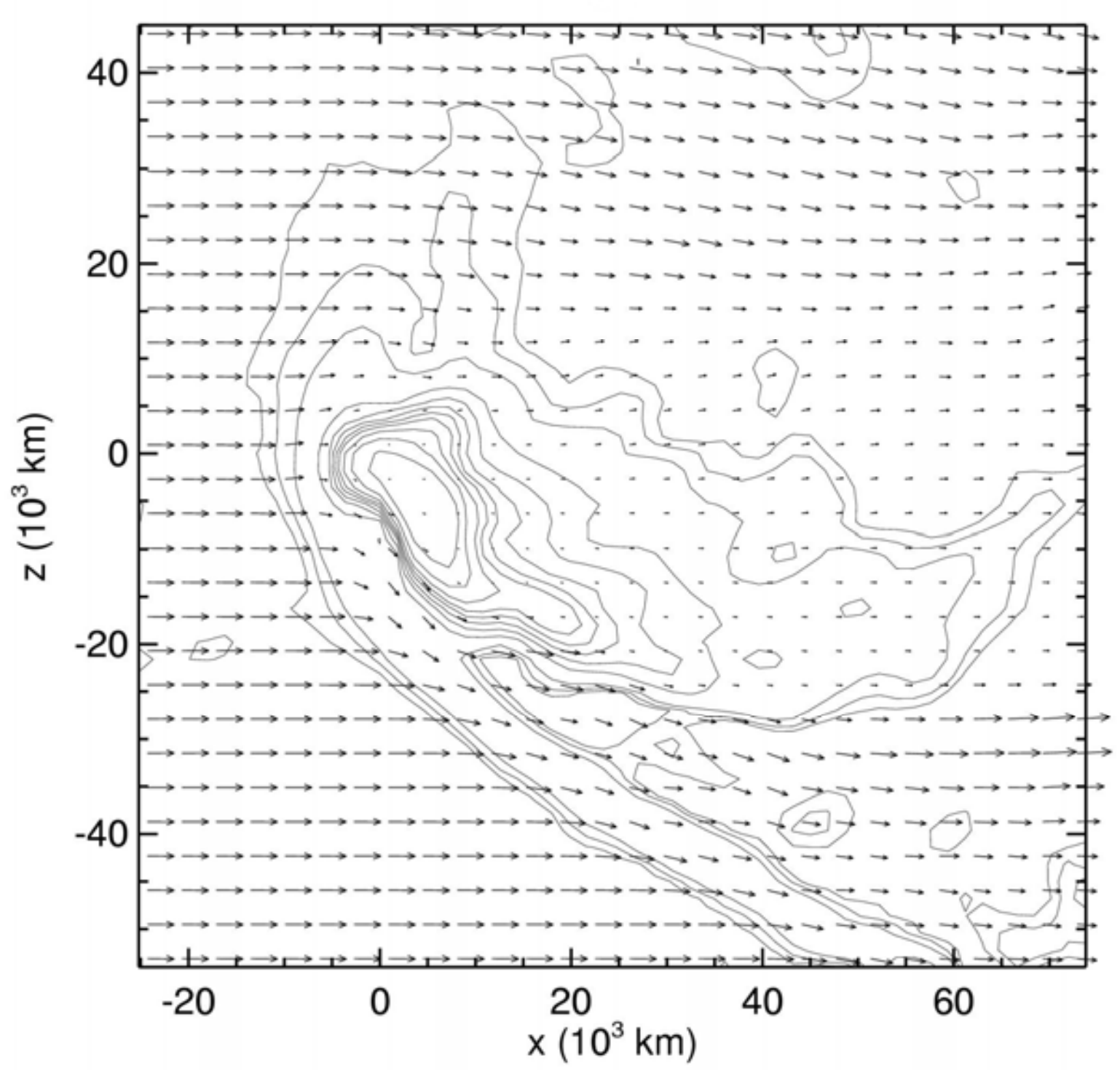

Figure 5. Ion bulk flow velocity (pickup ions plus solar wind protons) with contours of total ion density illustrating the high degree of asymmetry in the plasma flow.

Clearly, the upstream direction and magnitude of the IMF are critical drivers for the solar wind interaction with Pluto. Unfortunately, *New Horizons* did not have the resources to carry a magnetometer, so we will need to use our best knowledge of solar wind physics to infer the magnetic properties. Generally speaking, the IMF becomes increasingly tightly wound spirals on conical surfaces (for different latitudes) with increasing distance from the Sun. Thus, we expect the average field direction to be nearly perpendicular to the flow by 32 AU. Because of the alternating sector structure of the IMF, it is equally likely that the IMF will be pointed in either direction. Measurements of pick-up ions, either by SWAP or by PEPSSI [McNutt et al., 2007] will provide the critical indication of the sign of the magnetic field at the time of the encounter.

## 2.3 Solar wind Interaction at Low Atmospheric Escape

If Pluto's escape rate is less than ~$10^{27}$ $s^{-1}$, with few pick-up ions to slow it down, then the solar wind is expected to penetrate close to Pluto and impinge directly onto its ionosphere. Strictly speaking, the "obstacle" that deflects the solar wind is this interaction produced by the electrical currents induced in the ionosphere. Such an interaction would be similar to the solar wind interaction with Venus or the Saturnian magnetospheric interaction with Titan (Luhmann et al. 1991). The photochemistry of Pluto's upper atmosphere and ionosphere modeled by both Krasnopolsky and Cruikshank (1999) and by Ip et al. (2000) show peak ionospheric densities of a few x $10^3$ $cm^{-3}$ at altitudes of ~1000 km above the surface of Pluto. They find the main ionospheric ion to be $H_2CN^+$.

The solar wind deflection and any pick-up ions will be significantly harder to measure on the *New Horizons* flyby in the case of a Venus-like interaction. However, the broad nature of the bow shock and the large gyroradii of any ions beyond the ionosphere should still produce detectable signatures several radii away from Pluto.

## *2.4 Solar wind Interaction at Aphelion*

It is not clear whether the atmosphere of Pluto completely collapses onto the surface when Pluto reaches aphelion. In the absence of a significant atmosphere Pluto will have a "Moon-like" interaction, as is expected for its moon Charon. In this type of interaction, the solar wind is absorbed on the sunlit side and the IMF diffuses through the non-conducting bodies, generating an extremely hard vacuum in the cavity behind. It seems unlikely that at temperatures below 40K Pluto's icy outer layers could be electrically conducting. But should they have significant conductivity, the plasma interaction may be similar to the solar wind/asteroid interaction described by Wang and Kivelson (1996), Omidi et al. (2002), and Blanco-Cano et al. (2003). If the dimensions of the obstacle are small compared to the ion inertial length, then the interaction is mediated by the whistler mode. Comparison of Galileo observations of asteroid-associated perturbations with numerical models confirm the whistler-mode interaction. Magnetohydrodynamic shock waves are absent with small obstacles as compressional waves can only exist on size scales larger than the ion inertial length. Pluto represents a possible intermediate case if the interaction region is limited to an area close to the planet as the solar wind proton inertial length is roughly 2 $R_P$.

### *2.5 Heliospheric Pickup Protons*

In addition to the primary science of measuring the solar wind interaction with Pluto, SWAP may afford an excellent opportunity to measure heliospheric pickup protons on its way out through the heliosphere, *en route* to Pluto (and beyond).

The solar system moves continually through the local interstellar medium (LISM) – the part of the galaxy nearest to our solar system – causing a complex set of interactions between the outflowing solar wind and the matter in the LISM (see Figure 6). These interstellar interactions set up the plasma boundaries of our heliosphere: the termination shock where the solar wind abruptly slows, becoming subsonic, prior to bending back into an extended tail in the interstellar downstream direction; the heliopause separating the subsonic solar wind from the ionized plasma of the interstellar medium; and possibly a bow shock where the LISM plasma becomes subsonic prior to being deflected around the heliosphere.

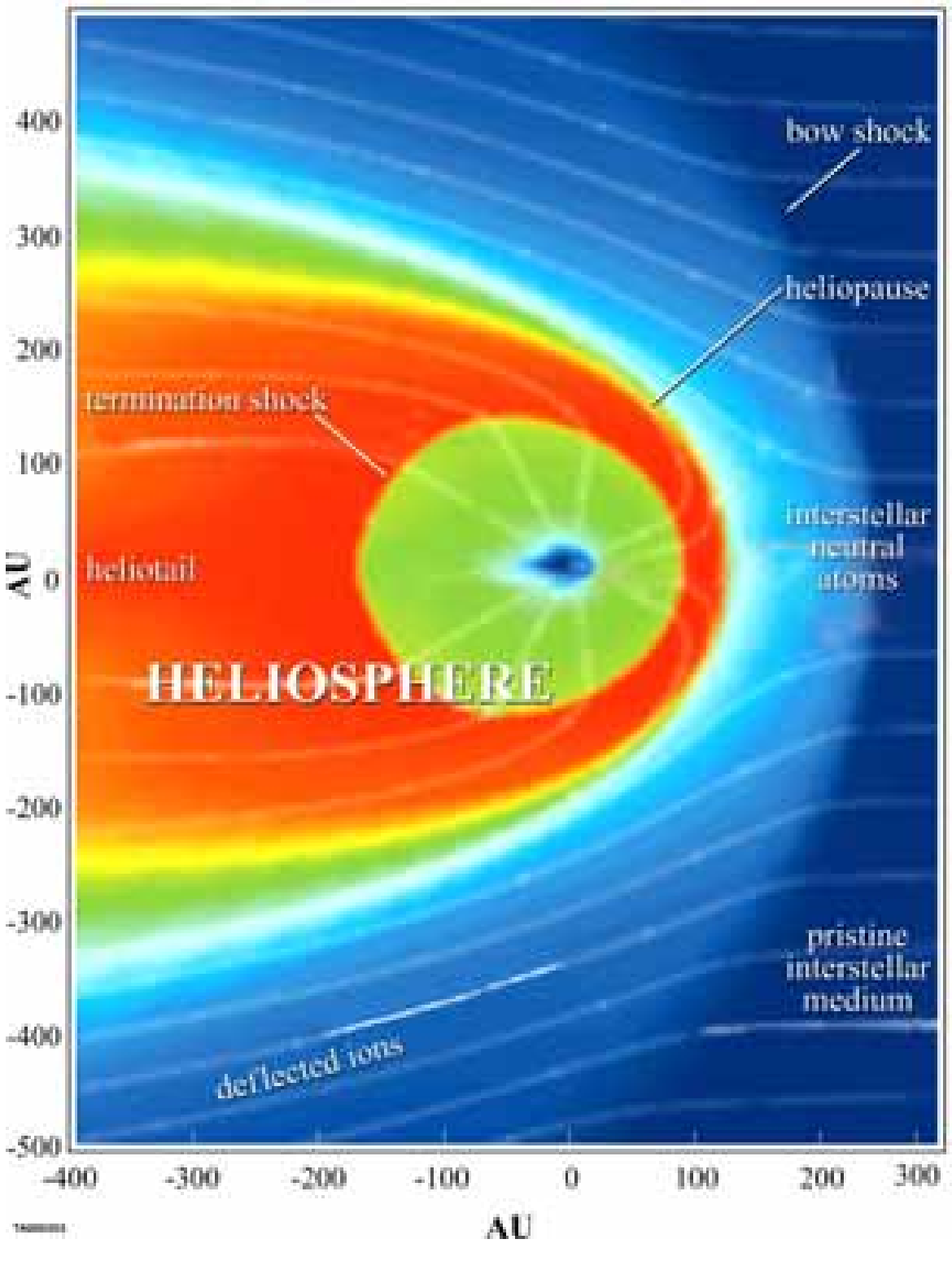

Figure 6. The SWAP instrument will measure protons produced from neutral matter that drifts into the heliosphere from the LISM.

Interstellar atoms, predominantly H, continually drift into and through the heliosphere, and due to their neutrality are unimpeded by the solar wind. The H atoms on trajectories toward the Sun move into regions of increasingly dense solar wind and higher levels of solar radiation, enhancing the probability of ionization through charge-exchange or photo-ionization. The vast majority of interstellar H atoms that penetrate inside of 4 AU become ionized and incorporated into the solar wind flow. Upon ionization, newborn ions begin essentially at rest in the reference frame of the spacecraft and Sun. Just like for planetary pickup ions, as described above, because the moving solar wind carries frozen-in magnetic field lines, the newborn ions encounter a motional electric field ($-\mathbf{v}_{sw} \times \mathbf{B}$) exerted by the solar wind and become picked-up and carried outward with the solar wind. As they move outward, the pickup ions also scatter due to magnetic inhomogeneities in the IMF, forming an almost spherical ring distribution in velocity space. As measured by spacecraft, these interstellar pickup-ion distributions are essentially flat for speeds less than 2 $v_{sw}$, then drop off sharply at speeds above this limit (Gloeckler et al., 1995).

The first interstellar pickup ions discovered (Möbius et al., 1985) were He+ created by ionization of interstellar neutral He that penetrated within 1 AU of the Sun. The composition and velocity space-resolved measurements by the SWICS experiment on Ulysses (Gloeckler et al., 1992) made it possible to explore pickup ions from 1.35-5.4 AU in great detail. The review by Gloeckler and Geiss (1998) provides an excellent summary of these pickup-ion observations. The observations include the most abundant pickup ion, H+, second most abundant, He+, and several other interstellar pickup-ion species, N+, O+, and Ne+ (Geiss et al., 1994). The SWICS team demonstrated the existence of pickup distributions of He++, which is produced largely by double charge exchange of atomic He with solar wind alpha particles (Gloeckler et al., 1997) and rare $^3$He+ pickup ions. In addition to the interstellar pickup ions, SWICS distributions showed that the majority of the C+ and a fraction of the O+ and N+ are produced by an additional "inner source" of neutral atoms located near the Sun (Geiss et al., 1995). The inner–source, velocity-space distributions are significantly modified as they cool over the solar wind's transition from the near-Sun source to several AU where they were observed (Schwadron et al., 2000).

Ulysses/SWICS also observed the ubiquity of pickup-ion "tails" in slow solar wind (Gloeckler et al., 1994; Schwadron et al., 1996; Gloeckler, 1999). These tails do not correlate strongly with the presence of shocks but with compressive magnetosonic waves, suggesting that pickup ions are subject to strong statistical acceleration through processes such as transit-time damping of magnetosonic waves in slow solar wind (Schwadron et al., 1996; Fisk et al., 2000).

Prior to Ulysses, it was expected that pickup-ion distributions should be fairly isotropic due to pitch-angle scattering from background turbulence and self-generated waves (Lee and Ip, 1987). Instead, pickup-ion distributions were observed to be highly anisotropic (Gloeckler et al., 1995) with scattering mean free paths ~1 AU; the most likely cause is the inhibition of scattering through 90° pitch angle (Fisk et al., 1997). Although this lack of scattering is not fully understood, it has been shown that the turbulence of the wind has a strong 2-D component (Matthaeus et al., 1990; Bieber et al., 1996) that is ineffective for pickup-ion scattering (Bieber et al., 1994; Zank et al., 1998). Pickup-ion models have been devised that take into account the long-scattering mean free path (e.g., Isenberg, 1997; Schwadron, 1998).

Interestingly, the same self-generated turbulence associated with pickup-ion scattering has been considered an energy source for heating the solar wind (Zank et al., 1996; Matthaeus et al., 1999). These theories for the turbulent heating of solar wind take into account both the pickup-ion driven turbulence, which is predominant outside of ~8 AU, and wind shear. Smith et al. (2001) and Smith et al. (2006) have rigorously tested the heating models using data from Voyager 2 and Pioneer 11, and improved models of the pickup-ion scattering and associated wave excitation (Isenberg et al., 2005) have lead to remarkable agreement between data and theory.

Charge-exchange between interstellar hydrogen and solar wind protons leads to a complex interaction near the nose of the heliosheath, where a so-called hydrogen wall is formed from slowed interstellar hydrogen atoms and charge-exchanged solar wind protons. These interactions cause the removal or filtration of a fraction of the penetrating interstellar hydrogen atoms (e.g., Baranov and Malama, 1995).

The solar radiation pressure and the rates of photo-ionization and charge exchange vary with solar latitude and over the solar cycle. Sophisticated models of interstellar neutral atoms have been developed to take these effects into account (e.g., Izmodenov et al., 1999).

Cassini was the first spacecraft to measure pickup ions in the interstellar downstream region of the heliosphere beyond 1 AU (McComas et al., 2004a). Both interstellar pickup H+ and He+ were identified with the familiar cutoff in velocity space at twice the solar wind speed. Observed enhancements in the pickup He+ were consistent with gravitational focusing by the Sun. Further, McComas et al. also reported observations of the interstellar hydrogen shadow caused by depletion of H atoms in the downstream region due to the outward force of radiation pressure (which exceeded the gravitational force at the time of observation) and the high probability of ionization for atoms that must pass close to the Sun to move behind it.

The trajectory of *New Horizons* takes the spacecraft out to Pluto, which is currently toward the nose of the heliosphere. Figure 7 shows simulated distribution functions of pickup protons and solar wind protons in the spacecraft frame. We have used a kappa distribution for the solar wind protons (e.g., Vasyliunas, 1968; Collier, 1993) (with a kappa ~ 3, which is typical), and a density that falls off as
$R^{-2}$ from a 1 AU density of 5 $cm^{-3}$ and temperatures consistent with observations (e.g., Smith et al., 2001). For the pickup protons, we have taken a steady-state distribution that includes convection and adiabatic cooling in the radially expanding flow (Vasyliunas and Siscoe, 1976). The proton-pickup rate at each location is the local ionization rate times the neutral density solved using the "hot" model (Fahr, 1971; Thomas, 1978; Wu and Judge, 1979), which accounts for gravitational focusing by the Sun, ionization loss, and the finite temperature of incoming neutral atoms. We have taken an interstellar density near the termination shock of 0.1 $cm^{-3}$, and neutral temperature of 11 000 K, a neutral inflow speed of 22 km/s, an ionization rate of 7 x $10^{-7}$ $s^{-1}$ (referenced at 1 AU). We have also taken the force of radiation pressure to be comparable to that of gravity. The top panel in Figure 7 shows that solar wind protons are typically comparable to pickup protons near 5 AU. The pickup protons are observable near the knee of the distribution, so long as the solar wind is not significantly hotter or has significantly broader tails than those shown. Since the plasma near the ecliptic plane is highly variable near 5 AU, the detection of pickup protons is possible, but often obscured by the solar wind proton distribution. As *New Horizons* moves out toward the nose of the heliosphere, the situation changes. The solar wind density falls off with the square of heliocentric distance (while the pickup proton density falls off only as $R^{-1}$) and cools. Both of these effects make it much easier to detect "clean" pickup-proton distributions.

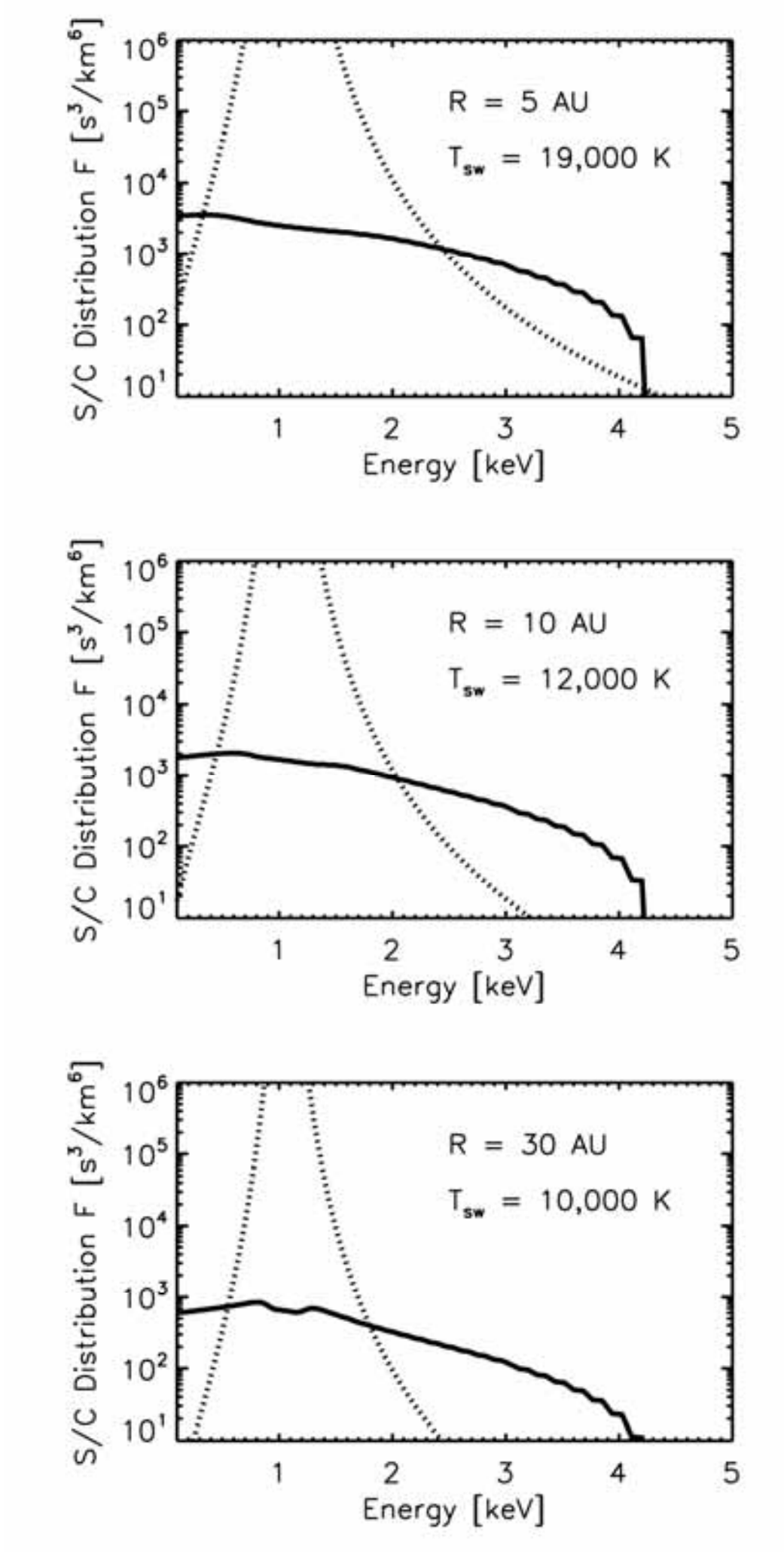

Figure 7. Pickup-proton (solid curves) and solar wind (dashed curves) distribution functions shown with distance from the Sun toward the nose of the heliosphere. The pickup-proton distribution functions become more and more prominent compared to the cooled solar wind distributions further out in the heliosphere.

The unique trajectory of *New Horizons* combined with SWAP's capability of pickup-proton measurement should enable it to address some fascinating issues:

The distribution functions measured by SWAP will allow investigation of possible sources (other than interstellar) of pickup ions. For example, it is thought that an Outer Source of pickup ions may be caused by the interaction of solar wind with dust grains from the Kuiper Belt (Schwadron et al., 2002). The Outer Source is likely to be far more variable in space and time than the interstellar source.

# 3.0 Instrument Description

The SWAP electro-optics consists of 1) a retarding potential analyzer (RPA); 2) a deflector (DFL); and 3) an electrostatic analyzer (ESA). Collectively, these elements select the angles and energies of the solar wind and pickup ions to be measured. Ions selected by the electro-optics are then registered with a coincidence detector system. Figure 8 schematically depicts the SWAP principle electro-optics. Ions enter through the RPA with all ions having energy per charge less than the RPA voltage being rejected by this hi-pass filter. Ions entering at angles from above horizontal in this figure can be deflected in to the subsequent electro-optics by applying a voltage to the deflector ring. Ions with E/q greater than the RPA voltage are then selected by the ESA, which rejects ions outside the selected E/q range as well as UV light and neutral particles.

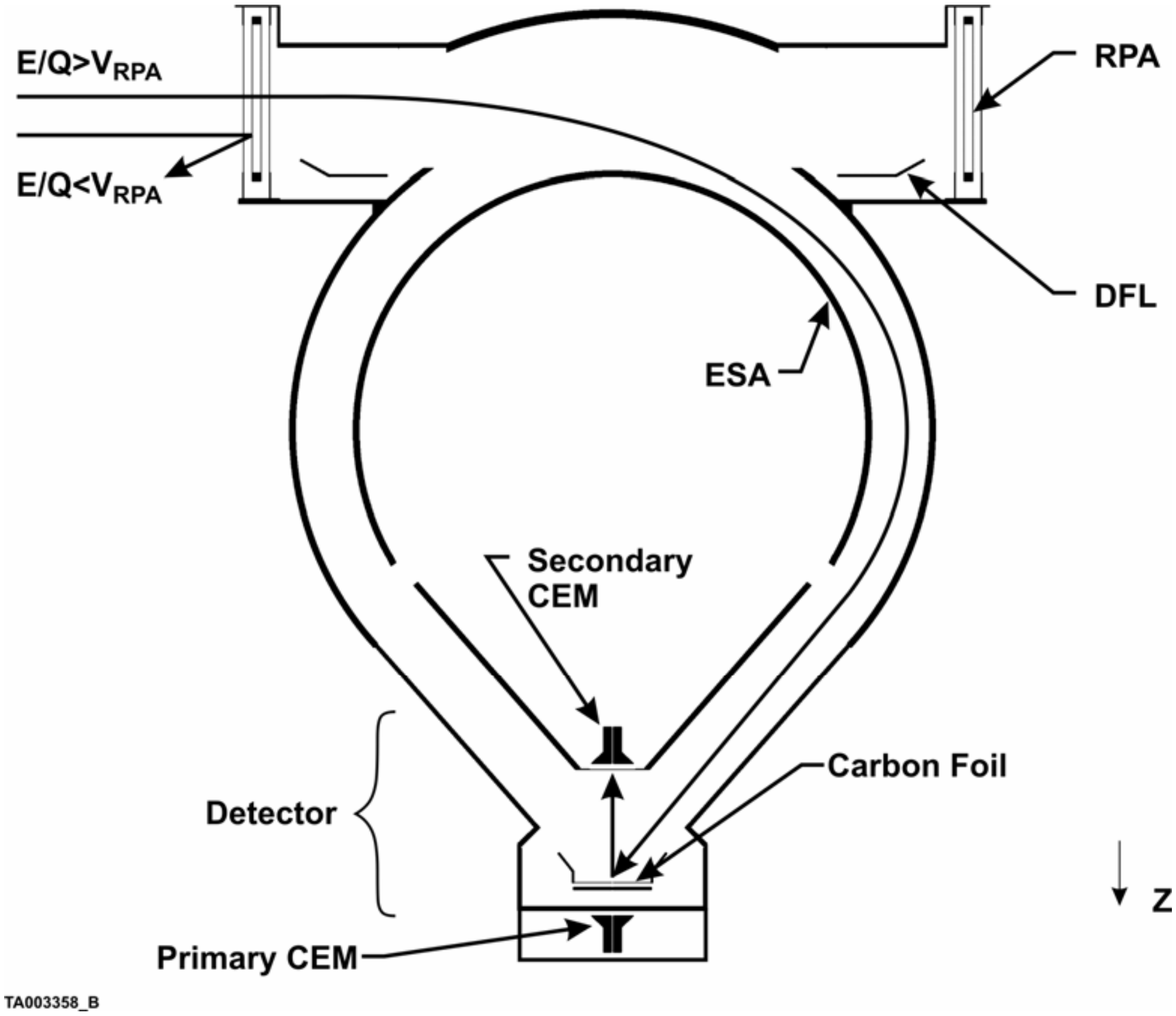

Figure 8. Schematic diagram of the SWAP electro-optics including RPA, deflector, ESA, and detector section.

Figure 9 schematically shows how SWAP's ESA and RPA are used together to select the E/q passband. When the RPA is off, the passband is determined solely by the ESA, which has an 8.5% FWHM resolution (top panel). At increasing RPA voltages for a given ESA setting, the passband is cutoff in a variable "shark-fin" shape, allowing roughly two decades decreased sensitivity (middle panel). Finally, differentiating adjacent RPA/ESA combinations, or better yet deconvolving multiple combinations, provides high resolution differential measurements of the incident beam.

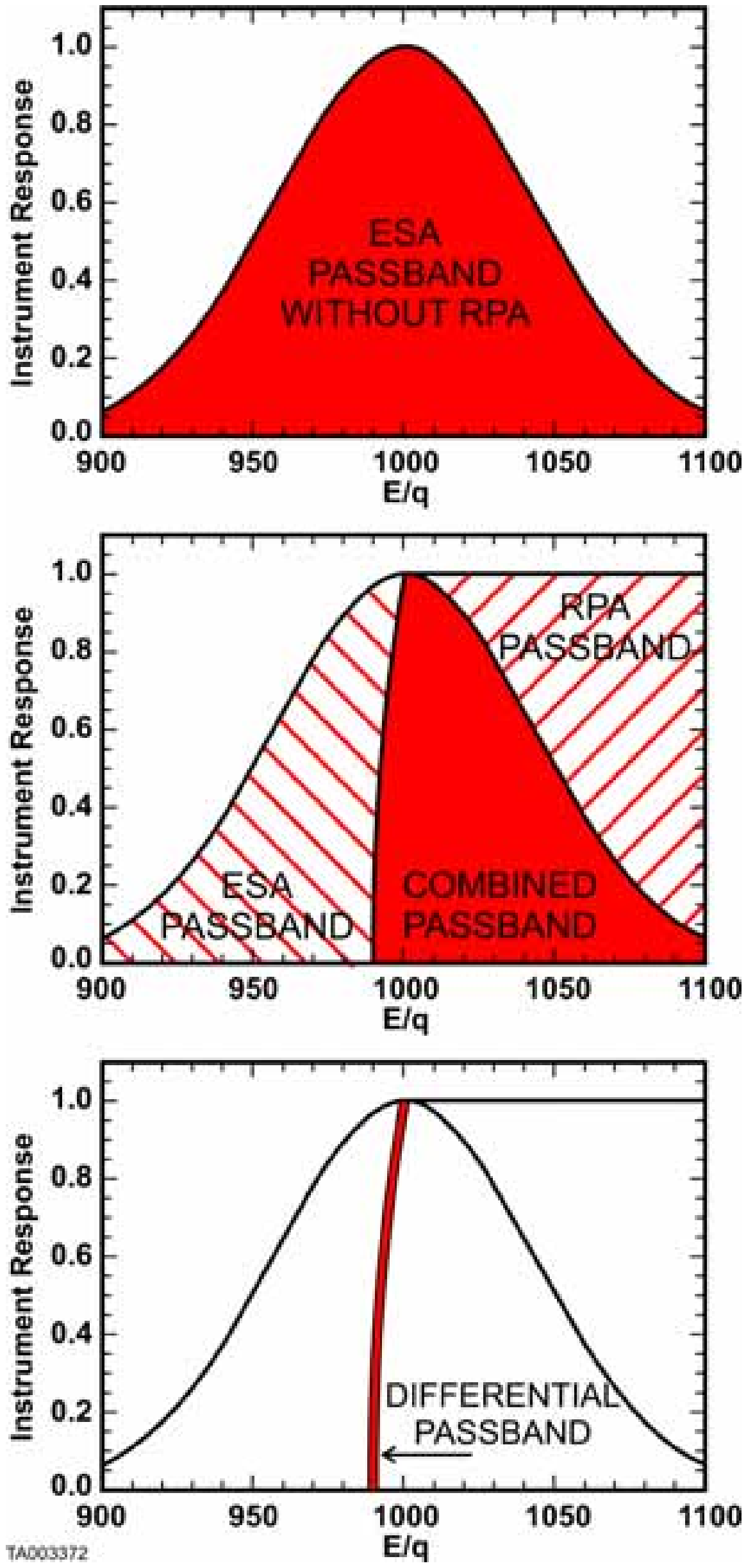

Figure 9. SWAP principle of operation. Different combinations of RPA and ESA settings provide for a variable E/q passband. Multiple combinations of settings can be differentiated to produce very high-resolution measurements.

As an example of how the RPA can be used to measure very small changes in beam energy, and hence solar wind speed, Figure 10 shows the count rate as a function of RPA voltages for a range of different ion beam energies taken with the flight electro-optics during instrument calibration (described in more detail below). A small subset of the full energy range of SWAP (990 to 1010 eV) is shown to highlight the energy resolution possible by this instrument. For each scan, the count rate is normalized to the rate when the RPA voltage is set at zero to take into account the differences in the ion beam flux. Differences in the ion energy as small as 1-2 eV are distinguishable at typical solar wind energies of ~1000 eV.

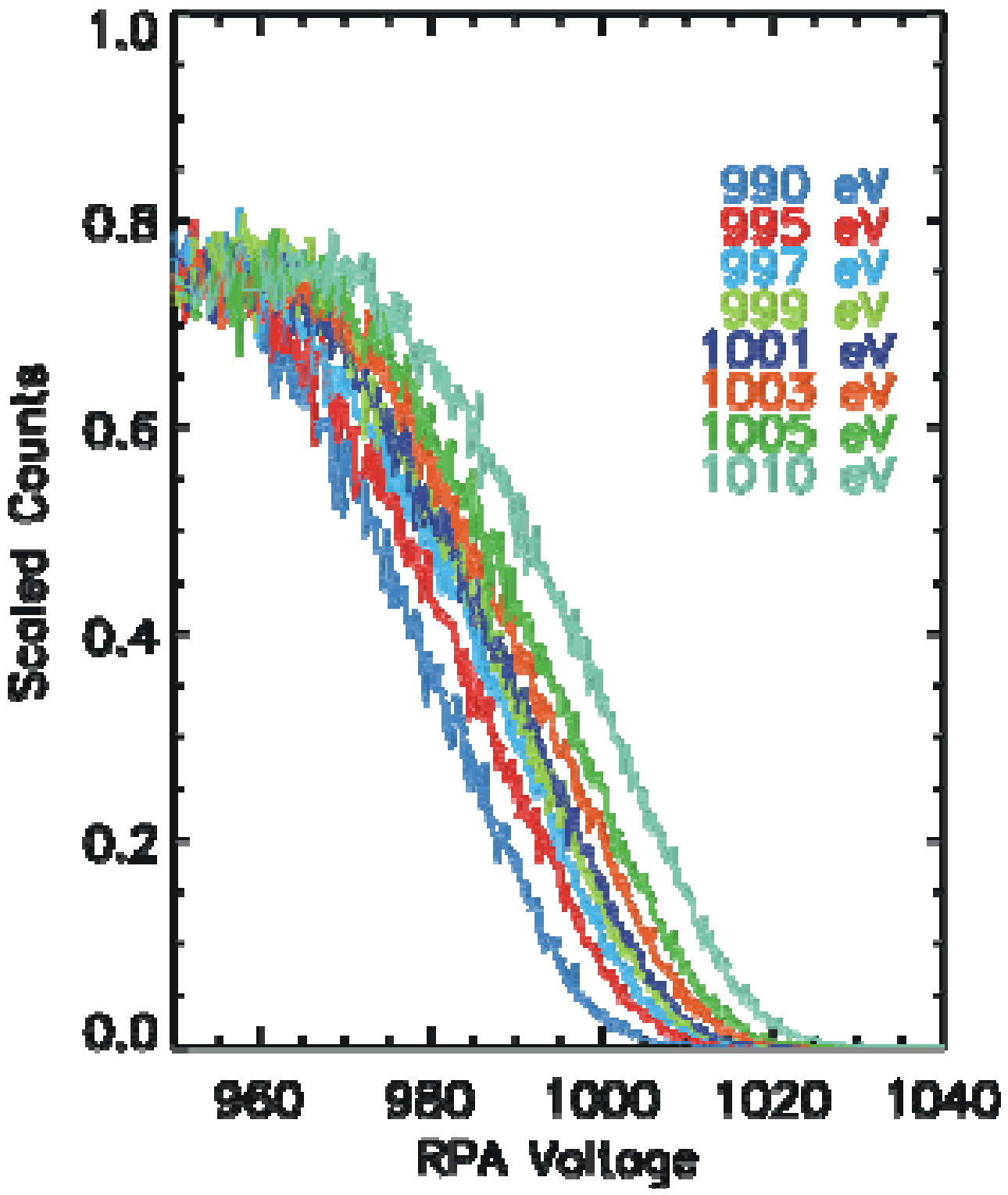

Figure 10. RPA resolution of ion beams with energies from 990-1010 eV. Each RPA scan shown was taken with the beam incident normal to the RPA. Deconvolution of such curves should make changes in beam energy as small as 1-2 eV possible.

The transmitted ions are post-accelerated into the detector section, which employs an ultra-thin carbon foil and two channel electron multipliers (CEMs) to make a coincidence measurement of both the primary particle and the secondary electrons generated when the primary particle passes through an ultra-thin carbon foil. Charge amplifiers (CHAMPs) service the two CEMs and transmit digital pulses when events are detected. High voltage power supplies (HVPS) provide power for the CEMs and sweep the voltages on the electro-optics. The control board processes the pulses from the CHAMPs, controls the sweeping of the high voltages, digitizes the housekeeping data, creates telemetry packets for transmission, accepts commands, and converts spacecraft power into the secondary voltages required by the instrument. Key properties of the SWAP instrument are given in Table II.

**Table II. Measured properties of the SWAP instrument**

| | |
|---|---|
| Field of View | 276° x 10° (deflectable >15° toward -Z) |
| Energy Range<br>ESA (bin centers)<br>RPA | <br>35 eV to 7.5 keV<br>0 to 2000 V |
| Energy Resolution<br>ESA ($\Delta E/E$)<br>RPA | <br>0.085 FWHM<br>0.5 V steps (high resolution requires deconvolution) |
| ESA factor (beam energy / ESA voltage) | 1.88 |
| Dynamic range | $\sim 10^6$ |
| Geometric Factor (Hot Plasma) | Coincidence: $2.1 \times 10^{-3}$ $cm^2$ sr eV/eV<br>Total: $1.3 \times 10^{-2}$ $cm^2$ sr eV/eV |
| Cold Beam Effective Area* (normal incidence) | Coincidence: $3.3 \times 10^{-2}$ $cm^2$ sr eV/eV<br>Total: $1.9 \times 10^{-1}$ $cm^2$ sr eV/eV |
| Time Resolution | Full energy range and 1) detailed peak measurements or 2) additional full energy sweep each 64s (128 steps with 0.39s accumulation times) |
| Mass | 3.29 kg |
| Volume | 0.011 $m^3$ |
| Power | 2.84 W |
| Telemetry | <1 - 280 bps |
| * Note: Calculation of expected count rates is presented in Section 5.1. | |

The *New Horizons* spacecraft points the optical instruments by rotating the spacecraft about its Z-axis. In order to make solar wind measurements largely independent of this scan angle, SWAP was designed to be essentially symmetric about Z and mounted into a cylindrical bracket as depicted in Figure 11. By placing it on the –Z corner of the spacecraft, the instrument's field-of-view is clear of any structures or obstructions for all scan angles.

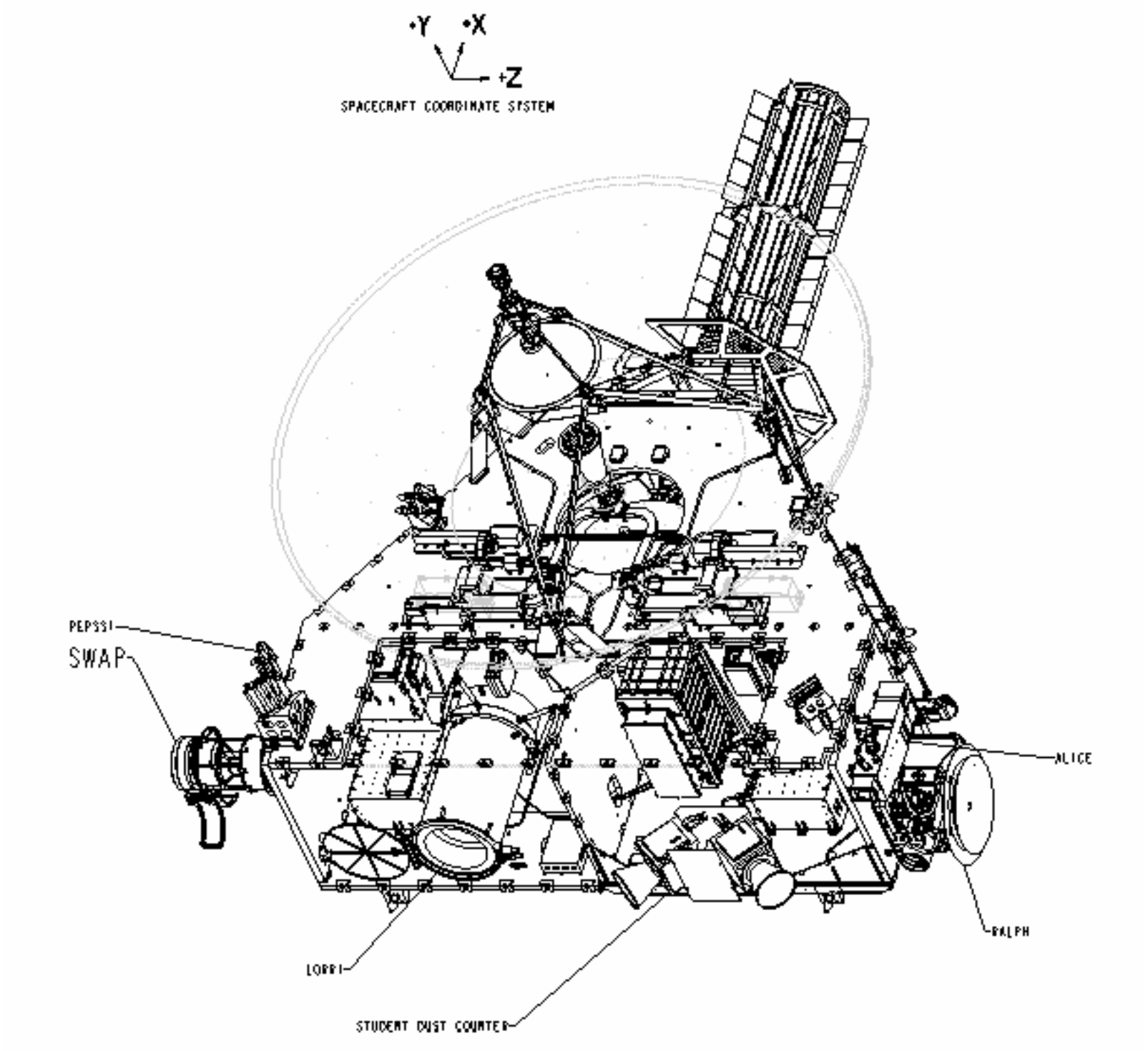

Figure 11. Location of SWAP on the *New Horizons* Spacecraft

## 3.1 Electro-Optic Design

The electro-optic design is driven by the unique needs of the *New Horizons* Mission to Pluto. A large aperture instrument is required since the density of the solar wind falls off as the square of the distance from the Sun. At ~32 AU, where we will encounter Pluto, the mean solar wind density is roughly 3 decades lower than the solar wind at 1 AU. Also, the *New Horizons* spacecraft does not have a scan platform for the optical instruments. Instead, the spacecraft slews about the Z-axis to provide the required scanning motion. Since SWAP's focus is to measure the solar wind, we required an electro-optic design that was largely unaffected by this rotation about the Z-axis. Therefore, we selected a top-hat design with an ESA that has a large angular and energy acceptance to maximize the effective aperture. Since the driving measurement is to look for small changes in the solar wind speed as *New Horizons* passes Pluto, we included an RPA to allow finer resolution energy measurements and to provide a variable electrostatic passband. This is accomplished by "crossing" the ESA and RPA such that the RPA admits particles only in the highest energy portion of the ESA passband as described above. Also, because the spacecraft will sometimes be required to tip out of the ecliptic plane, we incorporated a deflector to adjust the look direction of the instrument. Finally, because the mission is designed so the spacecraft rarely points the –Y-axis toward the Sun, we can provide an unobstructed field of view of >270° in the X-Y plane, centered on the +Y-axis, which is aligned with the high gain antenna (HGA). Nominally, the HGA points toward the Earth within a few degrees of the Sun for a large fraction of the Pluto encounter. All supports for the top-hat top-plate, RPA, ESA, and the aperture doors for protection during launch are restricted to the remaining 90°aligned with the – Y-axis. Figure 12 shows the key dimensions of the SWAP electro-optics.

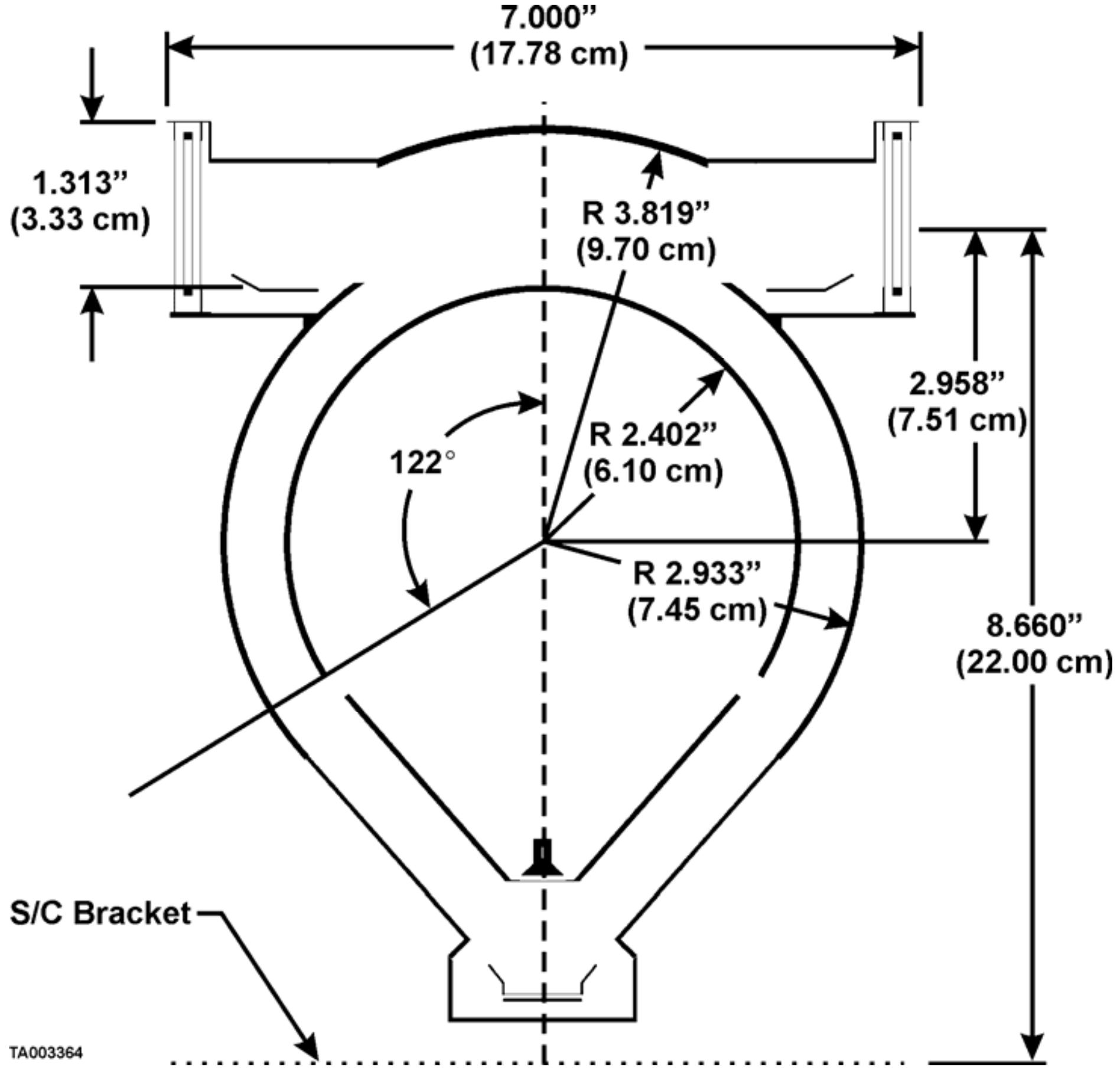

Figure 12. Key dimensions of the Electro-Optics

### *3.1.1 RPA Design*

The RPA consists of four concentric aluminum cylinders or screens, each of which is machined with ~90,000 close-packed holes to create a self-supporting grid structure. Each cylinder, approximately 65% transmissive, is 0.762-mm thick and has been machined into a grid-like structure by drilling 0.343-mm-diameter holes through nominal 0.394-mm-thick aluminum in a close-packed hexagonal configuration (Figure 13). From outermost to innermost, the outer diameters of the four RPA screens are 174.4, 169.6, 166.4, and 161.6 mm. The outside and inside cylinders are at ground potential (0 V). The two central cylinders are biased from 0 to +2000 V in 0.49 V steps and isolated from the rest of the structure by ceramic insulators.  There is a 2.032 mm gap between the biased and grounded RPA grids to provide sufficient high voltage clearance for the -2000 V potential. The two biased RPA grids are separated by 1.016 mm to stop penetration of lower energy equipotentials through the grid holes, thus creating a much more uniform electric field in the RPA region. The unobstructed Field of View (FOV) is 276° in the roll direction.

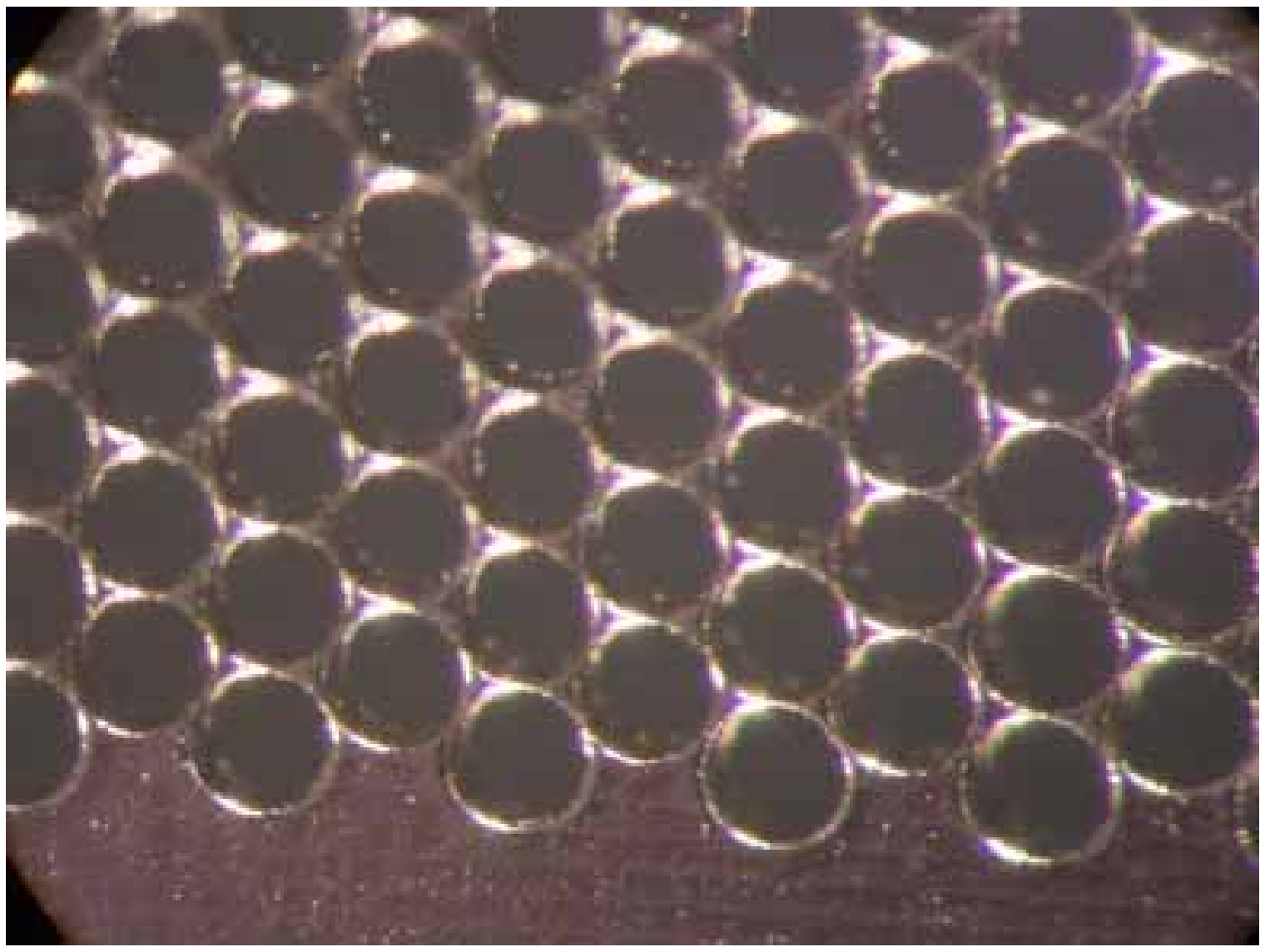

Figure 13. Photograph of 0.343 mm RPA holes on 0.381 centers

The RPA provides a low-pass filter with a relatively sharp energy cutoff so that we can make a fine sweep across the solar wind beam after locating it with a coarse ESA energy scan.  Ions that have sufficient energy to climb the electrostatic “hill” set by the voltage on the inner RPA grids are reaccelerated to their original energy as they pass from the inner RPA grids to the final grounded RPA grid.

### *3.1.2 Deflector*

SWAP incorporates a deflector that is used to deflect particles from above the central plane of the instrument (from further out in the –Z axis of the spacecraft) into the ESA.  The deflector is located just inboard of the RPA.  The voltage on the deflector ring is varied from 0 to +4000 V. It deflects up to 7000 eV/q particles up to 15° into the ESA (larger deflections for lower energies).

### *3.1.3 ESA*

The ESA provides coarse energy selection and protects the detectors from UV light. The dimensions of the top-hat ESA are shown in Figure 12. The outer ESA plate is serrated and blackened with Ebanol-C, a copper-black process that greatly reduces scattering of light and particles. The inner ESA plate, which is blackened but not serrated, is supported on insulators that attach it to a cantilevered support structure (Figure 14). On the other side of this structure a grounded cone completes the ESA design by providing a field-free region for particles to enter into the detector region. The voltage on the ESA is varied from 0 to -4000 V. It has a ratio of central E/q of particle transmitted to ESA voltage, or "K-factor", of 1.88 and selects up to 7.5 keV/q particles (passband central energy) with a $\Delta E/E$ of 8.5% FWHM.

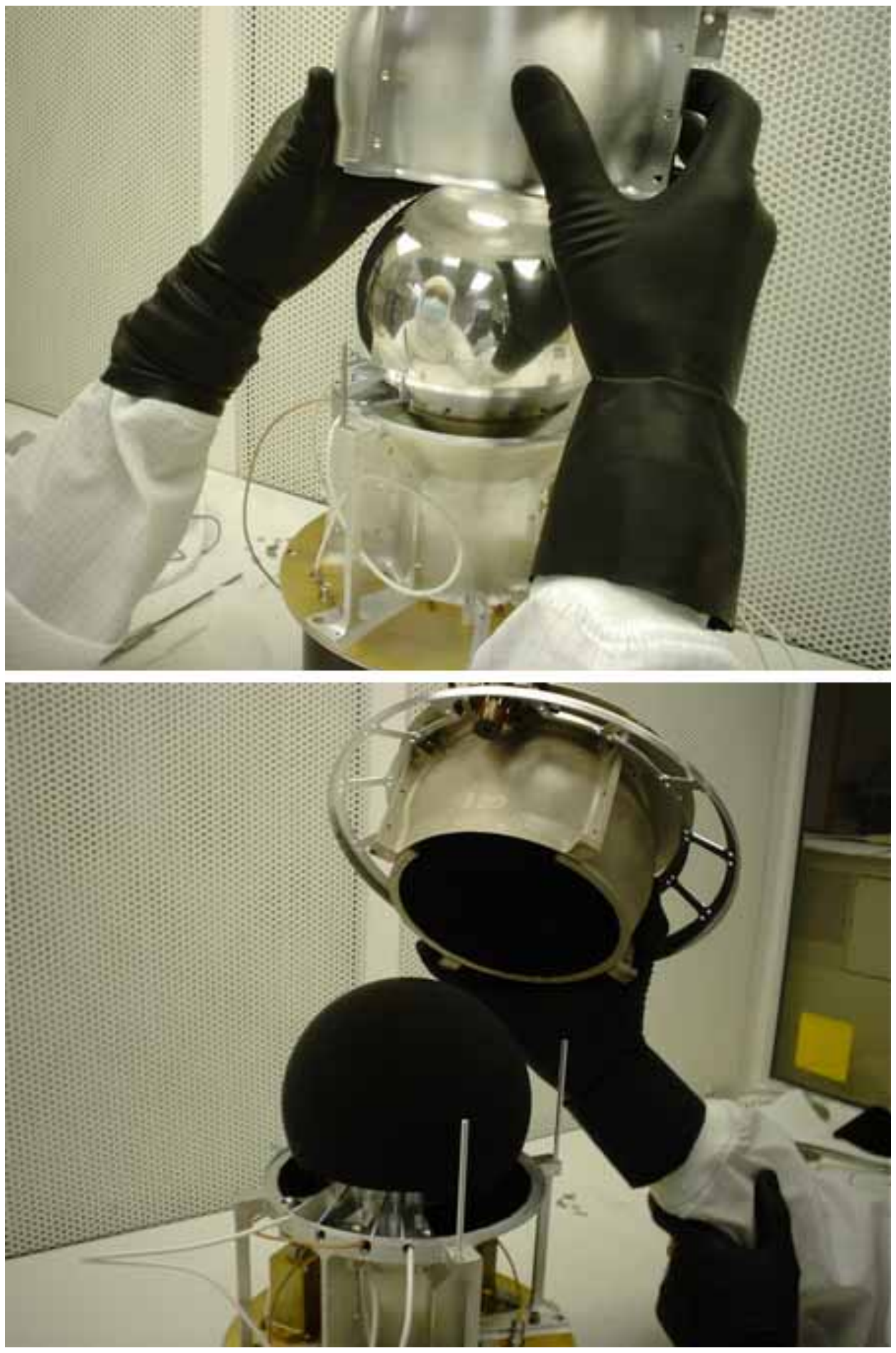

Figure 14. Cantilevered Inner-ESA dome with (bottom panel) and without (top panel) blackening.

## 3.2 Detector Design

SWAP employs a coincidence system to detect incoming ions. After ions of the desired energy and angle have been selected by the electro-optics system, they pass through a field free region between the ESA and detector region. Once a particle enters the detector region, it is accelerated by the CEM high voltage bars towards the Focus Ring on which is suspended an ultra-thin carbon foil (McComas et al., 2004b). The carbon foil is nominally 1 $\mu g/cm^2$ thick and is suspended on a 64% transmissive grid. The particle passes through the Focus Ring, which is at the Primary CEM (PCEM) HVPS output voltage, and travels on to the PCEM (Figure 15). Forward-scattered electrons from the carbon foil are also accelerated to the PCEM due to the ~100 V potential created by the PCEM strip current and a resistive divider. Backward scattered electrons are directed by the Focus Ring towards the Secondary CEM (SCEM) which collects them.

Counts from these two CEMs are registered by CHAMPs and their associated electronics. A count from either the PCEM or the SCEM starts a 100 ns coincidence window timer. Given the particle trajectories and the electron trajectories, it is possible for either the PCEM or the SCEM to trigger first, so no specific arrival order is required by the electronics. The CEMs have quad spiral channels with a resistance of 300 Mohm and dark counts less than 0.04 $s^{-1}$ (Figure 16).

The coincidence detector system reduces the background from CEM dark counts, penetrating energetic particles and UV noise, allowing SWAP to have a low enough noise floor to measure heliospheric pick-up ions. Two detectors provide the redundancy needed for a long-duration mission. SWAP can still make its primary science measurement using only one of its CEMs. This requirement for redundancy drove us to add a -1 kV Focus Ring supply that is slaved to the SCEM and diode OR-ed with the PCEM supply to the Focus Ring. If the PCEM is turned off, the -1 kV Focus Ring supply will accelerate back-scattered electrons from the carbon foil into the SCEM.

The long CEM lifetime needed for this mission required us to select all materials near the detector to be ultra-low outgassing.  We used only glass, metal, and ceramic materials for all parts of the detector that had venting access to the CEM detectors.  HV and LV cabling was brought to bulkheads, and the signals were conducted through ceramic feedthroughs to the exact location they were required (Figure 17).

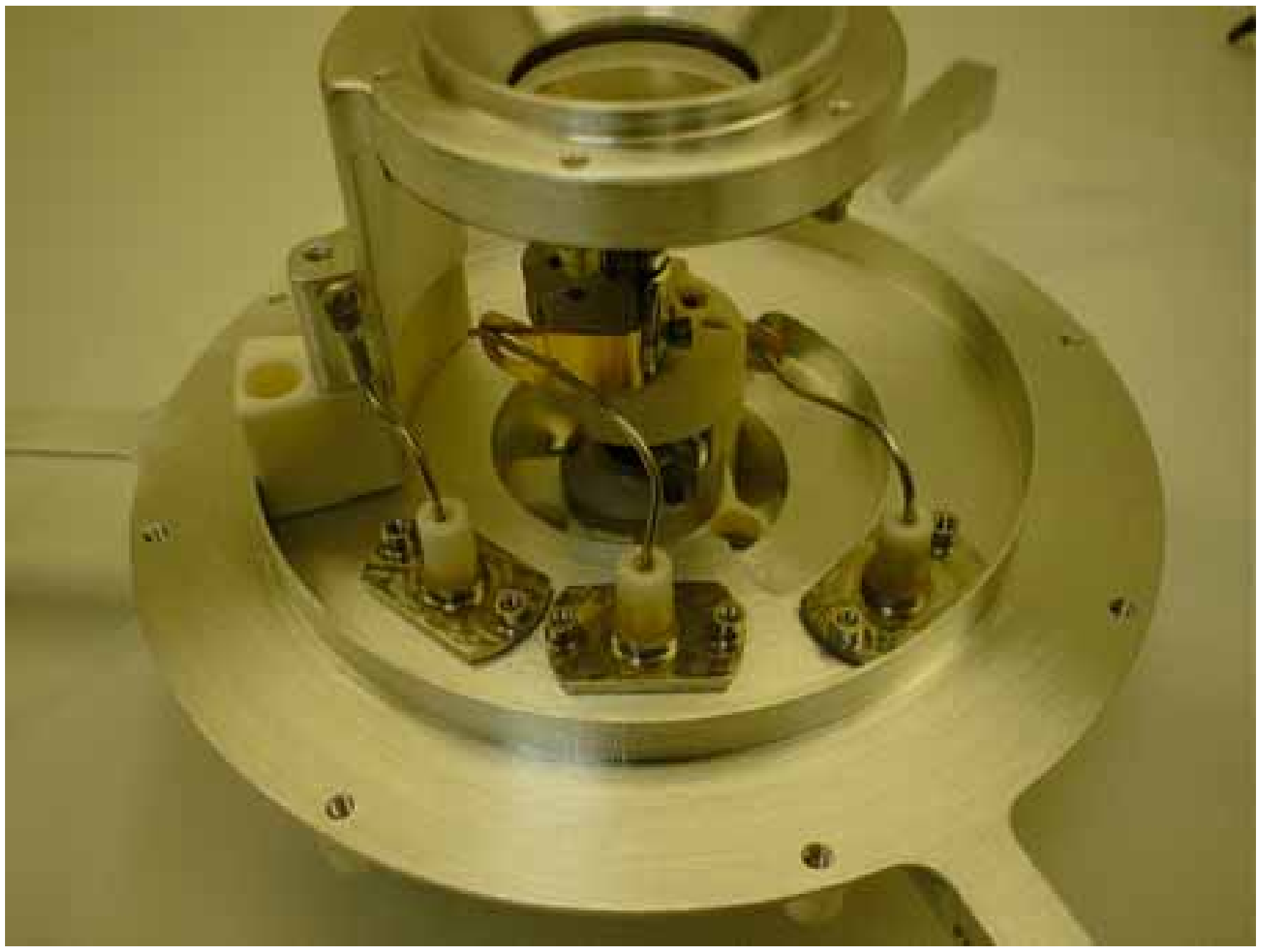

Figure 15. PCEM and Focus Ring with Carbon Foil, partially assembled

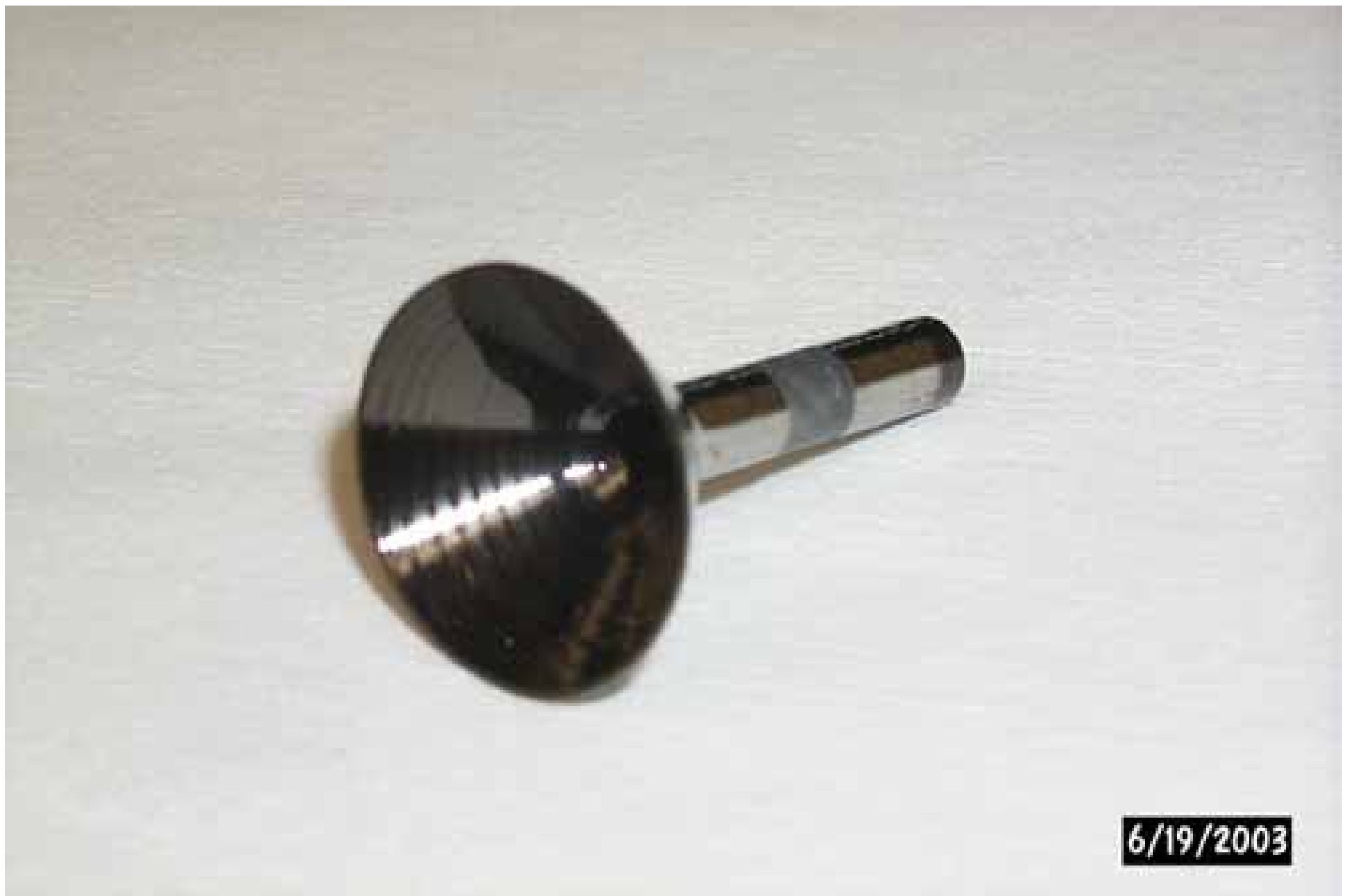

Figure 16. Quad Spiral Channel Electron Multiplier

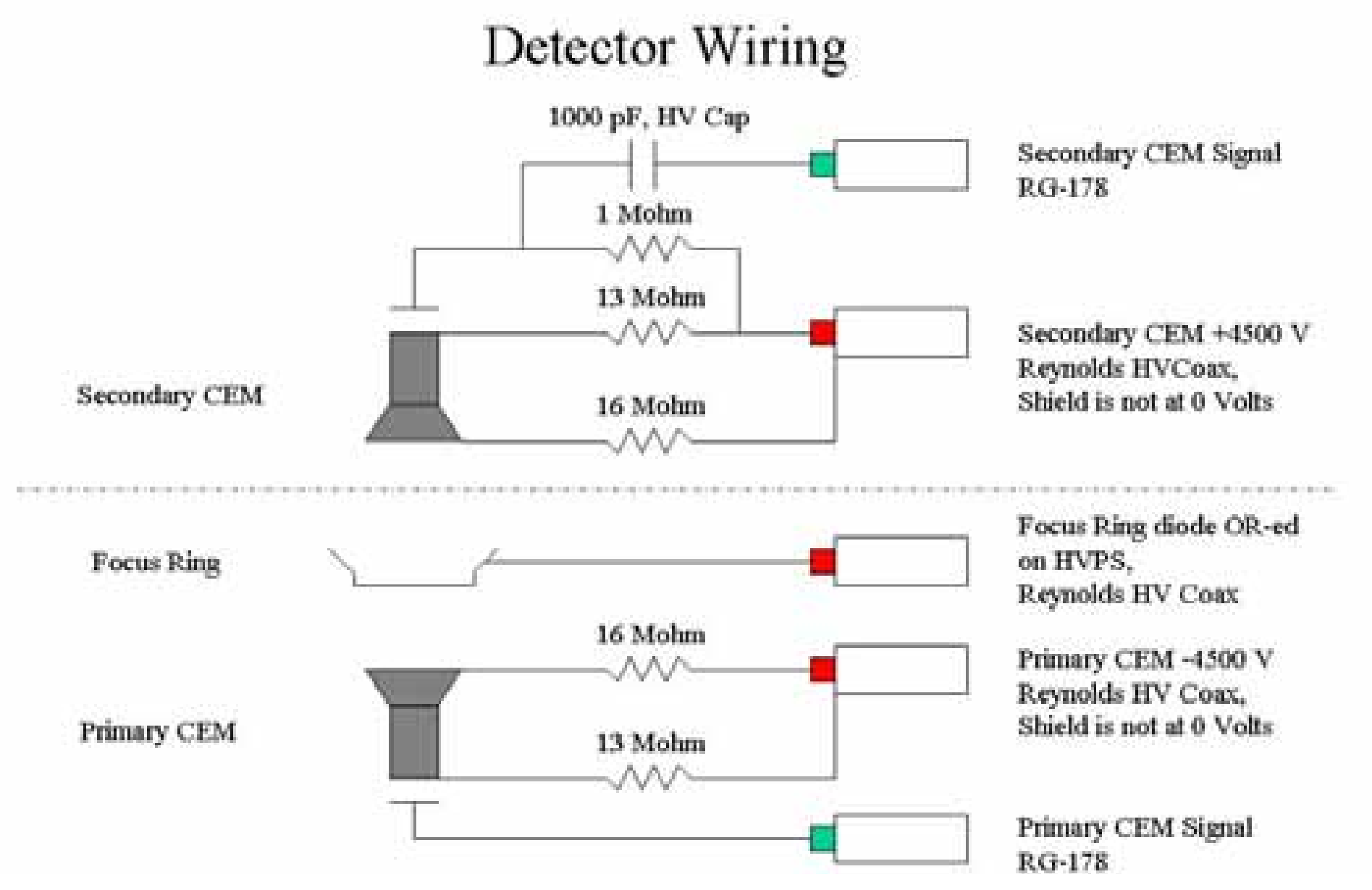

Figure 17. Detector Wiring Diagram

## 3.3 Mechanical Design

The SWAP mechanical design consists of three main subassemblies: optics and detector, aperture door, and electronics packaging and cabling (Figure 18). In addition to these instrument subassemblies, we also discuss SWAP's structural and thermal design in the following sections.

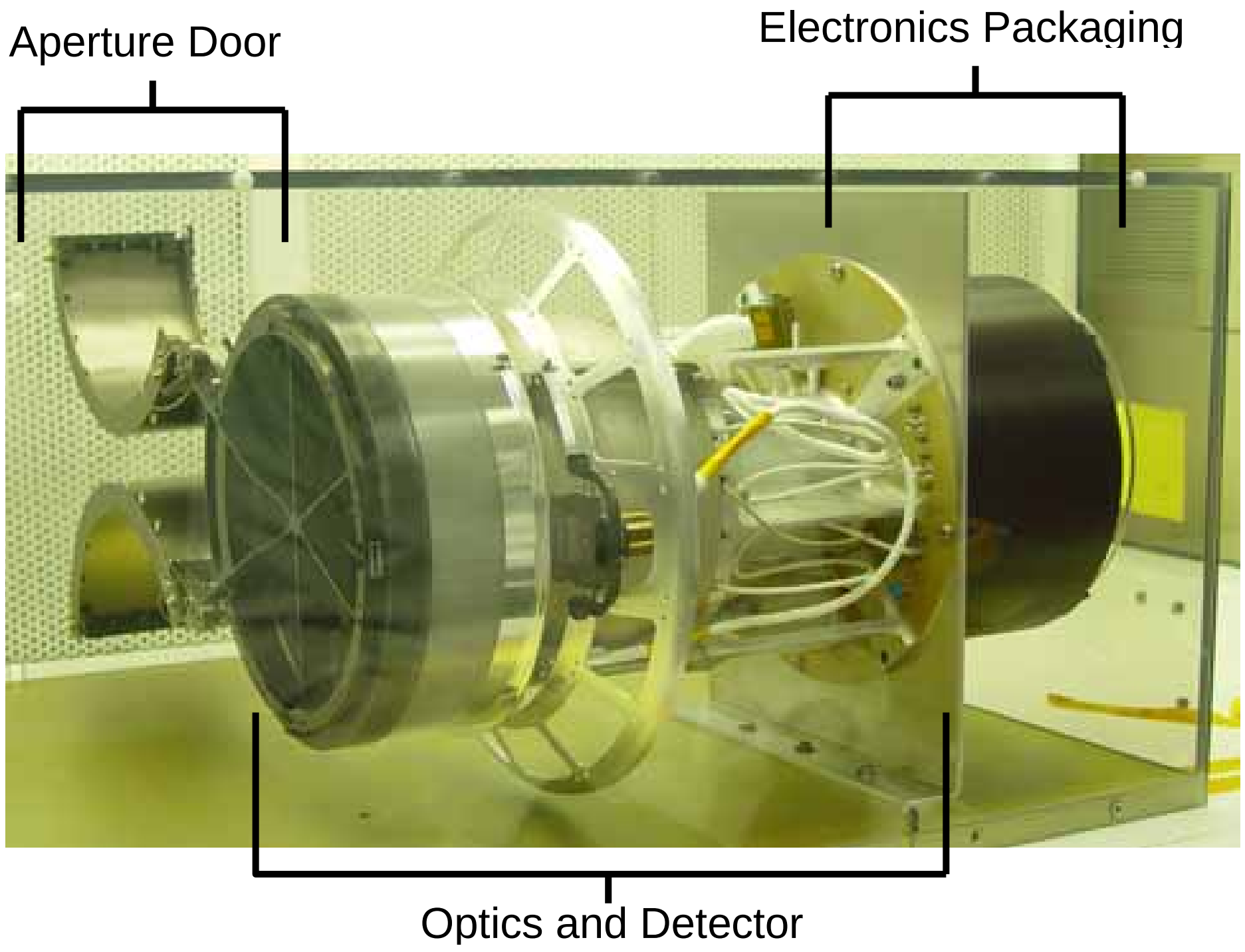

Figure 18. Photograph of SWAP instrument highlighting subassemblies

### 3.3.1 Optics and detector mechanical design

The critical requirements driving the optics and detector mechanical design were to mount the electro-optical components as specified by the ray-traced model, produce optically black surfaces inside the instrument, ensure contamination control for the detectors, provide for easy refurbishment, and accurately align all critical components.

We used a SIMION ray tracing model to define the electro-optics component (Figure 19) sizes and locations (RPA, ESA, CEMs), and required that the mounting of the various elements should minimally impact particle trajectories. We accommodated this by mounting the Inner ESA and Secondary CEM assemblies from a cantilevered support hidden in the 90° region where SWAP does not require particle viewing. The location of this cantilever matches up with the hinge assembly on the door.

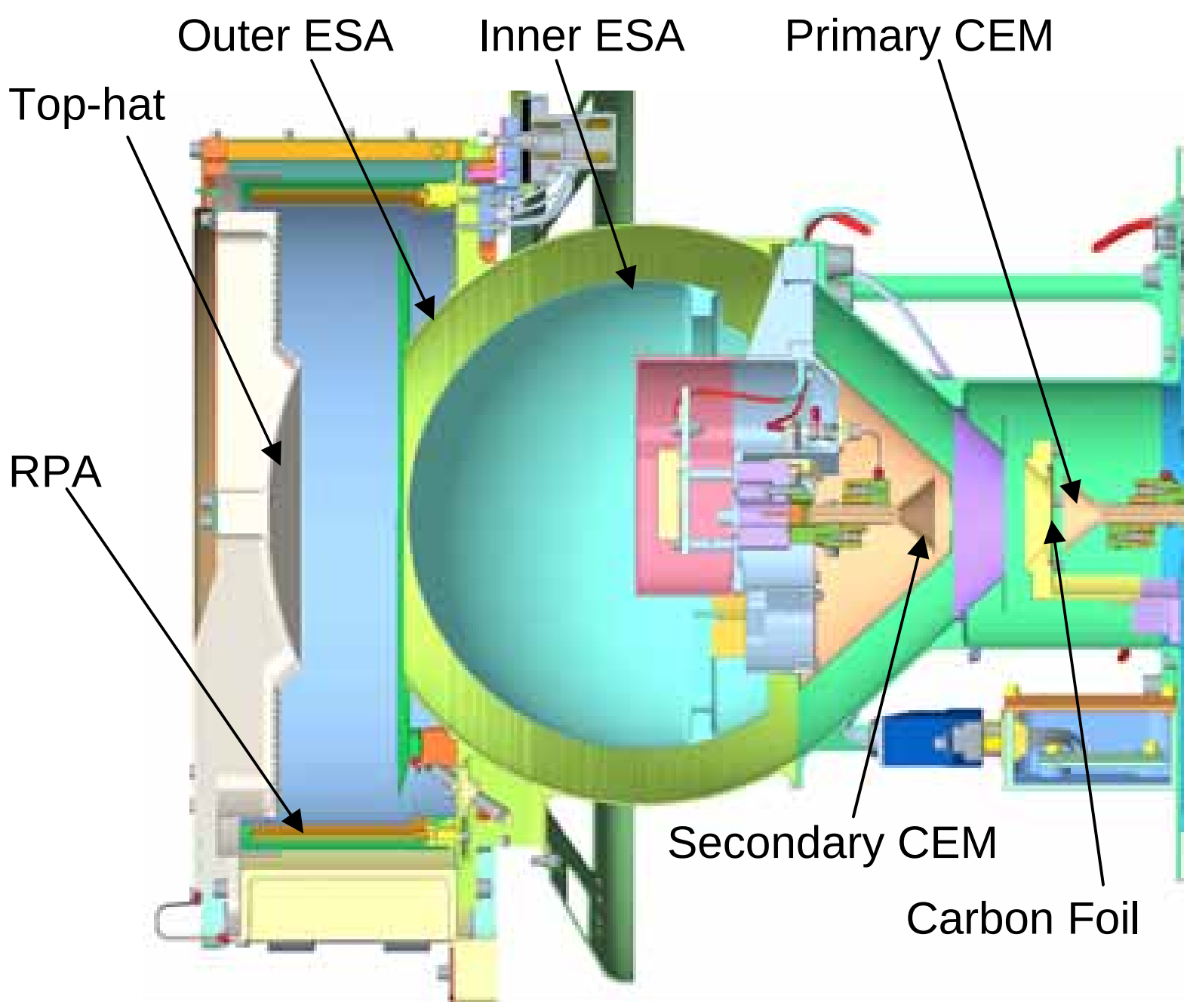

Figure 19. Mechanical configuration of the optics and detector assembly.

In order to accommodate the cleanliness requirement, the optics and detector areas included a split between ultra-clean and electronics volumes. We used ceramic for the electronics near the primary CEM, and fed the cabling into the primary CEM area using pass-thrus in the cantilever support so that it would not pass through the ESA gap.

We refurbished SWAP after spacecraft environmental testing, replacing the CEMs and the ultra-thin carbon foil with ones that had been stored in a clean environment after burn-in. To minimize the risk of dismantling the instrument, the CEM assemblies were designed so that they could be easily replaced. In order to meet the alignment requirements for the electro-optics, we designed features that would control the concentricity and placement of the optics and detectors in relation to each other both before and after refurbishment.

### *3.3.2 Aperture door design*

The aperture door's (Figure 20) main requirement is to protect the SWAP RPA from contamination and damage during ground and launch operations. Although we designed the door to open one time after launch, it was important that it easily reset for test on the spacecraft. We opened the door multiple times as part of the instrument verification and during the environmental test flow at the spacecraft level. The door could not be a glinting source because it stays with SWAP during the entire mission to Pluto. Spacecraft requirements dictated that external surfaces be conductively coupled to the instrument. Therefore, we coated the door surfaces with black nickel and provided for ground straps built into the design. The *New Horizons* Component Environmental Specification required verifying through test that the door's torque margin was greater than 2.25. Testing demonstrated a torque margin of 3.10, and the door successfully opened in flight.

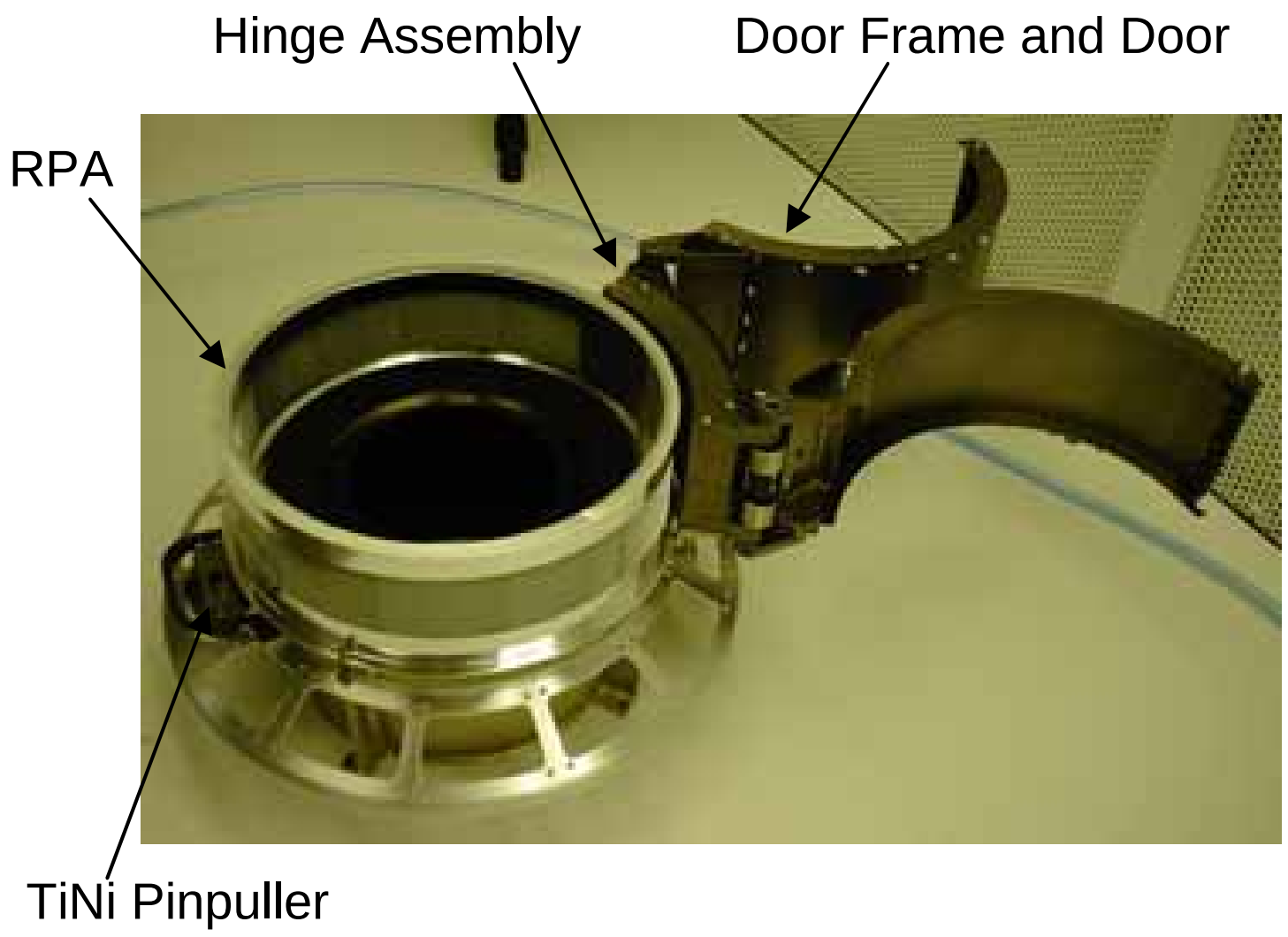

Figure 20. Door assembled on the SWAP outer ESA. The TiNi Pinpuller releases the door so it can spring open.

### *3.3.3 Electronics packaging and cabling design*

Figure 21 shows the as-built electronics packaging and cabling that are part of the SWAP instrument. Panel a) shows the HVPS and Control board with the electronics before the electronics housing was installed. Panels b) and c) show different views of the CHAMP electronics and cables that bring the HV to the RPA and ESA.

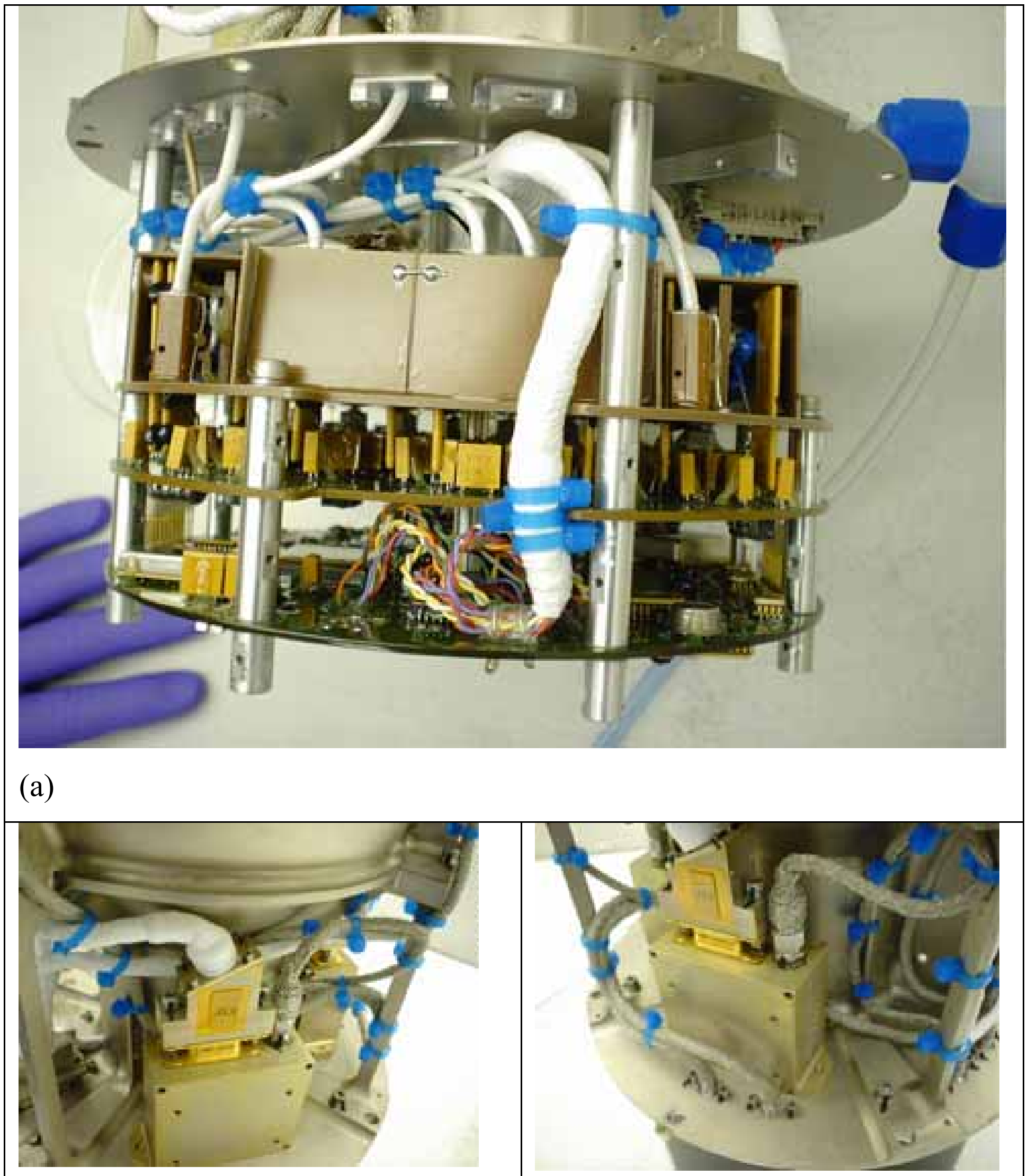

(a)

(b)

(c)

Figure 21. As-built electronics packaging and cabling.

To save weight and volume, the electronics are compactly packaged below the ESA in an "electronics volume", as shown in panel a). The electronics are laid out on three boards, which are connected via pin/socket connections. The top board contains the HV, the middle board includes the drivers for the HVPS, and the bottom board receives power from and provides telemetry to the spacecraft. The CHAMP boards were originally intended for the lower electronics, but engineering testing showed that they needed to be closer to the detectors to decrease noise. We routed the HV and signal cables through the electronics housing to their appropriate location and included cable routing tie points in the outer ESA design to prevent the cables from moving during launch loads or transportation.  The standoffs between the three bottom boards structurally attached the boards to the housing and allowed heat to conductively pass to the housing.

### 3.3.4 Structural Design

The Component Environmental Specification defined the following critical structural requirements for SWAP: 1) quasi-static load of 30 gs force applied separately along 3 orthogonal axes, 2) first mode structural frequency constraints of >70 Hz (thrust direction) and (>50 Hz) (lateral direction), 3) sine vibration up to 20 g (22-24 Hz, and 4) Random vibration with an overall amplitude of 10.4 Grms.

We performed a structural analysis during the final design phase to very that the SWAP structure would meet these requirements.  The analysis focused on the critical items such as the cantilevered ESA and outer ESA. The analysis also showed that the first natural frequency of the SWAP instrument, 180 Hz, was well within the requirement.  Finally, we performed vibration testing during environmental testing to ensure that SWAP would survive the prescribed levels.

### 3.3.5 Thermal design

SWAP has flight temperature limits of 0 to +40°C (operating) and -20 to +50°C (non-operating). We initially performed a thermal analysis to show that the electronics and other temperature-sensitive parts of the SWAP instrument could survive these temperature extremes.  The analysis also showed that the heat exchange with the spacecraft was less than 5 Watts (the limiting case was when the instrument was at its hot limit with the door open), as required by the spacecraft.  During environmental testing, we performed a thermal vacuum test to validate the functionality of SWAP at all temperatures within these extremes; hot and cold turn-ons of the instrument were also performed.

## 3.4 Electronics

### *3.4.1 CHAMPs*

The CHAMPs for SWAP convert a charge pulse from the CEMs into a logic pulse that can be registered and processed on the Control Board. The CHAMPs (Figure 22) are located as close to the detectors as practical in separate enclosures mounted to the top of the strong back. The SWAP design incorporates high-speed commercial hybrid CHAMPs, which work reliably to >1 MHz rates and have voltage-adjustable thresholds. The output from the CEMs is brought to the CHAMP through a short coaxial cable through a bulkhead connector that ties the shield to chassis ground. A 100-ohm resistor and back-to-back diodes protect the input from discharges, and the input is AC coupled to the charge amp.

Injecting test pulses into the front end demonstrates the end-to-end integrity of the microcircuits and cabling for CEM pulse processing. Test pulses are driven from the Control Board to the CHAMP, which is twisted with a return line. On the CHAMP, the signal is buffered by a gate with Schmitt triggering.

The threshold voltage for the CHAMP is set by a Digital-to-Analog Converter (DAC) on the Control Board, and this value can be set through standard software commands, allowing the threshold to be updated at any time. A resistor sets the output pulse width to 70 ns and the amplifier dead-time to 100 ns. The output pulses from the amplifier are buffered by two Schmitt trigger buffers and transmitted through a back-terminated series resistor to the cable that connects to the Control Board.

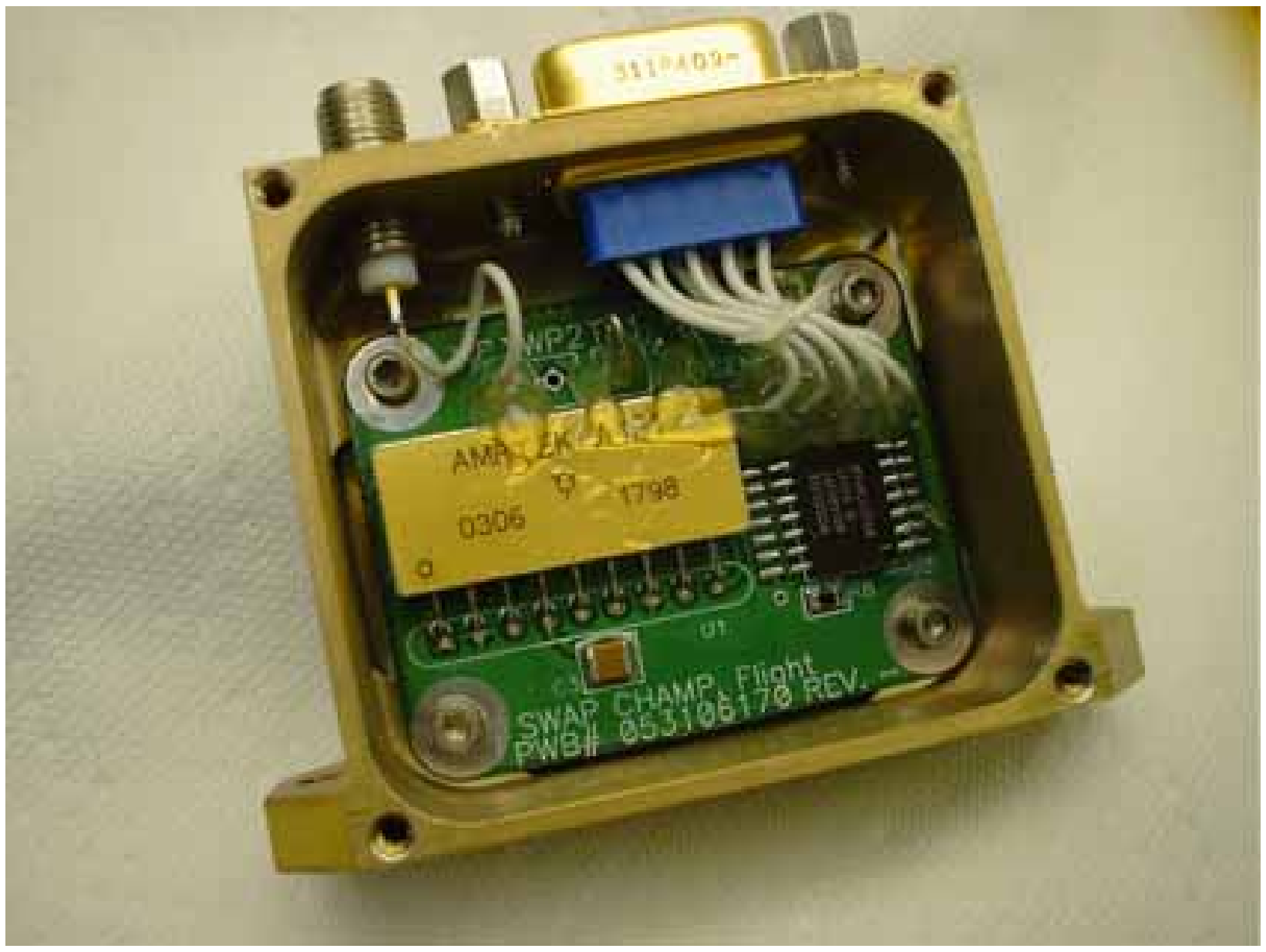

Figure 22. Charge Amplifier Photograph

### 3.4.2 HVPS

The HVPS (Figure 23) set the voltages on the optical surfaces (RPA, DFL, ESA, & Focus Ring) as well as supplying power to the Primary and Secondary Channel Electron Multipliers (PCEM and SCEM). Table III shows the primary HVPS properties. The two detector supplies are single string, but the two independent detectors provide redundancy. The Focus Ring is diode-ORed with the output of the PCEM and a -1000 V Focus Ring supply that is created and controlled in parallel with the SCEM. The optical power supplies are redundant and diode-ORed together. The secondary ground on the HVPS is “zap-trapped” to chassis through back-to-back diodes for protection. Mechanically, the HVPS are built onto two interconnected boards: the Driver Board and the Multiplier Board.

**Table III. Key HVPS Specifications.**

| | PCEM | SCEM | Focus Ring | RPA | DFL | ESA |
|---|---|---|---|---|---|---|
| Voltage Range [V] | 0 to -4500 | 0 to +4500 | 0 to -1000 | 0 to +2000 | 0 to +4000 | 0 to -4000 |
| Ripple | 0.5 Vrms | 0.5 Vrms | 10 Vrms | 0.1 Vrms | 0.5 Vrms | 0.5 Vrms |
| Settling Time | N/A | N/A | N/A | 100 ms to 0.1% | 100 ms to 0.1% | 100 ms to 0.1% |
| Accuracy | 5 V | 5 V | 20% | 0.5 V | 4 V | 4 V |

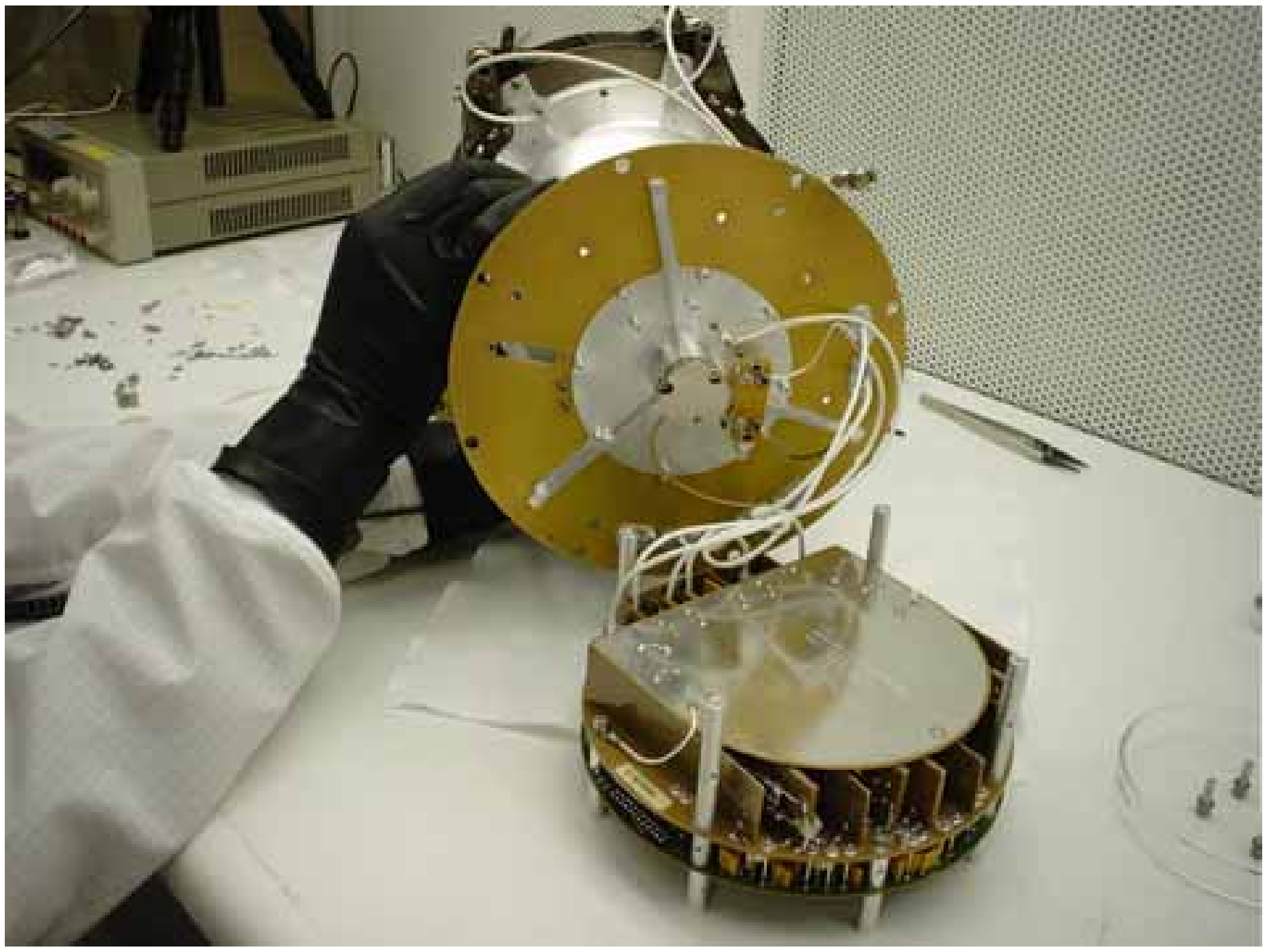

Figure 23. HVPS being assembled into the SWAP sensor

### *3.4.3 Control Board*

The SWAP Control Board provides the electrical interface between the *New Horizons* spacecraft and the SWAP instrument. All command, telemetry, power, safe/arm, and actuation interfaces reside on the Control Board. Software on an 8051 microcontroller responds to commands, controls the operation of the instrument, sequences the high voltage power supplies, collects the data, and formats telemetry for down-link.

SWAP communicates to the spacecraft through two redundant (A & B side) asynchronous RS-422 interfaces. SWAP accepts serial commands, produces serial telemetry, and synchronizes communication with the spacecraft through a Pulse Per Second (PPS) line. The command and telemetry data are transmitted at 38,400 Baud.

The Control Board contains the EMI filters and DC-DC converters, which create the +5 and -5 Volt secondary power rails that are isolated from the spacecraft primary bus. Two separate 1.5W DC-DC Converter provide power for the instrument. Since the DC-DC converters do not have provisions for synchronization, each converter has an independent EMI filter to eliminate low-frequency noise from the beat between the two converter oscillators. The Control Board has a set of power MOSFETs, which allow us to switch low voltage power to the PCEM HVPS, SCEM HVPS, Optical HVPS Bank A, Optical HVPS Bank B, and the Housekeeping circuits.

The micro-controller provides all of the on-board processing required to operate the instrument, responds to commands, and produces telemetry. The CPU executes at 4.9152 MHz, and a minimum of 12 clocks are required to execute a single instruction, so the top speed of the processor is 0.4 million instructions per second (MIPS). From this low rate, the clock can be divided down further to reduce the instruction rate to 0.05 MIPS. Due to spacecraft-level power constraints every effort was made to reduce the operating power required for SWAP. This maximized SWAP's ability to remain on during the Pluto encounter sequence when the spacecraft is power-limited and only runs two optical instruments at a time.

Boot code for the micro-controller resides in a radiation hardened 32k x 8-bit Programmable Read-Only Memory (PROM). This ensures that the instrument can always boot and establish basic communication with the spacecraft even if other memory devices have endured temporary upsets or even permanent degradation. Two separate 128k x 8-bit Electrically Erase-able Programmable Read-Only Memory (EEPROM) devices provide redundant storage for a 64k x 8-bit storage area for program code an a 64k x 8-bit storage area for Look-Up Tables (LUTs). Finally, a 128k x 8-bit Static Random Access Memory (SRAM) is used to provide 64k x 8 bits of code memory and 64k x 8 bits of data memory. During normal operations, if the program code in one of the Electrically Erasable Programmable Read-Only Memory (EEPROM) banks has a valid checksum, then it is loaded into RAM by the Boot PROM and executed. A Field-Programmable Gate Array (FPGA) controls the memory mapping and memory windows that the micro-controller needs to access these devices.

Signals received from the primary and secondary CHAMPs are processed on the Control Board. When the electro-optics have been set to the appropriate levels and sufficient settling time has elapsed, the software opens up an acquisition window and totals all of the primary and secondary CEM pulses that occur during the acquisition window. Whenever a secondary or primary event occurs, a 100 ns coincidence window is started. If a pulse is received from the other charge amplifier during the 100 ns coincidence window, then a coincidence counter is incremented. Due to the electro-optic design, primary and secondary events can be received in either order. There is no dedicated start or stop channel.

The Control Board sets the RPA, DFL, and ESA high voltage optical settings using a set of independent digital-to-analog converters (DAC). The analog settings, along with digital enables, and the low voltage power for the primary and redundant supplies are carried to the HVPS over a dedicated board-to-board connector. The HVPS returns the analog current and voltage monitors to the Control Board for analog housekeeping and real-time monitoring.

The health and safety of the instrument in general and the CEMs in particular are monitored extensively by the Control Board. The count rate of each detector is totaled and checked against a software limit every 0.5 s. The supply voltage and strip current for each CEM is monitored and verified by the software every 1 s. The strip current is also compared against a threshold level with 32 values. If the strip current exceeds this threshold, the software receives an interrupt and may immediately take action. This information, along with temperatures in the instrument and the current and voltage monitors of the low voltage power supplies are transmitted to the ground in housekeeping telemetry packets.

### 3.5 Modes of Operation

SWAP flight activities are conducted during the following operational phases: 1) Commissioning; 2) Jupiter Encounter; 3) Cruise/Annual Checkout; and 4) Pluto Encounter. To accommodate these different activities, SWAP has various modes that are controlled by the SWAP onboard flight software (FSW). A mapping of the operational phases to SWAP modes is described in Table IV, and a description of the modes and corresponding power and telemetry resource consumption is shown in Table V.

Mode changes are triggered by command with the exception of the BOOT to Low Voltage Engineering Mode transition, which is based on time since power up, and is automatic. Flight operations, and therefore the modes, drive the requirements for the FSW.

**Table IV. Operational Phases and SWAP Modes**

| SWAP Operational Phase | Description | BOOT | LVENG | LVSCI | HVENG | HVSCI |
|---|---|---|---|---|---|---|
| Commissioning: Low-Voltage | LV functional was performed and memory load and dump capabilities were tested | X | X | X | | |
| Commissioning: Door Opening | SWAP door was opened using the S/C actuator bus. SWAP was off. | | | | | |
| Commissioning: Initial High-Voltage | HVPS were turned on for the first time. High-voltage functional test was performed | X | | X | X | |
| Commissioning: Nominal Science | Fully functioning using HVSCI mode | X | X | | X | X |
| Jupiter Encounter | Fully functioning using HVSCI mode | X | X | | X | X |
| Cruise | Functional test to be performed; CEM gain test and EEPROM refresh | X | X | | X | X |
| Pluto Encounter | Fully functioning using HVSCI mode | X | X | | X | X |

**Table V. SWAP Modes and Resources**

| Name | Description | +30V S/C Power (W) | SWAP Internal (+5V/-5V) Power (W) | Telemetry Output [1] **(bits/sec)** |
|---|---|---|---|---|
| OFF | No power being applied to SWAP | 0 | 0 | 0 |
| BOOT | Bootup run from PROM image.<br>Checksum of non-volatile area and read/write tests of RAM performed. This mode has the ability to upload and commit new code and table images to EEPROM. Transition to LVENG is automatic based on time since power up. | 0.70 | 0.44 | 1200 |
| LVENG | Low-Voltage Engineering run from EEPROM image.<br>Safe mode (no HV), no sweeping takes place, for engineering and testing | 0.70 | 0.44 | 600 |
| LVSCI | Low-Voltage Science run from EEPROM image.<br>The instrument will begin sending test pulses through the CHAMPs in a pattern that will simulate a normal science sweep. This mode emulates HVSCI without HV. | 0.74 | 0.48 | 291 |
| HVENG | High-Voltage Engineering run from EEPROM image.<br>HV can be set according to commands rather than being controlled by the sweep algorithm; used during initial turn-on, calibration and ramping of HVPS prior to nominal science use in HVSCI mode. | 1.33 | 0.98 | 290 |
| HVSCI | High-Voltage Science run from EEPROM image.<br>HV is on; ESA/RPA/DFL supplies are swept according to tables | 1.33 | 0.93 (avg) | 291 |

[1] These are typical values for these modes. Other rates can be commanded as needed.

## 3.6 Flight Software

### *3.6.1 Overall Capabilities*

The FSW runs on the 80C51 microprocessor at 4.9152 MHz, equating to approximately 400,000 instructions per second. It is written almost entirely in the C programming language. The FSW consists of two different code images stored in nonvolatile memory. At power on, the FSW runs directly from a programmable, read-only memory (PROM) code image. This code gathers diagnostic information on the SWAP hardware, provides the ability to reload EEPROM code or tables and determines which copies of EEPROM code or table to use for science acquisition. Once the EEPROM code and tables have been chosen, the EEPROM code is copied into RAM for execution.

The FSW addresses these requirements areas: data interfacing and synchronization with the spacecraft, instrument engineering and safety, and science support. The software is synchronized with the 1 pulse-per-second (1PPS) signal from the spacecraft. The spacecraft sends a mission-elapsed time (MET) to each of the instruments for a common time tag. The one-second period is subdivided into twenty 50-ms time periods, which allow for scheduling of software processes and managing the S/C timing requirements for command and telemetry.

The software processes for science provide support in 1) managing the setting of HVPS optical power supply levels during science-data acquisition (during HVSCI mode, the ESA, RPA and DFL are set by means of lookup tables), 2) starting and collecting of counter data from the PCEM, SCEM, and coincidence electronics, and 3) performing calculations onboard using the counter data and telemeter science-data products.

In nominal operations, commands are required for the high-voltage ramp after a power on, but once the ramp has been completed, a small number of commands are used for initiating science acquisition that set a configuration for acquisition, the deflection angle and the telemetry rates.

The FSW outputs three types of telemetry: instrument state information, which can be read and acted upon by the spacecraft software for onboard anomaly recovery; engineering telemetry (housekeeping, messages and memory dump); and science telemetry (real-time, summary and histograms). The telemetry is summarized in Table VI.

**Table VI. SWAP Telemetry**

| Name | Description | Typical Packet Rate | Typical Bit Rate (bps) |
|---|---|---|---|
| Instrument State | SWAP information used onboard by S/C autonomy software. Instrument state includes heartbeat and safety flags. | 1 Hz | Varies – not considered part of SWAP |
| Housekeeping (Engineering) | Instrument status are contained in this packet (e.g., opcode echo, FPGA status, non-safety-critical monitors, software variable status, etc.) | 0.5 Hz in LVENG; once every 64 seconds in other modes | 292 in LVENG; 9.1 in other modes |
| Memory Dump (Engineering) | Memory dump for diagnostics | Low | N/A |
| Message (Engineering) | Warning message that is issued to provide additional detail if anomalous activities occur. | Low | N/A |
| Real-time (Science) | Contains the most detailed temporal data from SWAP consisting of RPA, ESA and DFL commanded DAC levels corresponding primary CEM, secondary CEM and coincidence counts. | 1 Hz during commissioning; 1 set of 64 consecutive packets per hour otherwise | 280 during commissioning and Pluto Encounter; 5.0 otherwise |
| Summary (Science) | Summary accumulation over a set period of time, typically an hour. A calculation is performed using the coincidence data from the previous 64-second period to generate values that are related to density, velocity and temperature and output in telemetry. The minimum, maximum and variance of these parameters are also output in telemetry. | 1 per hour | 0.19 |
| Histogram (Science) | All of the count data are accumulated over a set period of time, typically a day. The data are accumulated in a normalized energy array of 2048 elements. The normalization occurs such that for each 64-second acquisition period, the center of the array corresponds to the solar wind peak during those 64 seconds. The entire array is brought down in telemetry. Because of its size, the array must be trickled out in 64 packets. This packet contains the histogram header (APID 0x586) and the beginning of the histogram data. It is followed by 63 Science Histogram Data (APID 0x587) packets to create a complete histogram set. | 1 set of 64 (1 header + 63 data) packets per day | 0.15 |

### *3.6.2 Science-Data Collection*

Science data are collected in the HVENG and HVSCI modes. HVENG was used extensively during commissioning for initial HV ramp-up and instrument response characterization; however, HVSCI is the primary SWAP science mode. In HVSCI, the optical power supplies are stepped every 0.5 seconds and the PCEM, SCEM and coincidence counter data are collected with each step. During each 0.5-second period, approximately 100 milliseconds are allowed for the optical power supply settling time and 390 milliseconds are allocated to the counter sample period. The duration of the latter is set with digital hardware. An overall cadence of 64 seconds consisting of 128 0.5-second steps defines the 64-second science-acquisition frames and hence all science FSW activities. Two different methods of sweeping during the 64 seconds can be performed – called the coarse-fine and coarse-coarse sweeps – and are user selectable. A typical coarse-fine sweep consists of a 32-second coarse sweep, which covers the entire energy range with 64 logarithmically-spaced optical power supply values, followed by a 32-second (also 64 0.5-second steps) fine sweep. A coarse-coarse sweep consists of two 32-second coarse sweeps performed in one 64-second period. For both sweep types, the optical power supply values are read from one of several user-selectable tables.

For the coarse-fine sweep, the peak value of the coincidence counter during the coarse sweep is calculated to determine the center of the fine sweep so that finer resolution sweeping around the peak response can be performed.

A 64-second acquisition frame period running a coarse-fine sweep from SWAP calibration is shown in Figure 24, which plots the PCEM, SCEM and coincidence count rates, and the ESA and RPA voltages. A relationship between energy and coincidence rate can be derived from such a plot given that with each 0.5-second step, a different incident energy is defined for each ESA-RPA voltage combination. In this experiment, DFL was set to 0 throughout this data set since the angle relative to the SWAP aperture was 0°. The figure shows that for SWAP, the energy is swept from high to low, with the coarse sweep taking place from time = 0 to 32 seconds and a fine sweep from time = 32 to 64 seconds.

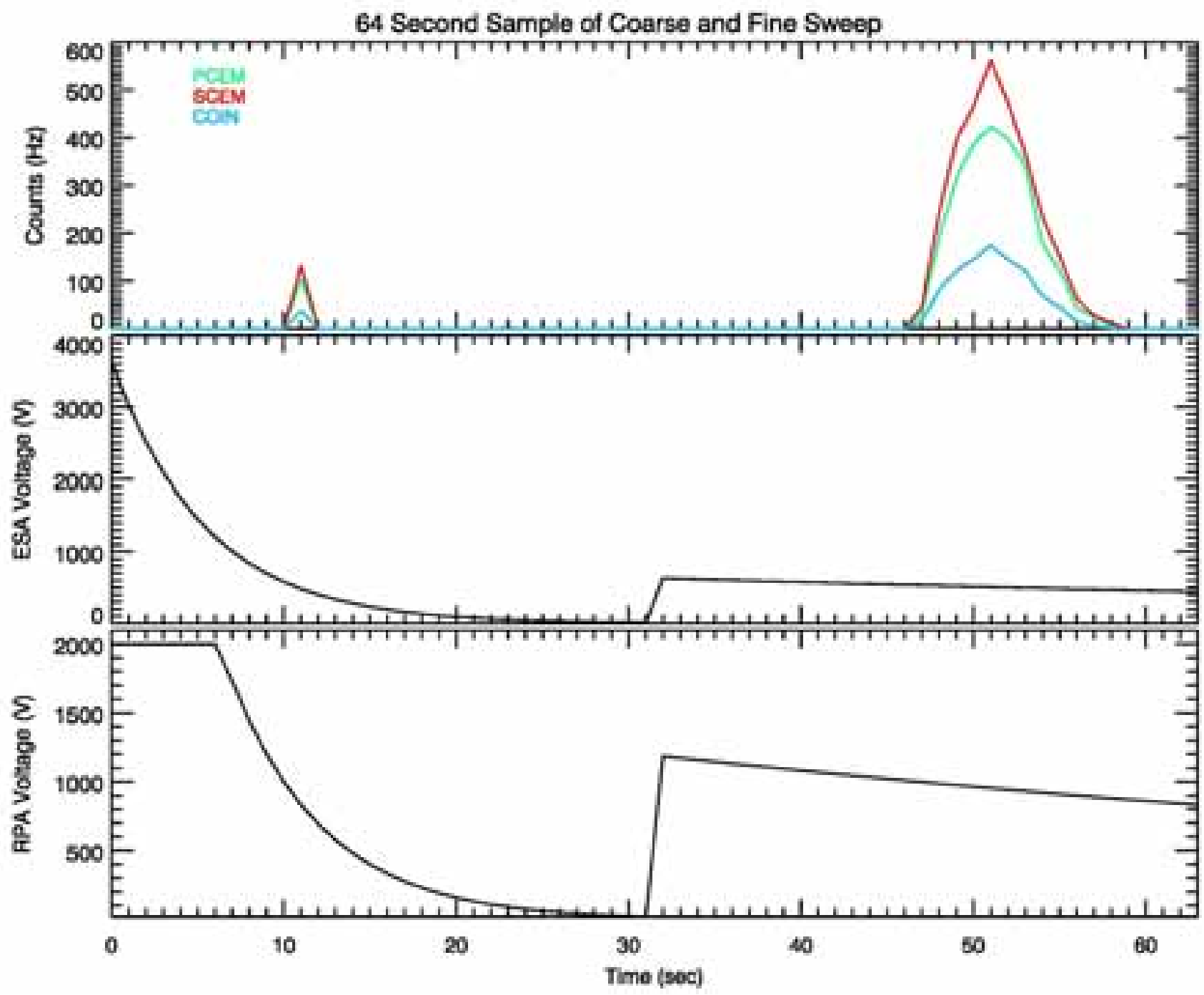

Figure 24. A coarse and fine sweep acquired during SWAP ground calibration activities.

To control acquisition while minimizing commanding to SWAP, science data collection instructions are stored in user-selectable tables in two identical EEPROMS. The hierarchy of tables consists of:

- Plan (64 tables) – schedules which sweep table is active for each 64-second acquisition frame.
- Sweep (16 tables) – defines which energy step and ESA/RPA table is used for each 0.5-second sample. This determines, e.g., whether a sweep is coarse-fine or coarse-coarse.
- ESA (4 tables) – provides a mapping between energy step and the digital-to-analog (DAC) value to program the ESA optical power supply.
- DFL values (1 table) – maps energy step and angle to DFL DAC setting.
- RPA values (4 tables) – maps energy and angle to RPA DAC setting.

With the above tables, the user chooses a plan number and angle setting to define science acquisition.

SWAP is typically run for long periods of time – i.e., days at a time. Although, the real-time science packet can be configured to telemeter each of the 0.5-second samples, because of telemetry bandwidth limits, the summary and histogram packets can be output instead. They are derived directly from the same data that are contained in the real-time science packet.

The summary is composed of a series of ongoing calculations based on the fine portion of a coarse-fine sweep. An array of coincidence counts is collected and processed during each 64-second acquisition frame. Figure 25 shows an idealized coincidence count rate response collected during a 32-second fine sweep in which the values related to pseudo-velocity, pseudo-density and pseudo-temperature of a fine sweep are defined. For each of the summary values (angle, pseudo-density, pseudo-velocity and pseudo-temperature), a running log of the sum, minimum, maximum and variance are calculated. While the sum, minimum, and maximum are represented in telemetry as 32-bit numbers, the variance is calculated by summing the squares of each of the values and accumulated into a 64-bit number. Because the summary packet defaults to transmitting once every hour, SWAP can use the summary to collect data even during low-telemetry rate periods. The rate can be altered via SWAP command for testing or other purposes.

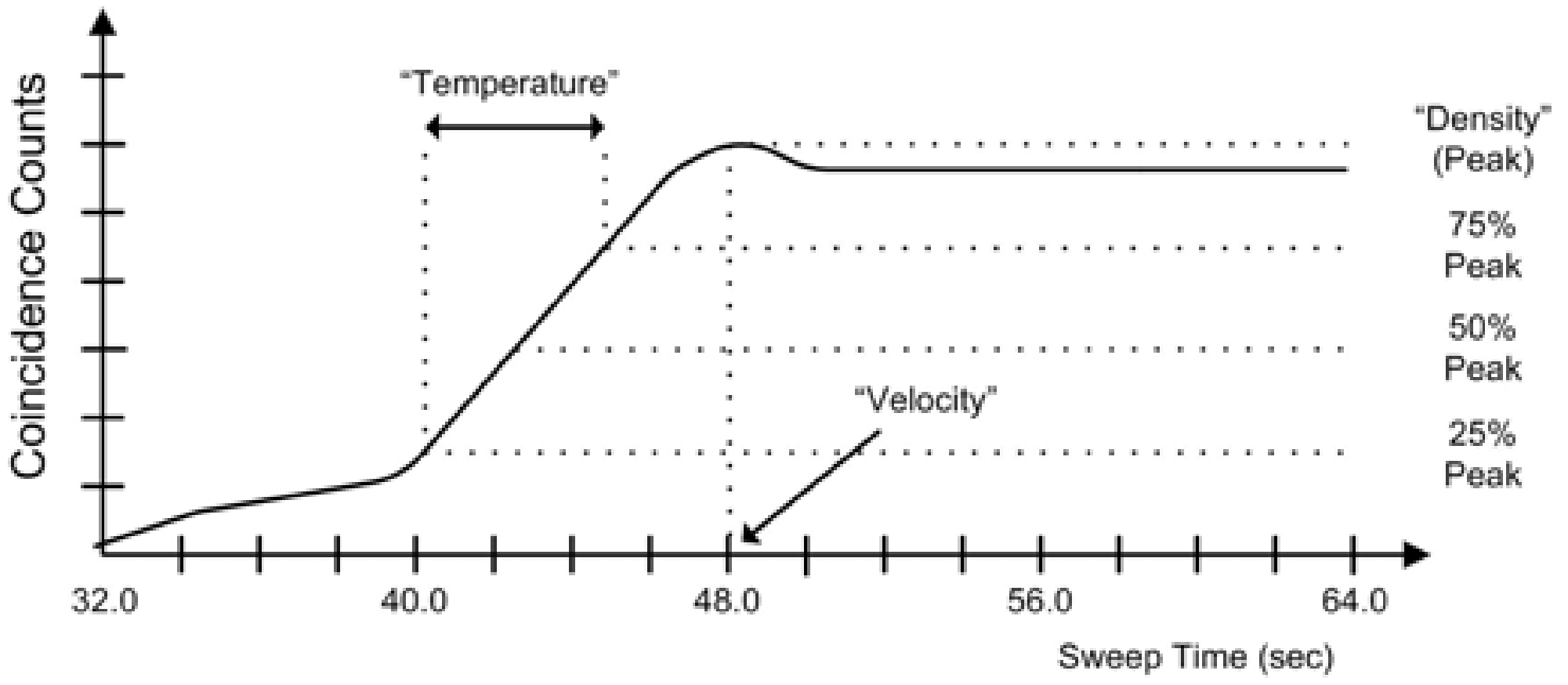

Figure 25. Illustration of SWAP Summary Calculation

Figure 26 summarizes the histogram calculation. The top half of the figure shows a time line of two consecutive 64-second frames. The red and blue plots show the coincidence counters during the coarse sweep and the black plots show the response from the fine sweeps. The red plot shows that the peak was found at an energy index of 512, which results in the FSW running a subsequent fine sweep from energy indices 480 to 543 in order to center the coarse peak in the middle of the fine sweep.

The coarse sweep data are shifted and accumulated into a normalized histogram array that is composed of 2048 32-bit energy bins. It is "normalized" because each bin corresponds to a relative energy index where 1024 always corresponds to the peak of the coarse array. In Figure 26, for Frame i, the fine peak energy index is 512. This number is used to calculate the shift required to move the wider-spaced coarse sweep so its energy index 512 aligns with histogram array index 1024. Note that the wide spacing of the coarse array is transferred also, so there are gaps between consecutive coarse samples accumulated into the array. In Frame i+1, the fine peak energy index is 508, so a slight shift between the frames is noted and results in the second coarse data set not aligning exactly over the first coarse data set.

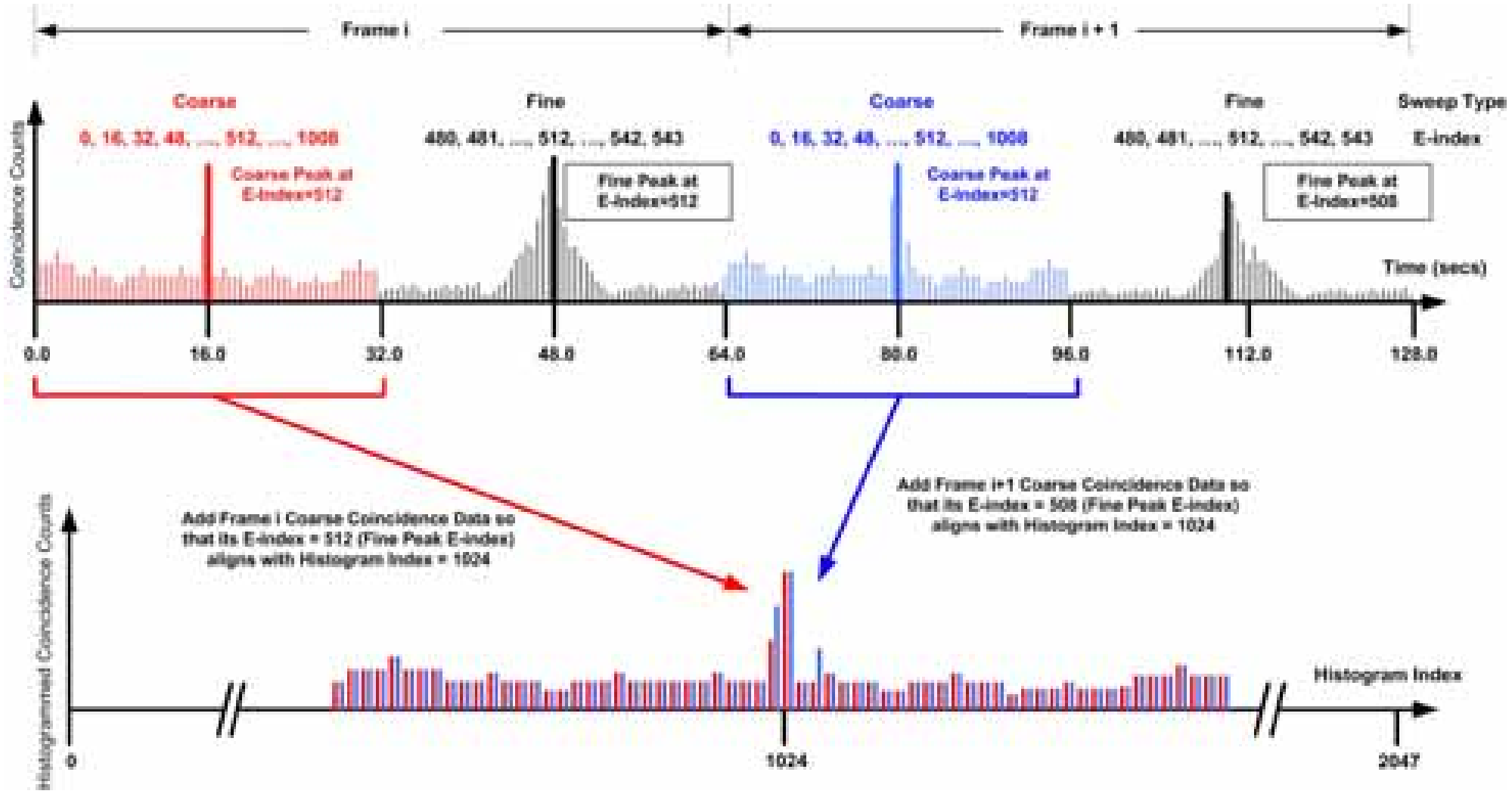

Figure 26. Illustration of Histogram Calculation. The bold vertical lines on the upper plot show the maximum coincidence value for that 64-sample (32-second) span.

The number of times a particular bin (also with indices 0 through 2047) has been accumulated is telemetered along with the histogram array so that the activity in each energy bin can be reported. The histogram packet defaults to being sent out approximately once per day. Because of the size of one histogram array, the packet is divided and sent out over the course of 64 seconds. The histogram is used so that SWAP can collect data even during times where only low-telemetry rates are allowed. Like the summary packet, the rate can be altered via SWAP command.

# 4.0 Instrument Performance

We characterized SWAP's performance through both laboratory testing and electro-optics modeling. The purpose of the laboratory testing was to calibrate the instrument and determine a ground truth. The computer models allowed us to examine more combinations of incident-beam and instrument parameters than possible in laboratory testing and to allow for analysis of other parameter combinations as needed after launch.

## 4.1 Laboratory Testing

SWAP was extensively calibrated at the SwRI Ion Calibration Facility. This facility produces ion beams from <500 eV to 51 keV. The ion species can be selected for mass per charge from 1 to >40. Figure 27 shows the SWAP instrument mounted on the facility's 4-axis positioning system, with the doors that protect the sensor for launch in their open configuration. Full functionality of the instrument was demonstrated during calibration, including its ability to open the doors while under high vacuum. The four-axis positioning system allows vertical and horizontal displacement in the plane normal to the ion bean and rotation about two orthogonal axes that cross in the beam path: the outer rotation is about a vertical axis, and the inner, nested, rotation about a horizontal axis.

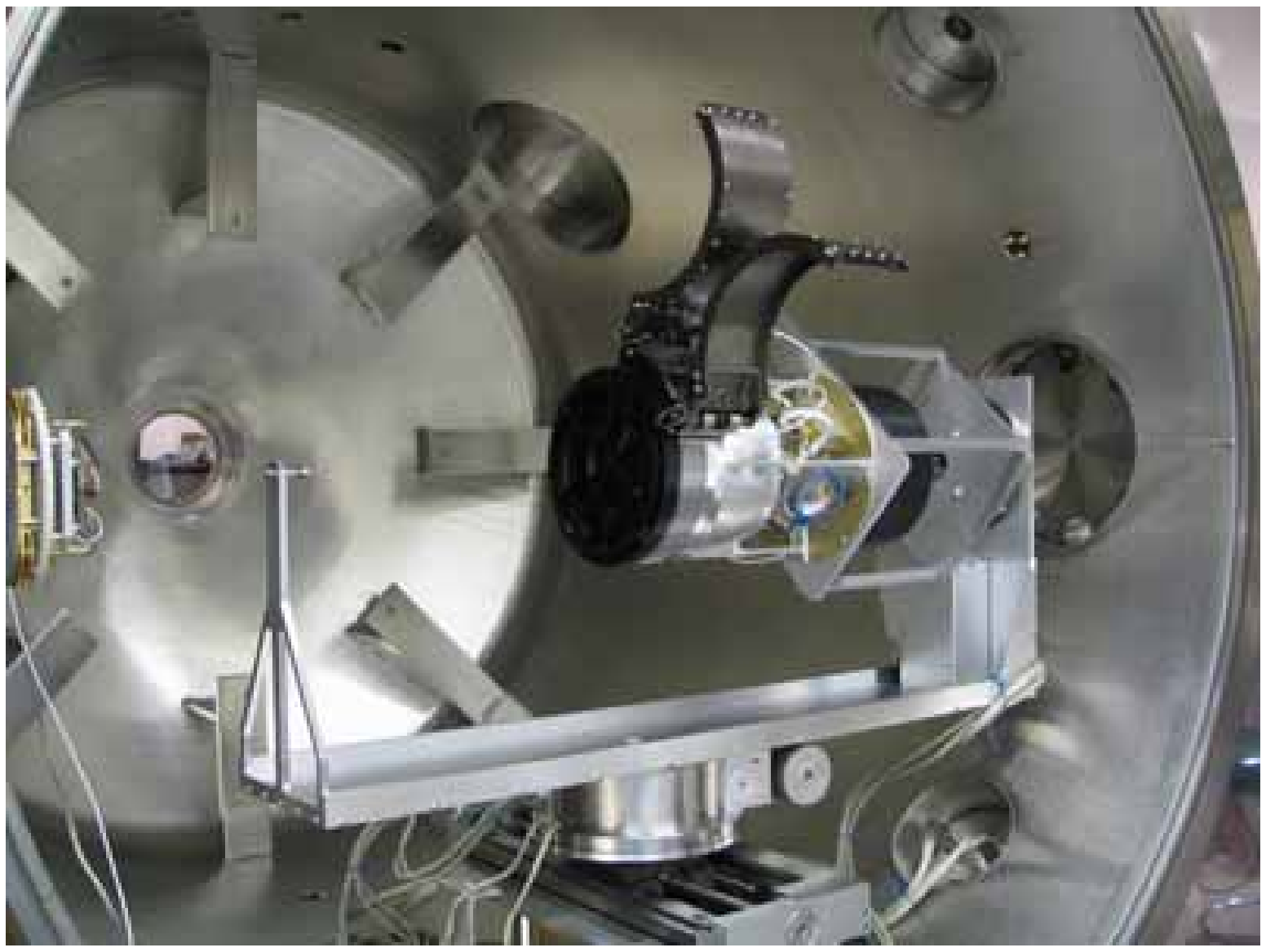

Figure 27. SWAP instrument mounted on 4-axis positioning system

The SWAP instrument was mounted with its center axis aligned to the inner-stage rotation axis, which is equivalent to roll maneuvers the *New Horizons* spacecraft will perform. Because of SWAP's 17-cm diameter aperture, outer-stage rotations resulted in translation in the input aperture. We used the horizontal stage to return the aperture to its original position. The positioning system's range of motion is large enough to fully illuminate the instrument's FOV.

In addition to testing performed on individual components and on a high-fidelity prototype, the full flight instrument underwent a number of calibrations. Calibrations were performed before and after instrument level environmental testing, after refurbishment activities following spacecraft environmental testing and a limited engineering beam test immediately prior to delivery to the spacecraft for final integration. The instrument was calibrated over the full angular response and energies from 500 eV to 8 keV. Due to the nature of electrostatic optics, the response of the instrument to other energies can be easily scaled. Because the solar wind protons have a nominal energy of ~1 keV, a large fraction of the calibration data was taken at this energy. Results presented here are from the instrument in the final delivered configuration.

### 4.1.1 RPA

Figure 28 shows the measured energy response of the RPA to a beam of 1 keV protons. The primary, secondary, and coincidence rates are shown in blue, green and orange, respectively. The RPA grids are curved, rounding the cutoff of the transmission as compared to a planer RPA. This is seen as the reduction of counts from 950 V to 1000 V. However, as shown in Figure 10, above, by making small steps in the RPA voltage, we are able to differentiate beams that are only ~1-2 eV apart at 1 keV.

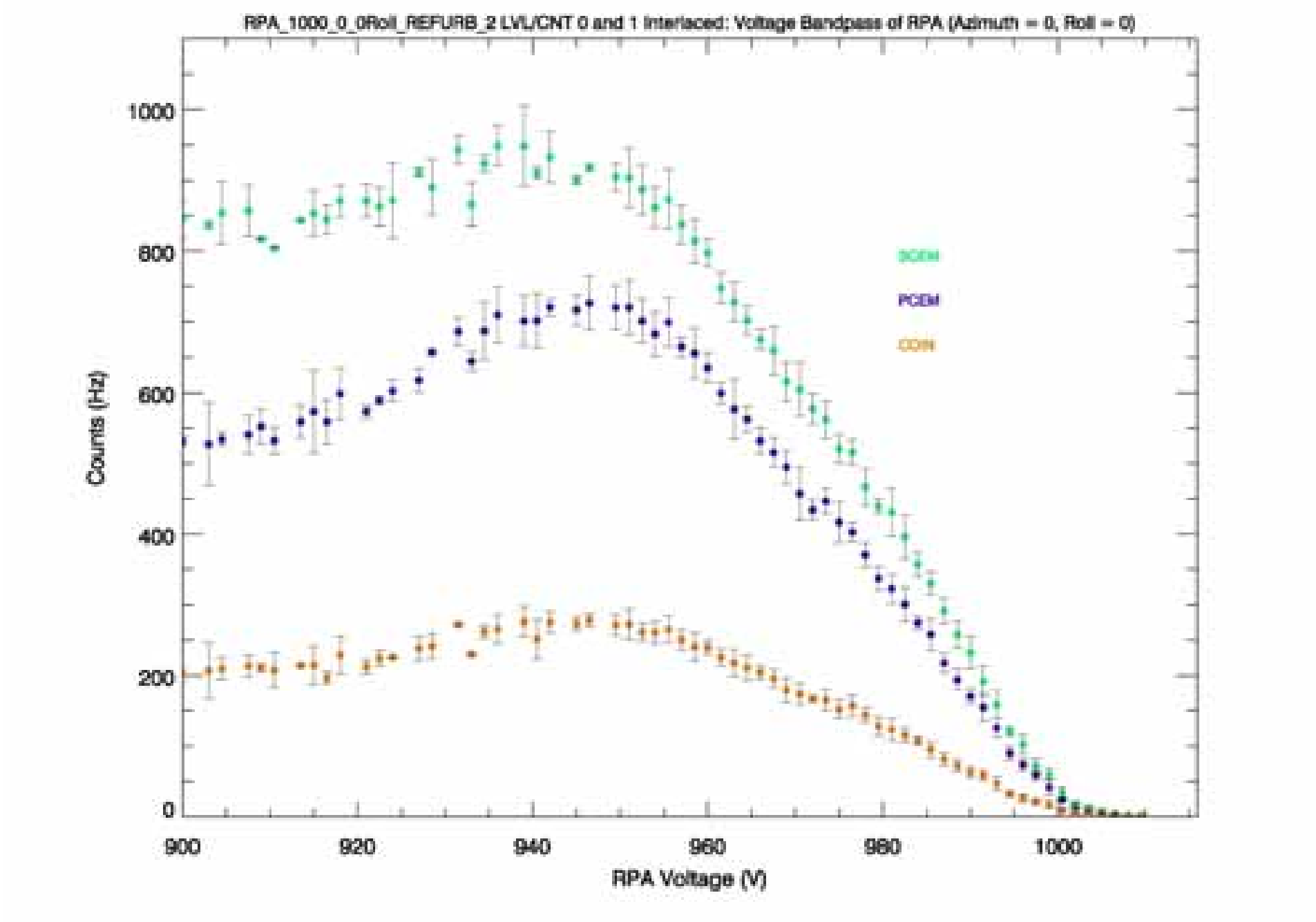

Figure 28. Measured energy response of RPA to a beam of 1 keV protons

Figure 29 shows the general shape of the RPA response over the full voltage range. Normalized coincidence is plotted as a function of the RPA voltage/beam energy. Because the individual RPA grids have a finite thickness, they act as a series of electrostatic lenses. The features in this figure, including maximum near 0.945 of the 1 keV beam energy, are due to the focusing properties, as discussed below in Section 4.2.

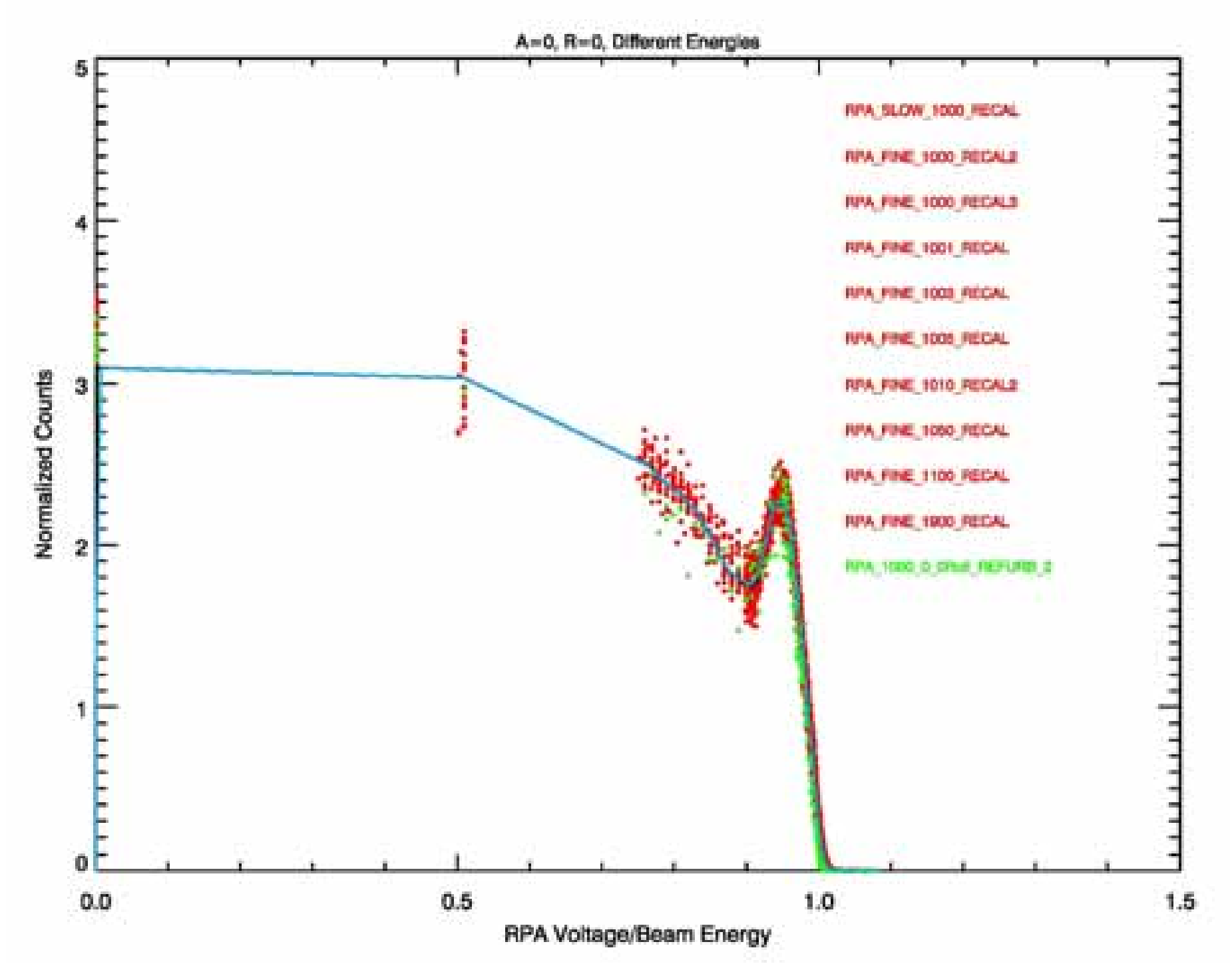

Figure 29. General shape of RPA response over full voltage range

### *4.1.2 Deflector System (DFL)*

The FOV of the SWAP instrument is 276° about the spacecraft roll axis by 10° in the plane normal to the roll axis. SWAP uses an electrostatic deflection (DFL) plate to increase the FOV out of plane by up to 15° in azimuth (α). The required DFL voltage, normalized to the beam energy goes as

$$\frac{V_{deflector}}{E_{beam}} = 0.0007\alpha^2 + 0.0209\alpha + 0.1009 \qquad (1)$$

where $\alpha$ is the angle from the plane.

For non-normal incident particles, the RPA response needs a calibration correction. For an ideal RPA, non-normal incident particles with $E/q > (V_{RPA} / \cos^2 \alpha)$ would be passed. The unique focusing properties of the SWAP RPA lead to a more complicated response function. Figure 30 shows the normalized response of the RPA as a function of the azimuth angle. The RPA voltage is scaled by the incident beam energy. Protons with energy of 1000, 1010, and 1900 eV are shown in red, green and blue, respectively. We set the DFL voltage according to Equation 1, leading to an empirically derived calibration function for SWAP:

$$f = \cos^2(2\alpha), \alpha \le 4.0$$
$$f = \cos^2(3.5\alpha - 20) - 0.03, 4.0 < \alpha \le 10 \quad (2)$$
$$f = \cos^2(5.5\alpha - 65) - 0.085, \alpha > 10$$

For the SWAP RPA, non-normal incident particles with E/q > ($V_{RPA}$ / $f(\alpha)$) would be passed.

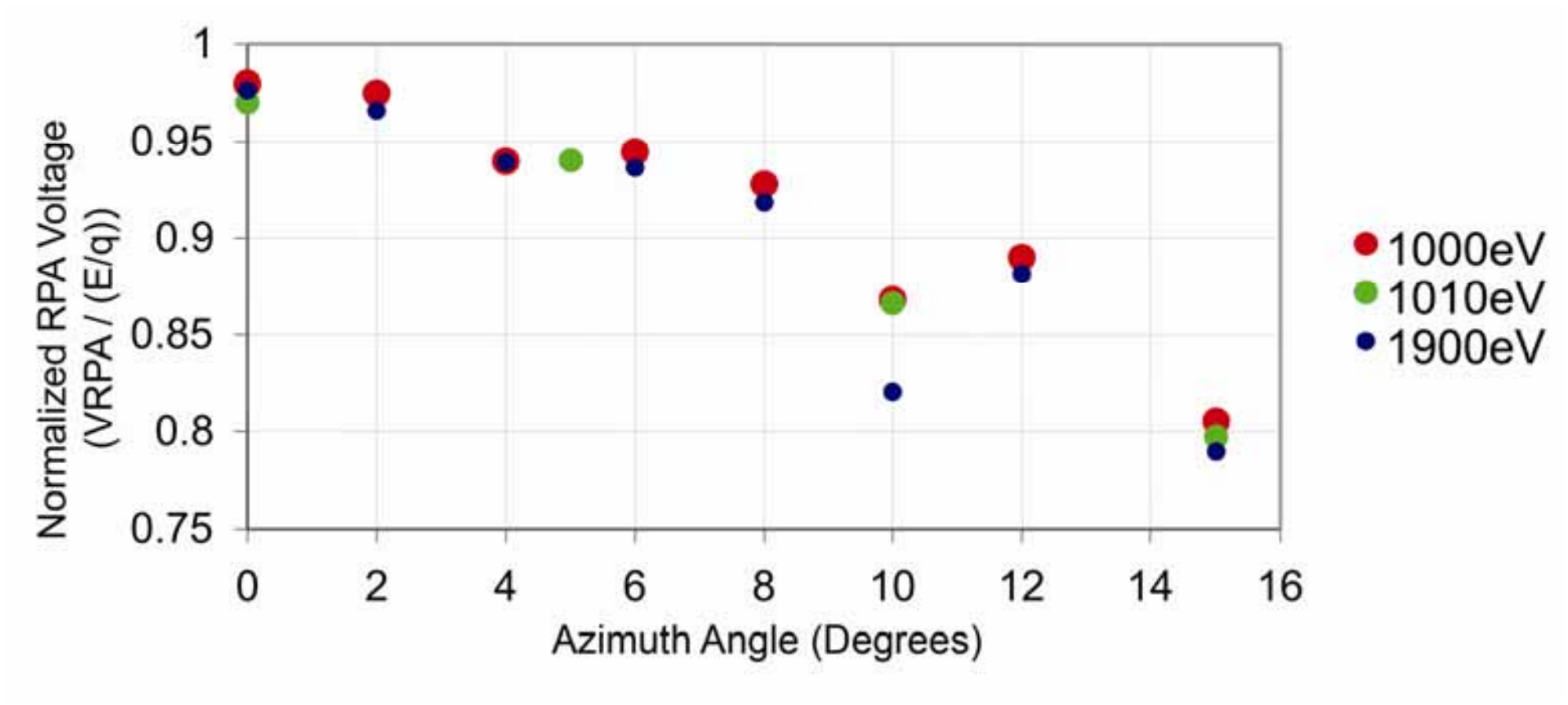

Figure 30. The normalized peak response of the RPA is shown as a function of the incident azimuth angle. The RPA voltage is scaled by the incident beam energy. Protons with energy of 1000, 1010, and 1900 eV are shown in red, green and blue, respectively. The DFL voltage was set according to Equation 1. For an ideal RPA, non-normal incident particles with E/q > (VRPA cos2 α) would be passed. The unique focusing properties of the SWAP RPA lead to a more complicated response function, leading to an empirically derived calibration function for SWAP as shown in Equation 2.

### 4.1.3 ESA

Figure 31 shows an example of the energy-angle response of the ESA to a 1keV proton beam. The RPA and deflector system's voltages were at zero for the data shown here. For this ESA, the analyzer constant is 1.88, the energy resolution ΔE/E is 8.5%, and the angular FOV, undeflected, is 10°.

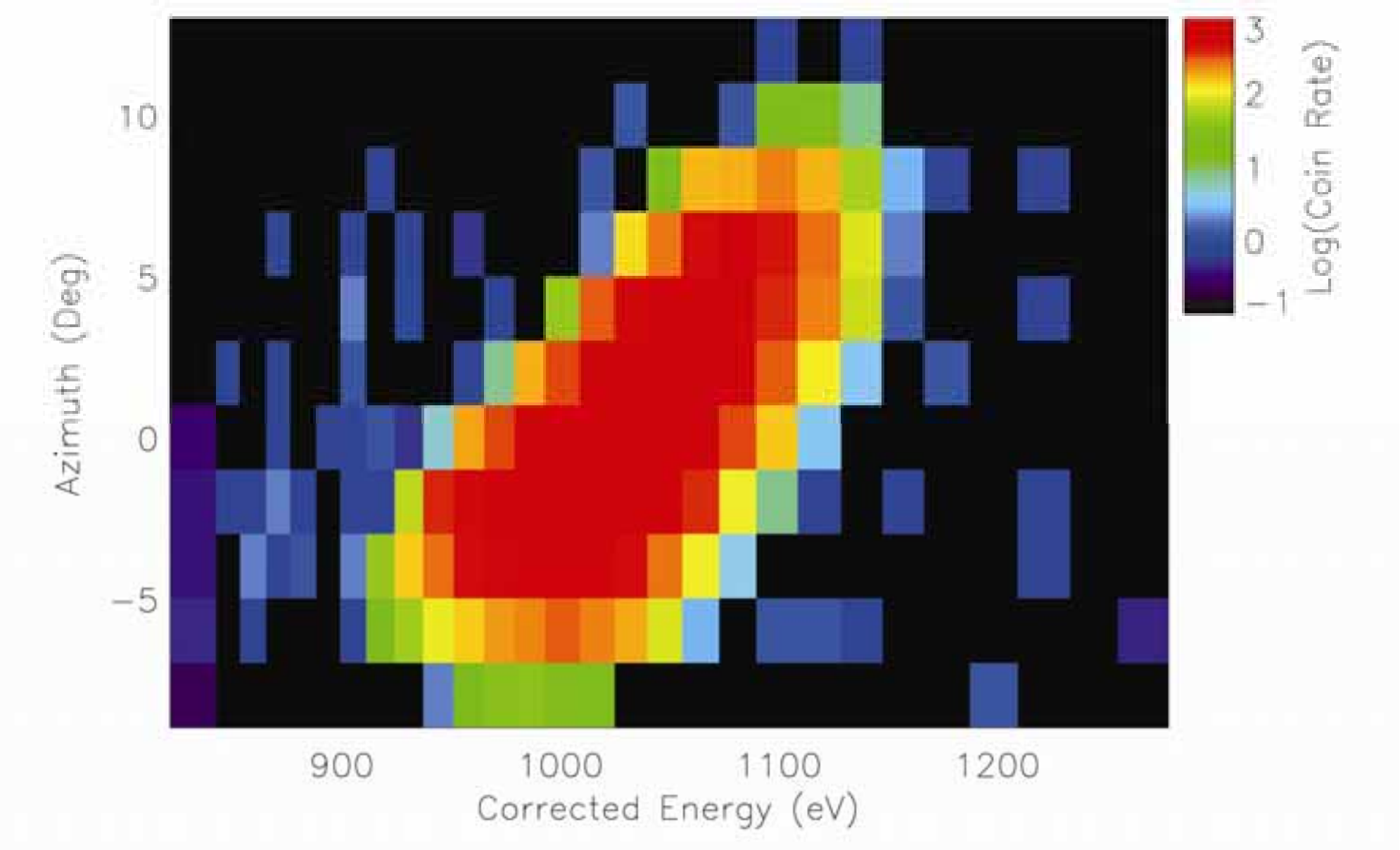

Figure 31. The energy-angle response of the SWAP instrument is shown for a 1 keV proton beam. The RPA and deflector voltages were set to zero for this data.

In addition to performing the coarse energy measurement, the ESA also blocks out UV light to the detectors. A Krypton line source, with emission at 123.6 nm and approximately the same intensity of the Sun at 1 AU, was used to test SWAP for light leaks. We used this worst-case test since SWAP operates further out in the heliosphere where the UV flux is lower by a factor of $R^2$. For all angles tested with the UV source, the count rate never exceeded 1 Hz for either primary or secondary rates, and no coincidence events were observed.

### *4.1.4 Detectors*

Ions that pass through the sensor then pass through a thin carbon foil and are measured by the PCEM. Secondary electrons liberated from the carbon foil are attracted to the SCEM. Events that are measured by both the PCEM and SCEM within 100 nanoseconds time are recorded as coincidence (COIN) events.

We determined the operation voltage for PCEM and SCEM by sweeping the voltage while illuminating the instrument with a constant intensity 1 keV proton source. Figure 32 shows the response of the PCEM, SCEM, and the COIN rate as a function of voltage. For the data shown here, the voltage applied to the PCEM and SCEM were equal. From these tests, we selected a nominal operation voltage of 2100 V at start of mission, which is well out on the saturation part of the CEM gain curves. The high voltage power supplies for the CEMs were designed to output up to 4.5 kV, allowing us to increase the voltage to the CEMs if their gain degrades over the mission lifetime.

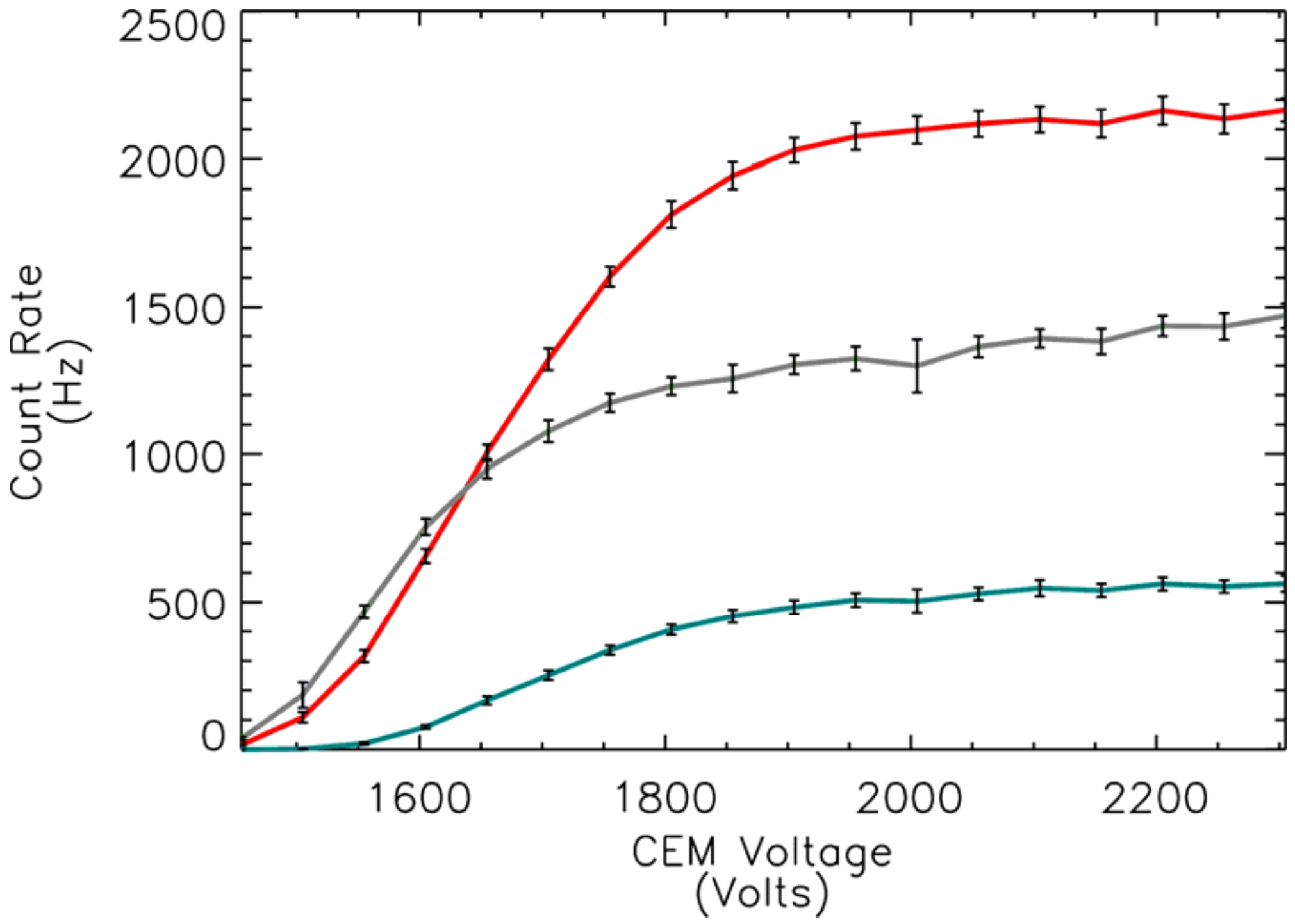

Figure 32. PCEM, SCEM, and coincidence rate is shown as a function of CEM voltage. The magnitude of the PCEM voltage is set equal to the SCEM voltage for the data shown here.

## 4.2 Instrument Modeling

We obtained instrument-modeling results with ion-optics simulations using SIMION 7.0 software (Dahl, 2000). SIMION is an ion-optics simulation program that models problems with 2D or 3D electrostatic potential arrays. It creates potential arrays based on a user-given geometry input (usually in the form of a text file containing commands). The potentials of points outside the electrodes are determined by solving the Laplace equation by finite difference methods with a maximum of 50-million array elements. The time-step integration for the calculation of ion trajectories uses a fourth-order Runge-Kutta method, with ions flown individually or in groups. Parameters such as position, velocity components, energy, etc. can be recorded at different locations in the instrument model.

The SWAP model consists of a 3D geometry with cylindrical symmetry. The limitation of 50-million array elements prevents reproducing the finer structure of the RPA design with the instrument design in the same potential array. Therefore, the RPA grids are modeled separately with a much finer resolution. We discuss the SWAP model (instrument with "ideal" RPA) in Section 4.2.1, and describe the RPA model in Section 4.2.2.

### 4.2.1 Instrument Model with Ideal RPA

For the ESA electro-optics we used a 2D model cylindrical symmetry, sometimes called 2.5D. Nevertheless, ions move in 3D space inside the model. Our model geometry (Figure 33) fully conforms to the as-built mechanical design of the flight instrument, with as much detail as the resolution permits. This mechanical detail helps us account for effects such as fringing fields. The presence of doors breaks the symmetry and is not accounted for in the model, but we can neglect the effects on the electro-optics within the field-of-view of the instrument.

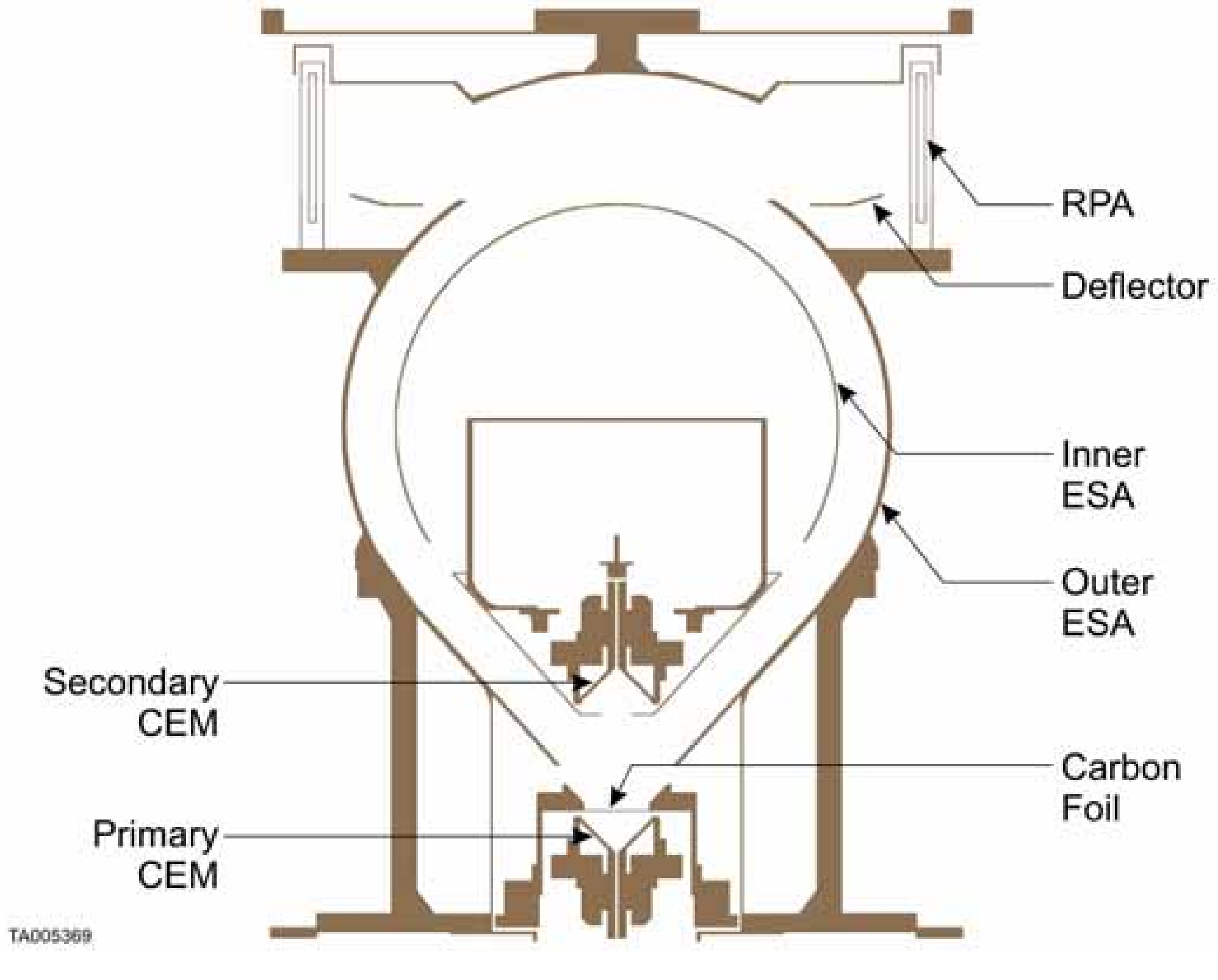

Figure 33. Cross section of the electro-optics model for the ESA. The model has a main axis of symmetry.

Figure 34 shows the focusing properties of the ESA for normal incidence ions in two orthogonal planes. In both representations, the ions are focused inside the ESA part and then accelerated and re-focused when they reach closer to the detectors. However, the impact positions on the primary CEM are nicely spread in the funnel in order to avoid a hot spot.

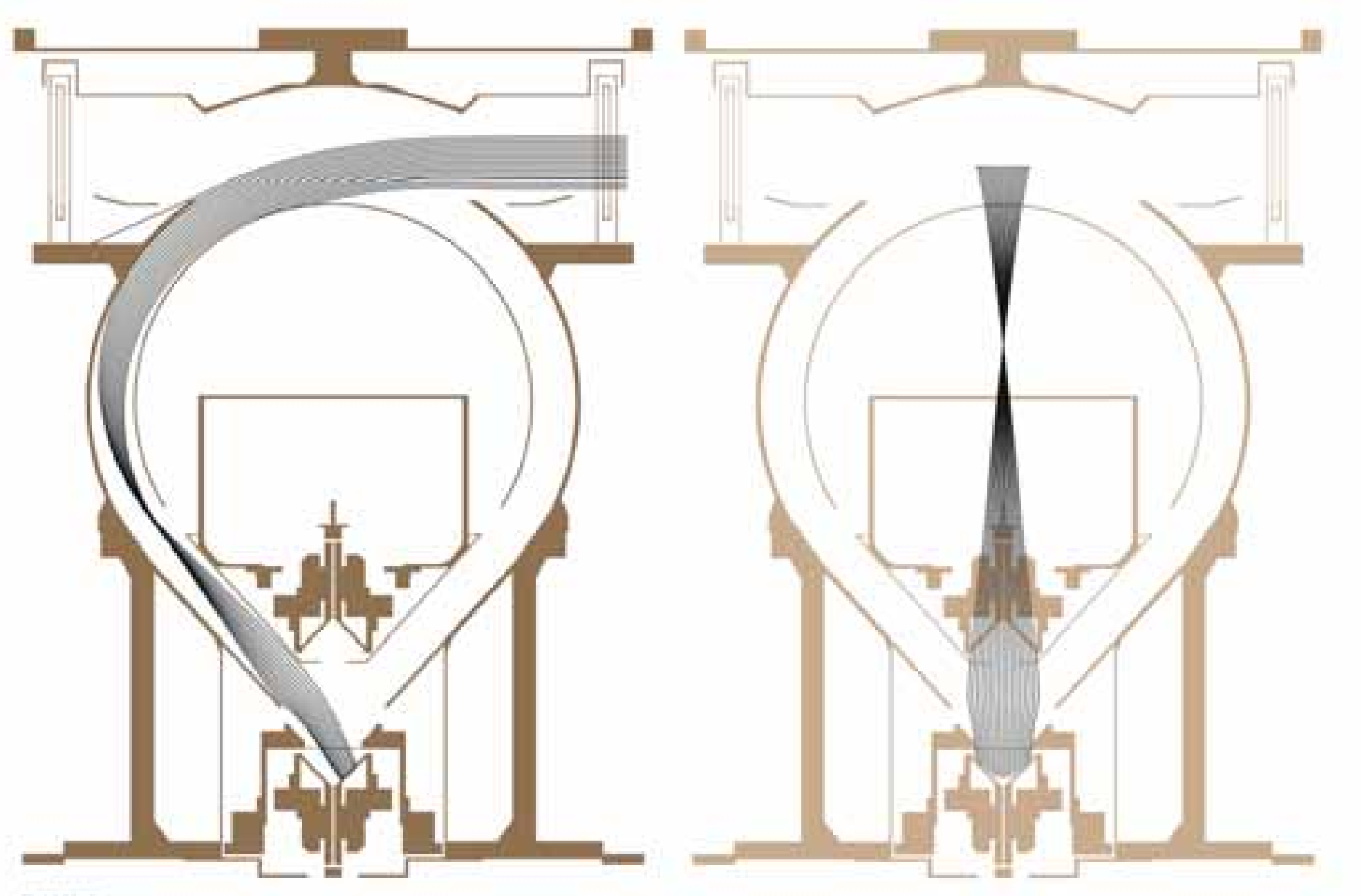

Figure 34. Focusing properties of the ESA shown in two orthogonal projections.

The instrument was thoroughly characterized with ion beam calibrations (above) and the results agree very well with those from the modeling. Table VII compares the energy-per-charge resolution, the k factor and the angular acceptance.

**Table VII. Comparison of the instrument characterization between measurements with ion beam and model**

| | Calibration with ion beam | Instrument model |
|---|---|---|
| E/q resolution ($\Delta E_{FWHM}/E$) | 8.5 % | 9.5 % * |
| k [V/eV] | 1.88 | 1.86 |
| Angular acceptance at HM [°] (undeflected) | -4.9 to +5.2 | -4.6° to 8.1° * |
| *Model is broader because collimating effects of RPA are not included. | | |

Figure 35 shows the angle-energy response. The angle and energy binnings have been intentionally set at the same values for better comparison. The right portion of the figure shows the normalized transmission, which is color-coded using a logarithmic scale. Again, there is excellent agreement between the modeling and the calibration measurements with an ion beam.

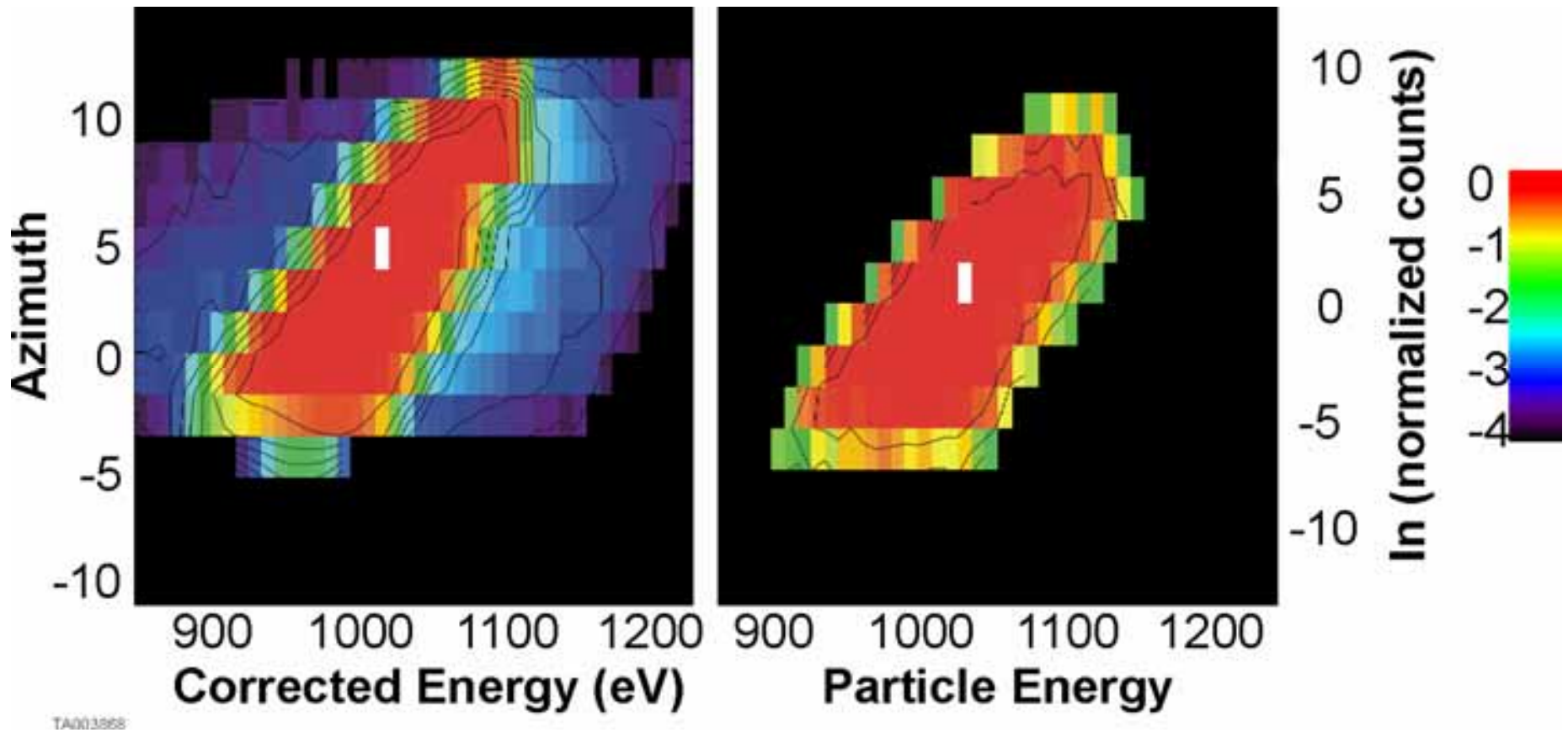

Figure 35. Comparison between the angle-energy transmission function of the instrument (left) and the model (right).

### 4.2.2 RPA Model

We modeled the RPA grids separately with a much higher resolution, matching the thickness and hole pattern of the mechanical drawings. The model includes the curvature of the grids and drilled holes, which are perfectly aligned for the four grids. This configuration does not account for a possible misalignment of the hole patterns, which may cause some differences between the model and actual hardware. Figure 36 shows an isometric view of part of the RPA.

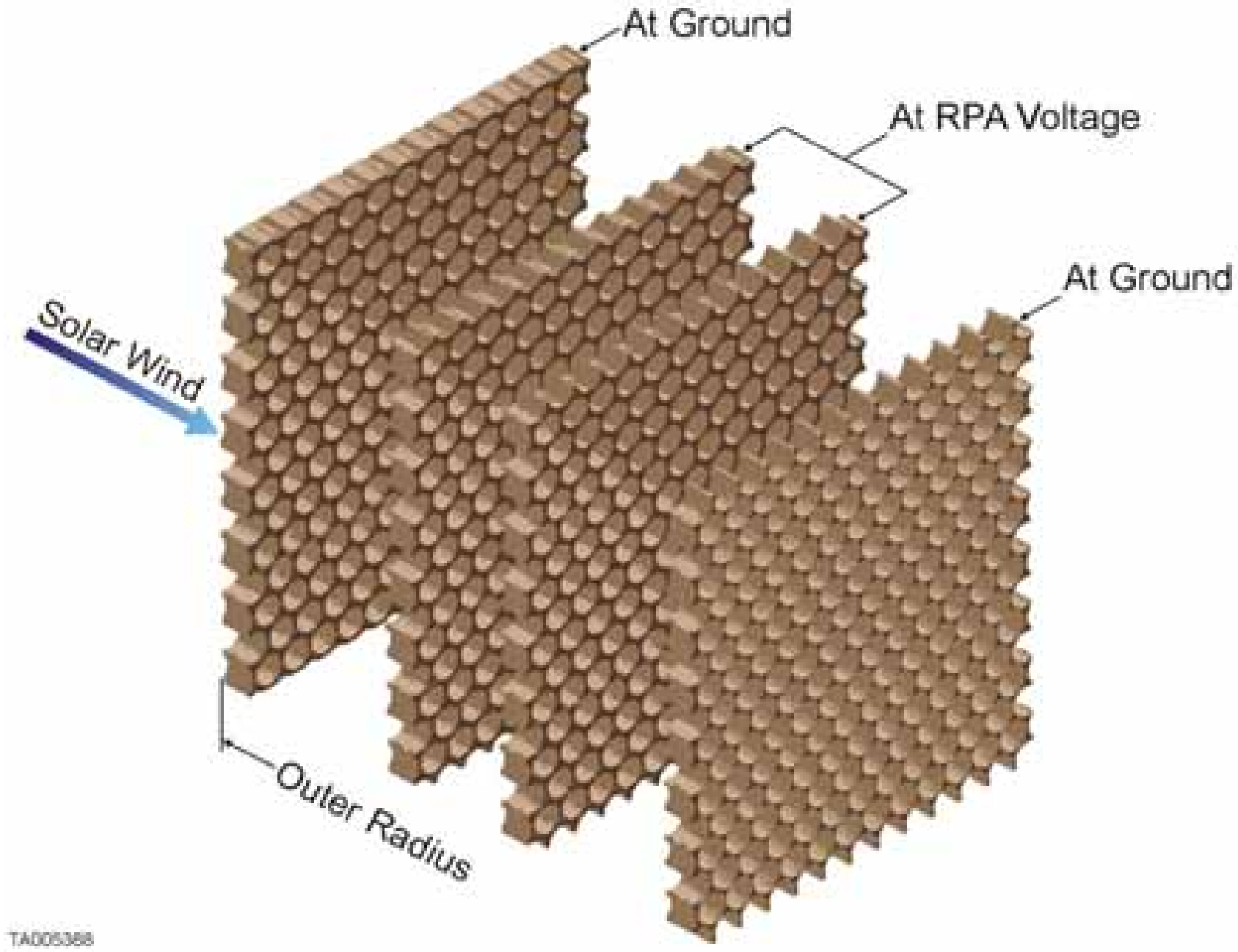

Figure 36. Isometric view of a part of the RPA. The voltage is applied on the two middle grids. The holes are aligned so that their axes are radial. The thickness of the grids and the diameter of the holes are in accordance with the mechanical drawings.

The ions are flown through the RPA grids, and the Cartesian coordinates and velocity-vector components are recorded at the exit of the RPA for those ions that make it through. Those ions are then flown inside the instrument model described in Section 4.2.1, and an ion that makes it to the carbon foil is considered to be detected.

In the following discussion, we present results using a 1 keV/q parallel beam, which is uniformly distributed over the aperture of the instrument at normal incidence (0° in azimuth and elevation). Figure 37 shows a transmission curve as a function of the RPA voltage for an ESA voltage of -524 V, with the ordinate in arbitrary units. The red curve depicts the results of the RPA model and is compared to the calibration data under the same conditions (black curve).

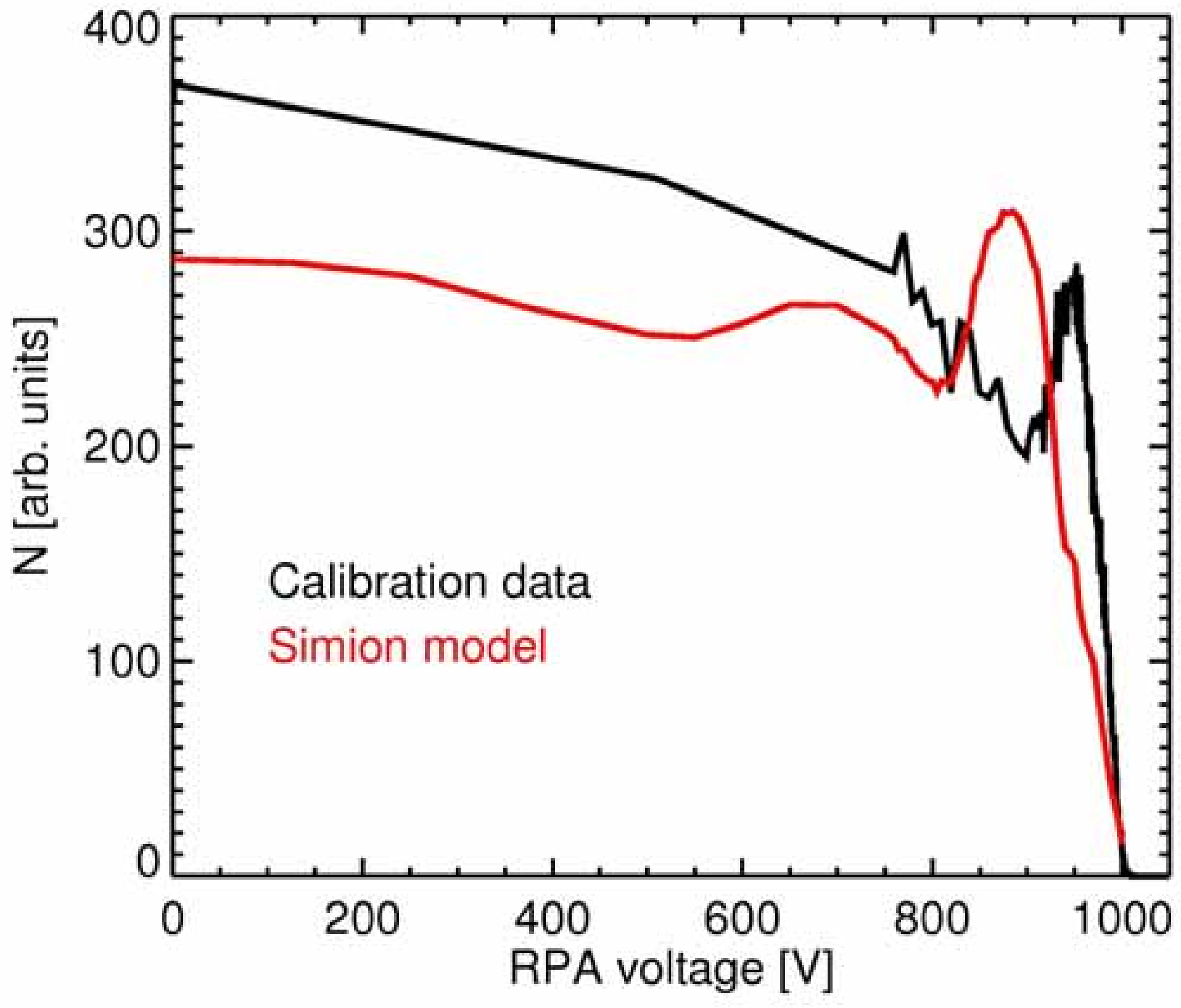

Figure 37. Relative transmission function as a function of the RPA voltage for a 1 keV/e ion beam and an ESA voltage of –524 V.

The RPA does not give a sharp cut-off at 1 kV because each hole of the RPA grids acts like a tiny lens, focusing or defocusing the ion trajectories. At a given E/q, the deflection increases with the RPA voltage. Because of the grid thickness, some ions hit the inside of the holes, which can also cause a non-uniform response below the cut-off. Because the model holes are radially aligned, the energy per charge and the RPA voltage have to be within a certain ratio for the trajectory to deviate from a row of holes to another neighboring row. Therefore, transmission varies as a function of RPA voltage for a particular energy per charge. Although the general responses are similar, a discrepancy exists between the calibration data and the simulations. The likely cause of this discrepancy is that the RPA grid holes are not aligned along a radial direction. In addition, the ion beam in calibration entered the instrument with a small angle with respect to the normal of the RPA grids (offset in azimuth and/or elevation angle) and the ion beam had a small divergence (~1.0° at 1 keV/e). These results highlight the importance of using the flight calibration data in concert with the instrument model for understanding the detailed response function.

We presented above the simulation results for one particular case, however the ESA, deflector, and RPA responses were simulated with numerous different initial conditions. The model aids in understanding the instrument response and provides additional interesting information that can be used for the analysis and interpretation of the data.

# 5.0 Science Operations and Data

This section describes the expected fluxes and predicted count rates, instrument operations, and data pipeline. We used the fluxes measured from previous missions and the instrument model described above to estimate the anticipated count rates in the solar wind and at Jupiter. The subsequent operations discussion describes the voltage settings, which vary according to the anticipated count rates.

Finally, the data pipeline section describes the data products and level 2 data analysis.

## 5.1 Fluxes and Predicted Rates

We predict SWAP count rates in the solar wind scaling from fluxes observed by Voyager 2. The proton flux decreases with distance from the Sun (Figure 38) primarily because the density decreases with distance. For our calculations we use a constant speed of 1 keV (~440 km $s^{-1}$). Fit functions for the density, and density standard deviations are listed below; the standard deviation for the speed is based on Figure 6 of Richardson et al., (1996).

$$n_p = 5.22 r^{-1.93} \quad [cm^{-3}] \tag{3a}$$

$$\sigma(n_p) = 7.4 r^{-2.15} \quad [cm^{-3}] \tag{3b}$$

We used the solar wind fluxes from Figure 38 and the instrument model described above to estimate primary and secondary CEM count rates in the solar wind as a function of heliocentric distance. The results are shown in Figure 39. The instrument model includes the instrument effective area and the ESA and RPA response curves. The model also includes the deflector and angular response function (FOV) and assumes a 1 keV beam centered in the ESA passband. The red line at 1 MHz indicates a reasonable maximum count rate for the detectors. Although the count rates are too high in the inner heliosphere with these voltage settings, the RPA and ESA voltages can be adjusted to reduce passband.

As described above, the ESA and RPA can be crossed to provide a reduced energy passband. Figure 40 shows the estimated count rates as a function of the center energy of the ESA ($kV_{ESA}$) passband for a 1 keV beam with a flux that is one standard deviation higher than the average flux at 5 AU. Each curve corresponds to sweeping the RPA and ESA voltages while keeping the RPA voltage to ESA energy ratio fixed. At a ratio of 1.03 the expected count rates at 5AU are reduced to near maximum allowable levels, and for a ratio of 1.05 the expected rates are <$10^5$ Hz. For the last phase of commissioning we used a ratio of 1.05 for safety.

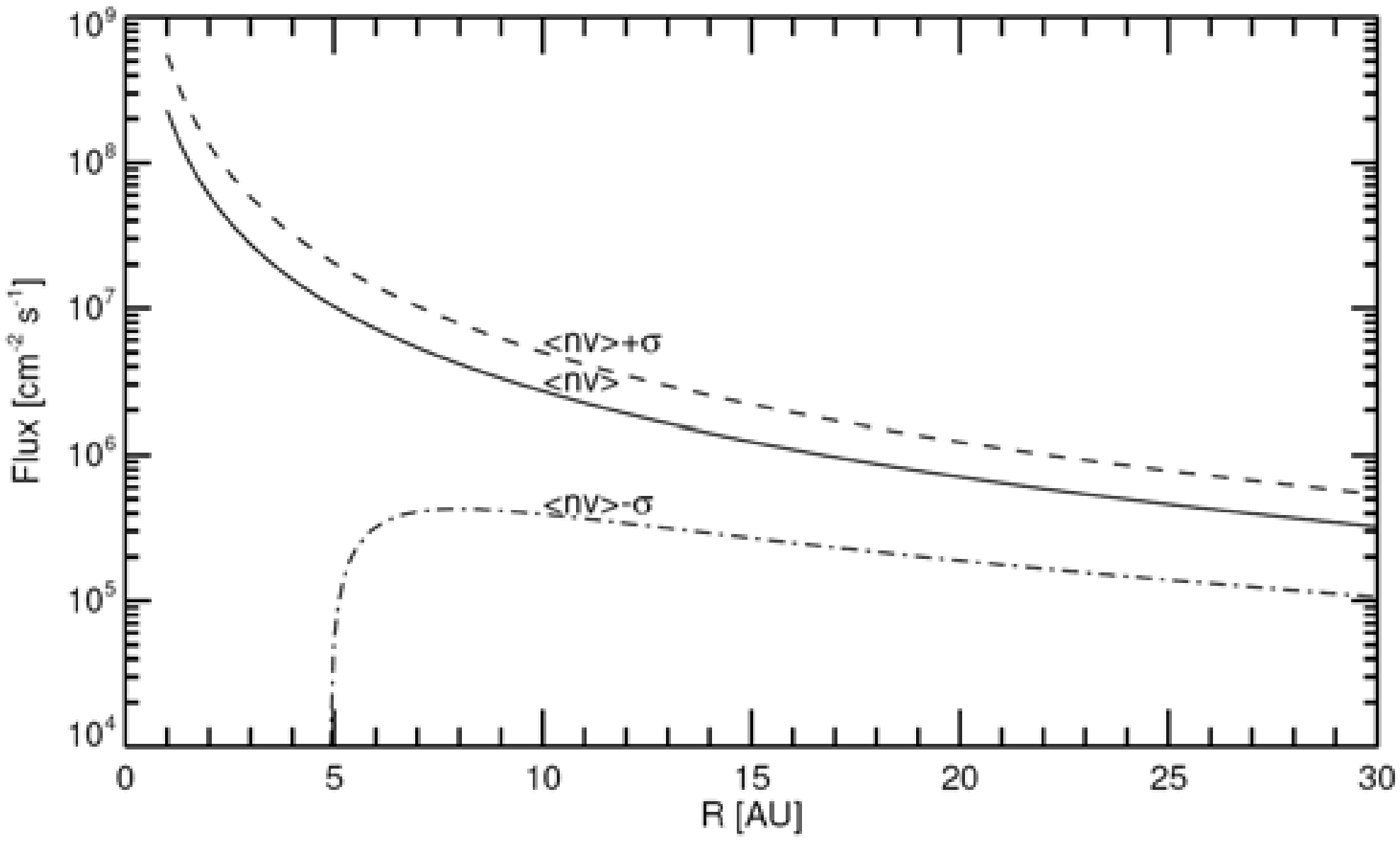

Figure 38. Flux of solar wind protons versus heliocentric distance. The solid curve shows average values, while the other two show ±1 σ probabilities.

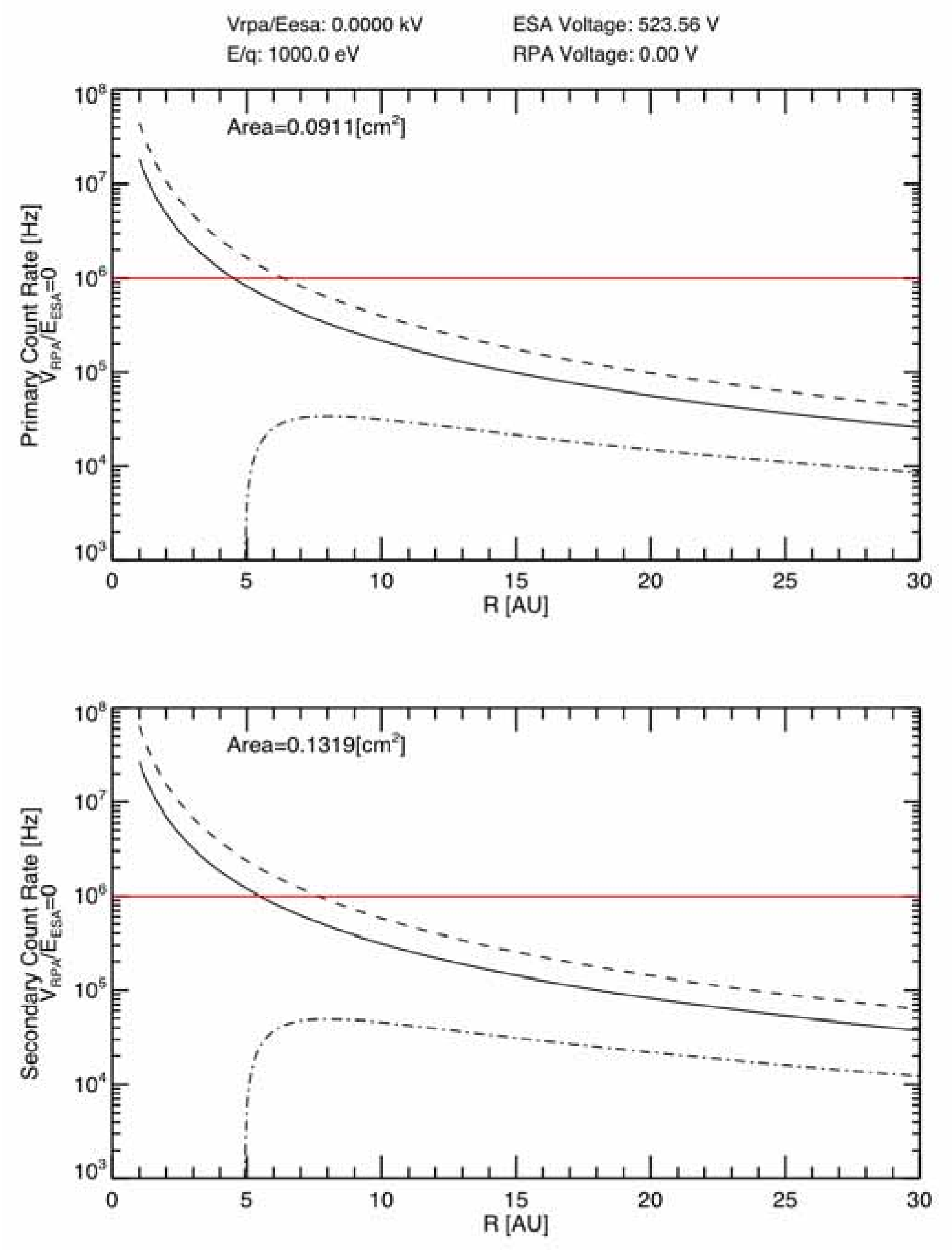

Figure 39. Estimated count rates on the primary (top) and secondary (bottom) CEM as a function of heliocentric distance for the fluxes given in Figure 38.

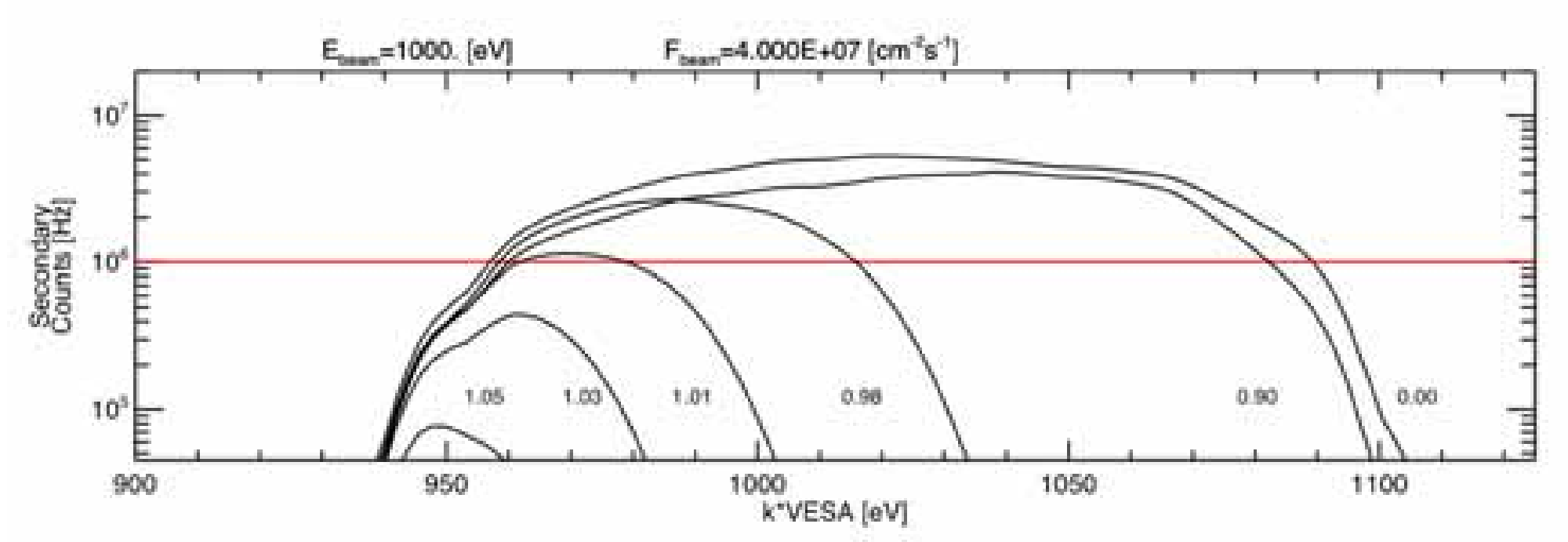

Figure 40. Estimated count rates on the secondary CEM as a function of the center of the ESA bandpass. Each curve corresponds to sweeping ESA and RPA voltages such that the ratio of the RPA voltage to the center energy of the ESA bandpass ($V_{RPA}/E_{ESA}$) is fixed. The flux used is higher than normal for 5 AU.

We also considered the fluxes expected at Jupiter. The fluxes in Jupiter's tail lobes should be to be quite low because of the very low density in these open field regions. This density has been shown to be as low as $10^{-5}$ to $10^{-6}$ $cm^{-3}$ (Gurnett et al., 1981) and direct measurements of these low fluxes by SWAP are unlikely. *Frank et al.* (2002) showed that the density in Jupiter's plasma sheet generally decreases with increasing distance and ranges from 0.4 to .01 $cm^{-3}$ between 20 and 100 $R_J$. Their speed measurements were sparse, but for those same distances they showed the speeds of less than 275 km $s^{-1}$. From these numbers, the maximum flux observed should be about $1\times10^7$ $cm^{-2}s^{-1}$, which is close to the solar wind flux at 5 AU. Voyager 1 and 2 observations of Jupiter's sheath indicate that the flux is less than $2x10^8$ $cm^{-2}$ $s^{-1}$ (Richardson, 2002). Fluxes measured with Cassini at Jupiter also indicate that the sheath flux is less than $2x10^8$ $cm^{-2}$ $s^{-1}$ (*personal communication F. Crary*). This flux is greater than the solar wind flux at 5 AU, but the temperature is quite high in the sheath; thus only a portion of the ion distribution will be observed at a given energy step. For a flux of $2x10^8$ $cm^{-2}$ $s^{-1}$, assuming all the ions occur at one energy step and using an RPA to ESA ratio of 1.05, the rate on the CEMs is about $4\times10^5$ Hz, which is also less than the maximum allowed rate.

## 5.2 Instrument Operations

For the Jupiter phase of the mission, we determined that the radiation levels during the encounter should be low enough to allow SWAP to operate for the entire encounter. SWAP's shielding is similar to Ulysses-SWOOPS, and the *New Horizons* spacecraft does not get as close to Jupiter's radiation belts as Ulysses did. Phillips et al., (1993) showed that Ulysses/SWOOPS operated as close as 16 $R_J$, compared to the New Horizon's closest approach of about 32 $R_J$. Inbound to Jupiter, we anticipate that real-time science mode data will be taken twice per hour, with 20 measurements recorded per hour in Jupiter's tail. The Jupiter tail observations continue until about 100 days after closest approach, which corresponds to about 2200 $R_J$ downstream. While not designed for measuring Jovian ion distributions, we are hopeful that some good new observations will be possible.

Past Jupiter, we plan cruise phase science to study how the solar wind and pickup ions evolve with radial distance from the Sun. For cruise phase real-time data will be recorded at a low rate and summary and histogram data, a moderate rate. In addition, the gain on the CEMs will be measured during each annual check out.

For the *New Horizons* primary mission, we will use SWAP's maximum sensitivity and operate at its highest rate throughout the Pluto and KBO flybys. These observations give a spatial resolution of <3000 km, adequate to determine if Pluto has a bow shock and if so, locate it with sufficient accuracy to make a good measure of the atmospheric escape rate for the likely orientation of the IMF at flyby.

### 5.3 Data Pipeline

SWAP data are provided at three levels in the data pipeline. Level 1 is the raw data decommuted and placed into files. Level 2 involves converting the level 1 data to more physical units, correlating data with thruster firings to look for any enhanced backgrounds, calculating some orbital and pointing information, and producing E/q spectrograms. For level 3, we further process the data to obtain speed, temperature, and relative density.

### 5.3.1 Level 1

Level 1 data are comprised of real-time science, summary, histogram, and housekeeping packets. We decommute and place the raw data into FITS files. Most of the SWAP data consist of binary tables with columns for given instrument parameters and rows for each measurement time. Since the histogram data consist of arrays, it is stored as an image in the FITS file.

We place all the real-time data for a given day into one file. A complete real-time observation consists of 64 real-time packets. In order to make sure that an instrument cycle is not split across two daily files, a parameter in the real-time packet is used, which indicates the beginning and end of the 64-second instrument cycle. We also create daily files for summary data, but the histogram data are handled differently since the histogram arrays are divided into 64 packets. A set of 64 histogram packets are combined and placed into one file.

We divide each FITS file into extensions. The first two real-time data extensions contain header information and the real-time data as a binary table. Additional table extensions contain housekeeping and thruster-firing data. We use this same organization for the summary data. The first two histogram extensions contain the array (image) with the number of samples placed into each bin, and another image holds the array of counts detected in each bin. The housekeeping, thruster, source sequence count (used to check for data gaps) and checksum (used to check for corrupted data) information is contained in additional extensions. For real-time and summary files, the source sequence and checksum information is not stored as separate extensions, but directly inside the tables.

### 5.3.2 Level 2

We also store SWAP level 2 data as binary tables and images in FITS files. The header information from level 1 is duplicated into level 2, since the header contains level 1 processing information and orbit-attitude information for the start time of the each file. The second extension for the real-time and summary level 2 files contains the corresponding data from those packets converted from raw to engineering units, e.g., converting the RPA DAC step number to voltage. The real-time files contain several additional extensions that hold spectrogram information. These extensions consist of the coarse scan ($1^{st}$ 32 steps) 2D spectrogram count arrays, the corresponding 2D error spectrograms, the y-axis labels (energy label) for the spectrograms, and the x-axis time label information. The real-time and summary files have housekeeping, quality, and thruster firing extensions. The histogram level 2 file format mirrors histogram level 1 format, but includes quality flags.

We base the flags that assess the quality of the data on housekeeping data, orbit and attitude, and how well the software is working. Currently, these flags are based on standard operating ranges for housekeeping data, but additional ones can be created later as needed.

We are developing SPICE code to calculate the time in UTC, position and velocity of the spacecraft in a variety of coordinate systems, field of view of the instrument, and the Sun location in instrument coordinates. The boresight of SWAP is nearly aligned with the +Ysc axis; therefore the SWAP frame kernel consists of only two small rotations from spacecraft to instrument coordinates. We are integrating SPICE code into the level 2 code, and an orbit and attitude table extension with all SPICE calculations for each measurement time is planned for the level 2 files. Flags will be created to indicate whether or not the Sun is SWAP's FOV since accurate solar wind properties can only be determined if the bulk of the solar wind beam enters the instrument.

We will produce energy-time spectrograms from real-time science mode data. These count rate spectrograms provide a way to examine the data at high time resolution over the full energy range of the instrument. Such spectrograms have proven useful for analyzing a wide variety of distribution types. They have been used to analyze the solar wind, comets, plasma boundaries, and shocks. As an example, Figure 41 shows a spectrogram used to analyze ion measurements for the comet Borrelly flyby. In this figure it is easy to locate the bow shock and see the slowing of the solar wind due to mass loading.

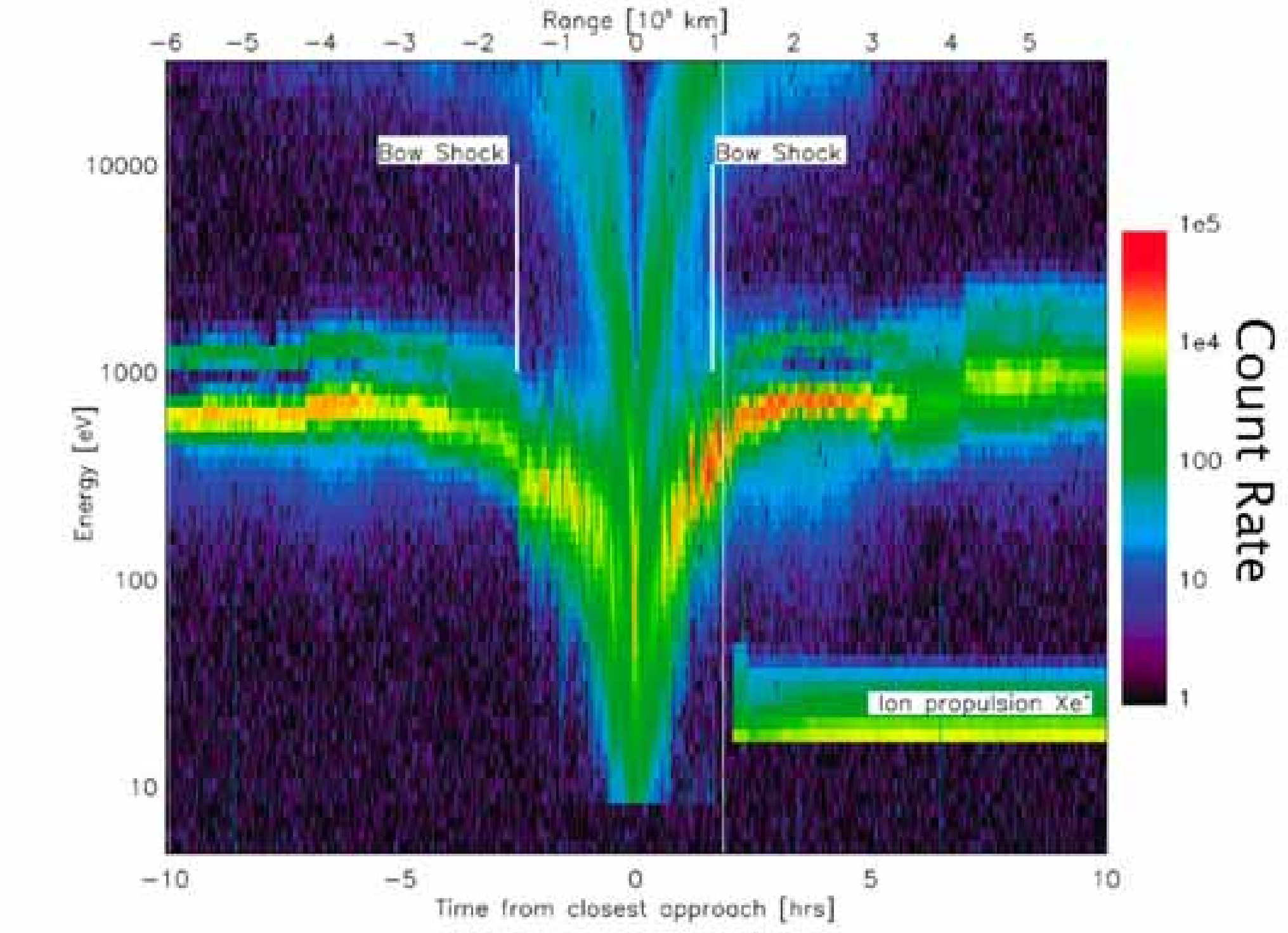

Figure 41. Comet Borrelly flyby E/q-time spectrogram (Figure 1 of Young et al. (2004)

### 5.3.3 Level 3

Speed, temperature, and relative density depend heavily on the calibration, orbit, and attitude information. We have planned these data products for level 3 since the energy distribution (spectrograms) and the orbit and attitude information, both of which are accomplished in level 2 analysis, must be determined first. Speed, temperature, and relative density from the energy distribution/spectrum for the solar wind measurements will be derived during solar wind intervals. When only pickup ions are observed, the energy distribution/spectrum is a more appropriate data product. We will use ESA and RPA response functions, angular response function, instrument solid angle, detector efficiencies, and effective area determined during calibration to establish plasma properties and distributions from the detected count rates. The effective area (Table II) is used for a cold beam such as the solar wind when the entire solar wind beam is in the field of view. Note that the effective area is for a normal incident beam and the transmission varies as a function of angle; therefore, all of the angular dependent calibration functions discussed in section 4 are also required to determine the flux. In hot plasma the geometric factor (Table II) is used to determine the flux in the measured field of view. Precise speeds and temperatures will be obtained from the fine RPA scans. Lab calibration results already indicate the ability to accurately determine the solar wind beam energy and speed with the RPA fine scans. For times when the deflector is operated, we will use the deflector's calibration data to analyze the data.

## 6.0 Early SWAP Observations

New Horizons was launched toward Jupiter and on to Pluto on January 19, 2006. Early operations included commissioning of all of the instruments on board. The SWAP cover doors were opened, low and high voltages turned on, and all aspects of the SWAP instrument were demonstrated to be functioning nominally in space. Because of the closeness to the Sun and high fluxes possible (Figures 38 and 39), we ran SWAP with a very high crossing ratio of 1.05% between the ESA and RPA. This crossing effectively limits the counts to ~1% of what would be otherwise measured for all E/q values less that 2 keV/q; above that value, the RPA is fixed at 2 kV, so the energy passband of the ESA is effectively fully open.

Figure 42 shows an initial color-coded spectrogram of the background-subtracted coincidence measurements of solar wind ions as a function of E/q for January 8-20, 2006, when New Horizons was ~4.9 AU from the Sun (~0.4 AU upstream from Jupiter). The lower trace is produced by solar wind protons, while the upper trace comes from the alpha particles (He++), which travel at approximately the same speed as the protons as thus have twice the E/q. Alpha particle densities in the solar wind are typically only ~4% that of the protons, so SWAP's enhanced sensitivity above 2 keV/q is nearly ideal for making high sensitivity measurements of these ions while still limiting counts due to the protons to safe levels. Solar wind speed is a function E/q, with 1 keV protons corresponding to typical, ~440 km/s solar wind and larger (smaller) E/q representing faster (slower) wind speeds.

A forward interplanetary shock passed over the New Horizons spacecraft at ~1800 on January 11 and a reverse shock passed by at ~1300 on January 14; the speed immediately following the latter shock was in excess of 600 km/s. Such forward-reverse shock pairs, followed by rarefaction regions (slowly decreasing speeds and consequent falling E/q of the proton and alpha beams) are typical of corotation interaction regions (CIRs) in the solar wind at these distances [Hundhausen and Gosling, 1976; Smith and Wolfe, 1976]. These and the continuing upstream measurements of the solar wind are critical for understanding the external plasma environment as New Horizons entered the Jovian magnetosphere late in February 2007.

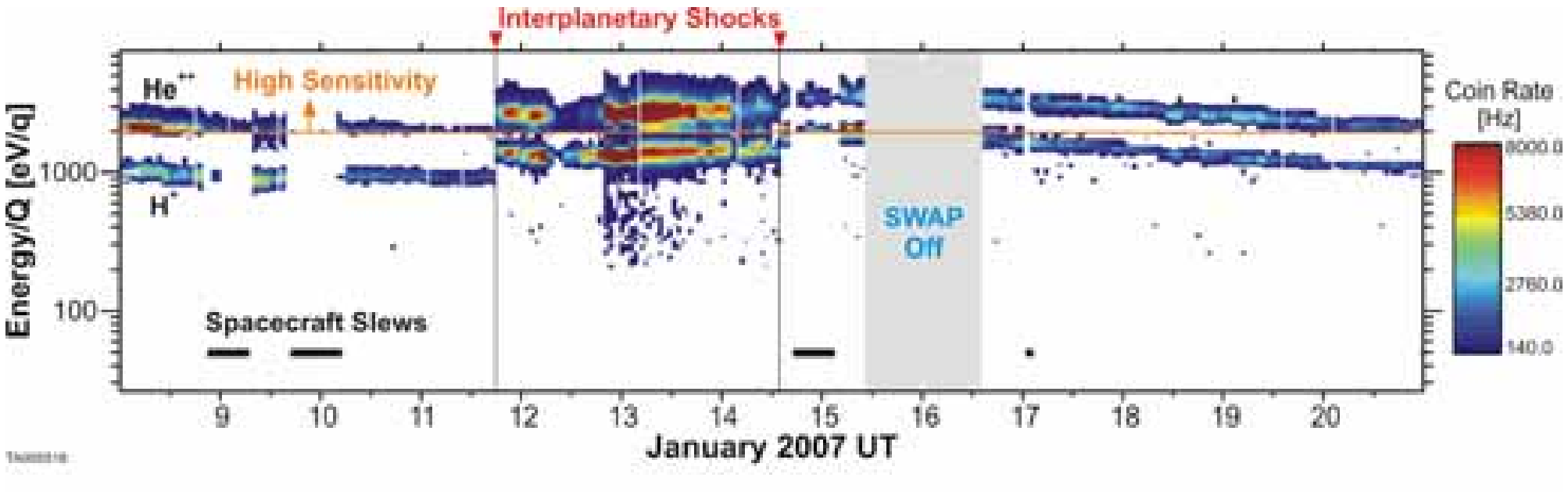

Figure 42. Early observations from SWAP in the solar wind at a heliocentric distance of ~4.9 AU. The upper (lower) trace shows solar wind alpha particles (protons) with the sensitivity significantly reduced below 2 keV/q for safety reasons (see text). Times when the proton and alpha particle beams are greatly reduced or disappear correspond to times when the instrument was temporarily off or the spacecraft pointed SWAP's aperture away from the Sun. The observed stream structure and forward reverse shock pairs is consistent with a CIR in the solar wind at these heliocentric distances.

# 7.0 Conclusions

The SWAP instrument on *New Horizons* will provide the best measurement of the global atmospheric escape rate from Pluto. This escape rate determines the type of interaction that Pluto has with the solar wind. For a large escape rate, with many ions leaving the atmosphere, the interaction should be comet-like, where ionization of atmospheric material mass loads the solar wind, causing slowing of the overall bulk flow of the solar wind and draping of the IMF. For lower escape rates, the interaction is likely more like the interaction that the solar and wind has with Venus. There, the interaction size is much smaller and the solar wind interacts directly with ions closely bound to the planetary atmosphere, causing a much smaller region of draped magnetic field. In either case, SWAP should also be able to identify and locate a bow shock, if one exists.

In addition to measuring the overall escape rate from Pluto's atmosphere and allowing us to develop an understanding of the solar wind interaction with Pluto for the first time, SWAP should also be able to measure hydrogen pickup ions around Pluto. In addition, in the unlikely event that Pluto has any significant internal or remnant magnetic field, SWAP should measure the effects that such a field would have on the solar wind plasma interaction. Finally, *en route* to Pluto and Kuiper belt objects beyond, SWAP should make unique observation of the variation of heliospheric pickup hydrogen ions and sample the Jovian magnetosphere.

The SWAP instrument was designed to meet the unique and challenging requirements of measuring the solar wind and its interaction with Pluto out at 30 AU, from a spacecraft that scans its orientation over large angles in order to image Pluto and Charon with body-mounted cameras as it flies by. In addition, because observations of the plasma interaction is not primary science for the mission, resources for SWAP were highly constrained and we had to remain focused on developing a simple instrument optimized only for measurements of the solar wind interaction with Pluto, while working within the detailed constraints of the *New Horizons* spacecraft and mission plan.

To satisfy all of these requirements and constraints, we designed the SWAP instrument to be cylindrical symmetric over 270° of rotation angle, and combined an RPA that acts a high pass filter, a deflector ring, and an ESA. Different combinations of the RPA and ESA produce a variable passband and deconvolution of measurements made while micro-stepping the RPA allows the measurement of very small changes in the solar wind speed. Ions that pass through the SWAP electro-optics are then measured in a coincidence detector section that both provides redundancy for making the measurements and suppresses the background for noise generated by anything other than real ions transiting through the instrument. The SWAP instrument has been well characterized by both ground calibrations and a detailed instrument model. SWAP has been successfully turned on in space and will start making more routine science observations once it is further out from the Sun.

SWAP will make unique and exciting measurements *en route* to, at Pluto, and beyond. SWAP will measure the slowing of the solar wind flow, mass loading of the solar wind plasma, the escape rate of atmospheric material from Pluto and the fascinating solar wind interaction with Pluto for the first time.

## Acknowledgments

We gratefully acknowledge all of the scientists, engineers, technicians, and support personnel who have made the SWAP instrument and *New Horizons* mission a reality! While it is not possible to list everyone who has contributed to the SWAP instrument, a partial list includes: Maureen Ahr, Barry Eggington, George Geleta, Pat Gonzales, Evan Guy, Helen Hart, Marc Johnson, Matt Maple, Greg Palacios, Robert Rendon, Syrrel Rogillio, Carlos Urdiales, Kenny Whitton, and Mike Young. We also gratefully acknowledge the work of Wendy Mills, who helped assemble and edit this paper and two reviewers and Alan Stern, who provided helpful comments on this manuscript. This work was carried out under the *New Horizons* mission, which is part of NASA's New Frontiers program.

## 8.0 References

Bagenal, F. and McNutt, R. L.: 1989, Pluto's interaction with the solar wind, Geophys. Res. Lett. 16, 1229-1232.

Bagenal, F., Cravens, T. E., Luhmann, J. G., McNutt, R. L., and Cheng, A. F.: 1997, Pluto and Charon: Pluto's interaction with the solar wind, University of Arizona Press, Tucson, Arizona, p. 523.

Baranov, V. B. and Malama, Y. G.: 1995, Effect of local interstellar medium hydrogen fractional ionization on the distant solar wind and interference region, J. Geophys. Res. 100 (A8), 14755-14761.

Bieber, J. W., Wanner, W., and Matthaeus, W. H.: 1996, Dominant two-dimensional solar wind turbulence with implications for cosmic ray transport, J. Geophys. Res. 101, 2511.

Bieber, J. W., Matthaeus, W. H., and Smith, C. W.: 1994, Proton and electron mean free paths: The palmer consensus revisited, Astrophys. J. 420, 294.

Blanco-Cano, X., Omidi, N., and Russell, C. T.: 2003, Hybrid simulations of solar wind interaction with magnetized asteroids: Comparison with Galileo observations near Gaspra and Ida, J. Geophys. Res. 108, 1216.

Bogdanov, A., Sauer, K., Baumgärtel, K., and Srivastava, K.: 1996, Plasma

structures at weakly outgassing comets-results from bi-ion fluid analysis, Planet. Space Sci. 44, 519-528.

Collier, M. R.: 1993, On generating Kappa-like distribution functions using velocity space Levy flights, Geophys. Res. Lett. 20, 1531-1534.

Dahl, D. A.: 2000, SIMION for the personal computer in reflection, Int. J. Mass Spectrom. 200, 3-25.

Delamere, P.A. and Bagenal, F.: 2004, Pluto's kinetic interaction with the solar wind, Geophys. Res. Lett. 31, L04807.

Elliot, J. L., Dunham, E. W., Bosh, A. S., Sliven, S. M., Young, L. A., Wasserman, L. W., and Millis, R. L.: 1989, Pluto's atmosphere, Icarus 77, 148.

Elliot., J. L., Ates, A., Babcock, B. A., Bosh, A. S., Buie, M. W., Clancy, K. B., Dunham, E. W., Eikenberry, S. S., Hall, D. T., Kern, S. D., Leggett, S. K., Levine, S. E., Moon, D. S., Olkin, C. B., Osip, D. J., Pasachoff, J. M., Penprase, B. E., Person, M. J., Qu, S., Rayner, J. T., Roberts, L. C., Salyk, C. V., Souza, S. P., Stone, R. C., Taylor, B. W., Tholen, D. J., Thomas-Osip, J. E., Ticehurst, D. R., Wasserman, L. H.: 2003, The recent expansion of Pluto's atmosphere, Nature 424, 165.

Fahr, H. J.: 1971, The interplanetary hydrogen cone and its solar cycle variations, Astron. Astrophys. 14, 263.

Fisk, L. A., Gloeckler, G., Zurbuchen, T. H., and Schwadron, N. A.: 2000, Ubiquitous statistical acceleration in solar wind, in R. A. Mewaldt, J. R. Jokipii, M. A Lee, E. Möbius, and T.H. Zurbuchen (editors), Acceleration and Transport of Energetic Particles Observed in the Heliosphere, American Institute of Physics, Melville, New York, pp. 229-233.

Fisk, L. A., Schwadron, N. A., and Gloeckler, G.: 1997, Implications of fluctuations in the distribution functions of interstellar pick-up ions for the scattering of low rigidity particles, Geophys. Res. Lett. 24, 93.

Frank L. A., Paterson, W. R., and Khurana, K. K.: 2002, Observations of thermal plasmas in Jupiter's magnetotail, J. Geophys. Res. 107 (A1), 1003.

Galeev, A. A., Cravens, T. E., and Gombosi, T. I.: 1985, Solar wind stagnation near comets, Astrophys. J. 289, 807.

Geiss, J., Gloeckler, G., Mall, U., von Steiger, R., Galvin, A. B., and Ogilvie, K. W.: 1994, Interstellar oxygen, nitrogen, and neon in the heliosphere, Astron. Astrophys. 282, 924.

Geiss, J., Gloeckler, G., von Steiger, R., Balsiger, H., Fisk, L. A., Galvin, A. B., Ipavich, F. M., Livi, S., McKenzie, J. F., Ogilvie, K. W., and Wilken, B.: 1995, The southern high speed stream: Results from the SWICS instrument on Ulysses, Science 268, 1033-1036.

Gloeckler, G., Schwadron, N. A., Fisk, L. A., and Geiss, J.: 1995, Weak pitch angle scattering of few mv rigidity ions from measurements of anisotropies in the distribution function of interstellar pickup h+, Geophys. Res. Lett. 22, 2665.

Gloeckler, G., Geiss, J., Balsiger, H., Bedini, P., Cain, J. C., Fisher, J., Fisk, L. A., Galvin, A. B., Gliem, F., Hamilton, D. C.: 1992, The solar wind ion composition spectrometer, Astron. Astrophys. Suppl. Ser. 92, 267-289.

Gloeckler, G. and Geiss, J.: 1998, Interstellar and inner source pickup ions observed with SWICS on Ulysses, Space Sci. Rev. 86, 127.

Gloeckler, G., L., Fisk, A., and Geiss, J.: 1997, Anomalously small magnetic field in the local interstellar cloud, Nature 386, 374-377.

Gloeckler, G., Geiss, J., Roelof, E. C., Fisk, L. A., Ipavich, F. M., Ogilvie, K. W., Lanzerotti, L. J., von Steiger, R., and Wilken, B.: 1994, Acceleration of interstellar pickup ions in the disturbed solar wind observed on Ulysses, J. Geophys. Res. 99, 17637.

Gloeckler, G.: 1999, Observation of injection and pre-acceleration processes in the slow solar wind, Space Sci. Rev. 89, 91-104.

Gurnett, D. A., Scarf, F. L., Kurth, W. S., Shaw, R. R., and Poynter, R. L.: 1981, Determination of Jupiter's electron density profile from plasma wave observations, J. Geophys. Res. 86, 8199-8212.

Hansen, C.J. and Paige, D.A.: 1996, Seasonal Nitrogen cycles on Pluto, Icarus 120 (2), 247-265.

Harnett, E. M., Winglee, R. M., and Delamere, P. A.: 2005 Three-dimensional multi-fluid simulations of Pluto’s magnetosphere: A comparison to 3D hybrid simulations, Geophys. Res. Lett. 32, L19104.

Hopcroft, M. W. and Chapman, S. C.: 2001, 2D hybrid simulations of the solar wind interaction with a small scale comet in high Mach number flows,

Geophys. Res. Lett. 28, 1115.

Hubbard, W. B., Hunten, D. M., Dieters, S. W., Hill, K. M., and Watson, R. D.: 1988, Occultation evidence for an atmosphere of Pluto, Nature 336, 452.

Hundhausen, A. J. and Gosling, J. T.: 1976, Solar wind structure at large heliocentric distances: An interpretation of Pioneer 10 observations, J. Geophys. Res. 81, 1436.

Isenberg, P. A.: 1997, A hemispherical model of anisotropic interstellar pickup ions, J. Geophys. Res. 102, 4719.

Isenberg, P. A.: 2005, Turbulence-driven solar wind heating and energization of pickup protons in the outer heliosphere, Astrophys. J. 623, 502-510.

Ip, W., Kopp, A., Lara, L. M., and Rodrigo, R.: 2000, Pluto's ionospheric models and solar wind interaction, Adv. Space Res. 26, 1559.

Izmodenov, V. V., Lallement, R., and Geiss, J.: 1999, Interstellar oxygen in the heliospheric interface: influence of electron impact ionization, Astron. Astrophys. 344, 317-321.

Kecskemety, K., and Cravens, T. E.: 1993, Pick-up ions at Pluto, Geophys. Res. Lett. 20, 543-546.

Krasnopolsky, V. A.: 1999, Hydrodynamic flow of $N_2$ from Pluto, J. Geophys. Res. 104, 5955–5962.

Krasnopolsky, V. A. and Cruikshank, D. P.: 1999, Photochemistry of Pluto's atmosphere and ionosphere near perihelion, J. Geophys. Res. 104, 21979.

Lee, M. A. and Ip, W.: 1987, Hydromagnetic wave excitation by ionised interstellar hydrogen and helium in the solar wind, J. Geophys. Res. 92, 11041.

Lipatov, A. S., Motschmann, U., and Bagdonat, T.: 2002, 3D hybrid simulations of the interaction of the solar wind with a weak comet, Planet. Space Sci. 50, 403-411.

Luhmann, J. G., Russell, C. T., Schwingenschuh, K., and Eroshenko, E.: 1991, A comparison of induced magnetotails of planetary bodies - Venus, Mars, and Titan, J. Geophys. Res. 96, 11199.

Matthaeus, W. H., Goldstein, M. L., and Roberts, D. A.: 1990, Evidence for the presence of quasi-two-dimensional nearly incompressible fluctuations in the solar wind, J. Geophys. Res. 95, 20673.

Matthaeus, W. H., Zank, G. P., Smith, C. W., and Oughton, S.: 1999, Turbulence, Spatial Transport, and Heating of the Solar Wind, Phys. Rev. Lett. 82, 3444-3447.

McComas, D. J., Schwadron, N. A., Crary, F. J., Elliott, H. A., Young, D. T., Gosling, J. T., Thomsen, M. F., Sittler, E., Berthelier, J. J., Szego, K., and Coates, A. J.: 2004a, The interstellar hydrogen shadow: Observations of interstellar pickup ions beyond Jupiter, J. Geophys. Res. 109, A02104.

McComas, D. J., Allegrini, F., Pollock, C. J., Funsten, H. O., Ritzau, S., and Gloeckler, G.: 2004b, Ultra-thin (~10 nm) carbon foils in space instrumentation, Rev. Sci. Instrumen. 75 (11), 4863-4870.

McNutt, R. L.: 1989, Models of Pluto’s upper atmosphere, Geophys. Res. Lett. 16, 1225-1228.

McNutt, et al.: 2007, Energetic Particle Spectroscopy Aboard New Horizons, SSR, this issue.

Möbius, E., Hovestadt, D., Klecker, B., Scholer, M., and Gloeckler, G.: 1985, Direct observation of he(+) pick-up ions of interstellar origin in the solar wind, Nature 318, 426-429.

Omidi, N., Blanco-Cano, X., Russell, C. T., Karimabadi, H., and Acuna, M.: 2002, Hybrid simulations of solar wind interaction with magnetized asteroids: General characteristics, J. Geophys. Res. 107, 1487.

Phillips, J. L., Bame, S. J., Thomsen, M. F., Goldstein, B. E., and Smith, E. J.: 1993, Ulysses plasma observations in the Jovian magnetosheath, J. Geophys. Res., 98 (A12), 21189-21202.

Richardson, J. D.: 2002, The magnetosheaths of the outer planets, Planet. Space Sci. 50, 503-517.

Richardson, J. D., Belcher, J. W., Lazarus, A. J., Paularena, K. I., and Gazis, P. R.: 1996, Statistical properties of solar wind, in D. Winterhalter, J. T. Gosling, S. R. Habbal, W. S. Kurth, and M. Neugebauer (editors), Proceedings of the Eighth International Solar Wind Conference, American Institute of Physics, Melville, New York, pp. 483-486.

Sauer, K., Lipatov, A., Baumgärtel, K., and Dubinin, E.: 1997, Solar wind-Pluto interaction revised, Adv. Space Res. 20, 295.

Schwadron, N. A., Geiss, J., Fisk, L. A., Gloeckler, G., Zurbuchen, T. H., and von Steiger, R.: 2000, Inner source distributions: Theoretical interpretation, implications, and evidence for inner source protons, J. Geophys. Res. 105 (A4), 7465.

Schwadron, N. A., Fisk, L. A., and Gloeckler, G.: 1996, Statistical acceleration of interstellar pick-up ions in co-rotating interaction regions, Geophys. Res. Lett. 23, 2871-2874.

Schwadron, N. A.: 1998, A model for pickup ion transport in the heliosphere in the limit of uniform hemispheric distributions, J. Geophys. Res. 103, 20643-20650.

Schwadron, N. A., Combi, M., Huebner, W., and McComas, D. J.: 2002, The outer source of pickup ions and anomalous cosmic rays, Geophys. Res. Lett. 29, 1993.

Smith, C. W., Isenberg, P. A., Matthaeus, W. H., and Richardson, J. D.: 2006, Turbulent heating of the solar wind by newborn interstellar pickup protons, Astrophys. J. 638, 508-517.

Smith, E. J. and Wolfe, J. H.: 1976, Observations of interaction regions and corotating shocks between one and five AU: Pioneers 10 and 11, Geophys. Res. Lett. 3, 137.

Stern, A. S., Trafton, L. M., and Gladstone, G. R.: 1988, Why is Pluto bright? Implications of the albedo and lightcurve behavior of Pluto, Icarus 75, 485-498.

Stern, et al.: 2007, New Horizons Mission,SSR, this issue,.

Thomas, G. E.: 1978, The interstellar wind and its influence on the interplanetary environment, Annu. Rev. Earth Planet. Sci. 6, 173.

Tian, F. and Toon, O.B.: 2005, Hydrodynamic escape of nitrogen from Pluto, Geophys. Res. Lett. 32, L18201.

Vasyliunas, V.M.: 1968, A survey of low-energy electrons in the evening sector of the magnetosphere with OGO 1 and OGO 3, J. Geophys. Res. 73, 2839-2884.

Vasyliunas, V. M. and Siscoe, G. L.: 1976, On the flux and the energy spectrum of interstellar ions in the solar system, J. Geophys. Res. 81 (7), 1247-1252.

Wang, Z. and Kivelson, M. G.: 1996, Asteroid interaction with solar wind, J. Geophys. Res. 101, 24479-24494.

Wu, F. M. and Judge, D. L.: 1979, Temperature and flow velocity of the interplanetary gases along solar radii, Astrophys. J. 231 (2), 594-605.

Young, D. T., Crary, F. J., Nordholt, J. E., Bagenal, F., Boice, D., Burch, J. L., Eviatar, A., Goldstein, R., Hanley, J. J., Lawrence, D. J., McComas, D. J., Meier, R., Reisenfeld, D., Sauer, K., and Wiens, R. C.: 2004, Solar wind interactions with comet 19P/Borrelly, Icarus 167, 80-88.

Zank, G. P., Matthaeus, W. H., Bieber, J. W., and Moraal H.: 1998, The radial and latitudinal dependence of the cosmic ray diffusion tensor in the heliosphere, J. Geophys. Res. 103, 2085.

Zank, G. P., Pauls, H. L., Cairnes, I. H., and Webb, G. M.: 1996, Interstellar pickup ions and quasi-perpendicular shocks: Implications for the termination shock and interplanetary shocks, J. Geophys. Res. 101, 457.